\documentclass[acmsmall, screen, final, nonacm]{acmart}

\PassOptionsToPackage{nosumlimits,nointlimits}{amsmath}
\PassOptionsToPackage{colorlinks,linkcolor={blue},citecolor={blue},urlcolor={red},breaklinks=true,final}{hyperref}

\usepackage[utf8]{inputenc}

\usepackage{etex}

\usepackage{microtype}

\usepackage[notcite,notref]{showkeys}

\usepackage{rotating}

\usepackage[normalem]{ulem}			    	    %
\usepackage{proof}					    %

\usepackage{stmaryrd}
\usepackage{bm}						    %
\usepackage{ushort}				        %
\usepackage{slashed}				    %

\usepackage{pifont}

\usepackage{textcomp}

\usepackage{optparams}				    %
\usepackage{xspace}
\usepackage{needspace}

\usepackage{enumitem}
\setlist[enumerate,1]{label=(\arabic*),font=\normalfont,align=left,leftmargin=0pt,labelindent=0pt,listparindent=\parindent,labelwidth=0pt,itemindent=!,topsep=3pt,parsep=0pt,itemsep=2pt,start=1}
\setlist[enumerate,2]{label=(\alph*),font=\normalfont,labelindent=*,leftmargin=*,start=1}
\newlist{citemize}{itemize}{1}
\setlist[citemize]{label=\labelitemi,wide} %

\newlist{cenumerate}{enumerate}{1}
\setlist[cenumerate,1]{label=\arabic*.~,ref={\arabic*},wide} %

\usepackage{ifdraft}  				%

\makeatletter\@draftfalse\makeatother %

\usepackage{ifthen}
\newboolean{full}
\setboolean{full}{true}

\numberwithin{equation}{section}

\usepackage[layout=footnote,final]{fixme}

\newcommand{\SG}[1]{\sgnote{#1}}

\FXRegisterAuthor{sg}{asg}{SG}%
\FXRegisterAuthor{sm}{asm}{SM}%
\FXRegisterAuthor{as}{aas}{AS}%

\newcommand\Exp[2]{\mathsf{E}_{{#1},{#2}}}
\newcommand\AExp[2]{\mathsf{A}_{{#1},{#2}}}

\usepackage{tikz}
\usetikzlibrary{arrows,shapes}
\usetikzlibrary{automata,positioning}
\usepackage{tikz-cd}			%

\tikzset{
  commutative diagrams/.cd,
  arrow style=tikz,
  diagrams={>=stealth},
  row sep=large, 
  column sep = huge
}

\pgfmathsetmacro\sprayRadius{3pt}
\pgfmathsetmacro\sprayPeriod{2cm}

\ifdraft{
 \usepackage{todos}
}{
 \newcommand{\todo}[1]{}
}

\DeclareFontFamily{OT1}{pzc}{}
\DeclareFontShape{OT1}{pzc}{m}{it}{<-> s * [1.200] pzcmi7t}{}

\DeclareMathAlphabet{\mathpzc}{OT1}{pzc}{m}{it}

\newcommand{\M}{\mathpzc{m}}
\newcommand{\takeout}[1]{}

\newcommand{\CA}{{\Cls A}}

\newcommand{\CE}{{\Cls E}}

\newcommand{\CL}{{\Cls L}}

\newcommand{\CR}{{\Cls R}}

\newcommand{\CT}{{\Cls T}}

\newcommand{\CD}{{\Cls D}}

\newcommand{\Lfun}{L}

\newcommand{\BBP}{\mathbb{P}}
\newcommand{\BBR}{\mathbb{R}}
\newcommand{\BBT}{\mathbb{T}}
\newcommand{\BBS}{\mathbb{S}}

\newcommand{\BBM}{\mathbb{M}}

\newcommand{\Set}{\Cat{Set}}

\newcommand{\Cat}{\mathbf}
\newcommand{\Cls}{\mathcal}

\newcommand{\Hom}{\mathsf{Hom}}

\newcommand{\id}{\operatorname{\mathsf{id}}}

\newcommand{\PSet}{{\mathcal P}}
\newcommand{\PFin}{{\mathcal P}_{\omega}}

\newcommand{\True}{\top}
\newcommand{\False}{\bot}

\newcommand{\impl}{\Rightarrow}

\newcommand{\eps}{\operatorname\epsilon}

\newcommand{\tensor}{\otimes}
\newcommand{\tpair}[2]{{#1\wideparen{\;}\kern-1pt #2}}
\newcommand{\argument}{\operatorname{-\!-}}%
\newcommand{\wave}[1]{\widetilde{#1}}

\def\case{}
\newcommand{\CASE}{\operatorname{\sf case}}
\newcommand{\OF}{\operatorname{\sf of}}
\renewcommand{\case}[3]{\CASE\kern1.2pt #1\kern1.2pt \OF\kern1.2pt #2;\kern1.2pt #3}

\newcommand{\caseOne}[2]{\CASE\kern1.2pt #1\kern1.2pt \OF\kern1.2pt #2}

\newcommand{\DO}{\operatorname{\sf do}}

\newcommand{\letTerm}[2]{\DO\kern1.2pt#1; #2}
\newcommand{\leteq}{\gets}

\newcommand{\letTermO}[2]{\DO_{\nu}\kern1.2pt#1; #2}
\newcommand{\LOOP}{\operatorname{\sf loop}}
\newcommand{\THEN}{\operatorname{\sf then}}

\makeatletter
\long\def\loopTerm@[#1][#2][#3]#4#5#6{
\DO\kern1.2pt#4 #1 \LOOP #3 #5 #2\THEN #6
}

\newcommand{\loopTerm}{
\optparams{\loopTerm@}{[;][\};][\{]}
}
\makeatother

\newcommand{\LET}{\operatorname{\sf init}}
\newcommand{\letLoop}[2]{\LET #1\kern1.2pt\LOOP\kern1.2pt\{#2\}}
\newcommand{\LetLoop}[2]{\LET #1\kern1.2pt\LOOP\kern1.2pt\bigl\{#2\bigr\}}

\newcommand{\IF}{\operatorname{\sf if}}
\newcommand{\ifTerm}[3]{\IF #1\kern2.2pt {\sf then}\kern1.2pt #2\kern2.2pt {\sf else}\kern2.2pt #3}
\newcommand{\ifTermO}[3]{\IF_{\nu} #1\kern2.2pt {\sf then}\kern2.2pt #2\kern2.2pt {\sf else}\kern2.2pt #3}

\newcommand{\WHILE}{\operatorname{\sf while}}
\newcommand{\whileTerm}[3]{\LET #1\kern1.2pt\WHILE\kern1.2pt #2 \kern2.2pt{\sf do}\kern2.2pt #3}
\newcommand{\whileTermS}[2]{\WHILE\kern1.2pt #1 \kern2.2pt{\sf do}\kern2.2pt #2}

\newcommand{\SEQ}{\operatorname{\sf seq}}
\newcommand{\seqTerm}[2]{\SEQ_{\nu}\kern1.2pt#1; #2}

\newcommand{\letin}[2]{\operatorname{\sf let}\kern2.2pt #1\kern2.2pt \operatorname{\sf in}\kern1.2pt #2}
\newcommand{\match}[2]{\operatorname{\sf let}\kern2.2pt #1\kern2.2pt \operatorname{\sf in}\kern1.2pt #2}
\newcommand{\letmon}[3]{\operatorname{\sf let}\kern1.2pt  #1\kern2.2pt \operatorname{\sf by} #2\kern2.2pt \operatorname{\sf in}\kern1.2pt #3}

\newcommand{\lettens}[4]{\operatorname{\sf let}\kern1.2pt  #3\kern2.2pt \operatorname{\sf by}  #1 (#2) \kern2.2pt \operatorname{\sf in}\kern1.2pt #4}

\newcommand{\lsem}{\llbracket}
\newcommand{\rsem}{\rrbracket}
\newcommand{\sem}[1]{\lsem #1 \rsem}

\newcommand{\tsem}[1]{\raisebox{.2pt}{\scalebox{.815}[.95]{$\boldsymbol($}}\kern-1.7pt[ #1 ]\kern-1.7pt\raisebox{.2pt}{\scalebox{.815}[.95]{$\boldsymbol)$}}}

\newcommand{\brks}[1]{\langle #1\rangle}

\newlength{\myboxwidth}
\DeclareMathOperator{\redpar}{{
\declareslashed{}{
\vrule height2pt depth 2pt
\kern1pt
\vrule height2pt depth 2pt
}{0}{0}{\rightarrowtail}\slashed{\rightarrowtail}}}

\let\comma\relax
\newcommand{\comma}{,\linebreak[1]}

\newcommand{\klstar}{\star}  					%

\renewcommand{\emptyset}{\mathop{\slashed0}} %

\newcommand\dash{\nobreakdash-\hspace{0pt}}

\newcommand{\by}[1]{\text{/\!/~#1}}

\def\epito{\twoheadrightarrow}
\def\monoto{\hookrightarrow}

\newcommand{\algstruc}{\alpha^\M}
\newcommand{\algstrucs}{\alpha^{\M_*}}
\newcommand{\algstrucv}{\alpha^{\M_v}}

\def\int{\mathbb{Z}}
\def\nat{\mathbb{N}}
\def\ev{\mathsf{ev}}
\def\blk{\boxtimes}
\def\N{\mathsf{N}}
\def\L{\mathsf{L}}
\def\R{\mathsf{R}}

\newcommand{\tmread}{\mathit{rd}}
\newcommand{\tmwrite}{\mathit{wr}}
\newcommand{\tmmove}{\mathit{mv}}

\newcommand{\update}{\mathit{update}}
\newcommand{\lookup}{\mathit{lookup}}

\def\defbbname#1{\expandafter\def\csname BB#1\endcsname{{\bm{\mathsf{#1}}}}}
\def\defbbnames#1{\ifx#1\defbbnames\else\defbbname#1\expandafter\defbbnames\fi}
\defbbnames ABCDEFGHIJKLMNOPQRSTUVWXYZ\defbbnames

\def\otr{\mathsf{tr}}
\def\tr{\overline{\mathsf{tr}}}
\def\norm{\mathsf{n}}

\providecommand{\mone}{{\text{\kern.5pt-}\kern-.4pt1}}
\providecommand{\mkah}{{\text{\kern.5pt-}\kern-.4ptk}}
\providecommand{\mmm}{{\text{\kern.5pt-}\kern-.4ptm}}

\theoremstyle{acmtheorem}
\newtheorem{theorem}{Theorem}[section]
\newtheorem{proposition}[theorem]{Proposition}
\newtheorem{lemma}[theorem]{Lemma}
\newtheorem{corollary}[theorem]{Corollary}

\theoremstyle{acmdefinition}
\newtheorem{notation}[theorem]{Notation}
\newtheorem{construction}[theorem]{Construction}
\newtheorem{remark}[theorem]{Remark}

\renewcommand{\qed}{\hfill \ensuremath{\Box}}
\newcommand{\noqed}{\def\qed{}\relax}

\acmVolume{00}
\acmNumber{00}
\acmArticle{00}
\acmYear{2018}
\acmMonth{00}

\setcopyright{acmcopyright}
\acmJournal{TOCL}
\acmYear{2020} 
\acmVolume{1} 
\acmNumber{1} 
\acmArticle{1} 
\acmMonth{1} 
\acmPrice{15.00}
\acmDOI{10.1145/3372880}

\begin{document}

\markboth{S. Goncharov, S. Milius, A. Silva}{Towards a Uniform Theory of Effectful State Machines}

\title{Towards a Uniform Theory of Effectful State Machines} 

\author{Sergey Goncharov}
\email{sergey.goncharov@fau.de}
\orcid{0000-0001-6924-8766}

\author{Stefan Milius}
\email{stefan.milius@fau.de}

\affiliation{%
  \institution{Friedrich-Alexander-Universit\"at Erlangen-N\"urnberg}
}

\author{Alexandra Silva}
\email{alexandra.silva@gmail.com}
\affiliation{%
  \institution{University College London}
}

\begin{abstract}
  Using recent developments in coalgebraic and monad-based semantics,
  we present a uniform study of various notions of machines,
  e.g.~finite state machines, multi-stack machines, Turing machines,
  valence automata, and weighted automata. They are instances of
  Jacobs' notion of a \emph{$\BBT$-automaton}, where $\BBT$ is a
  monad. We show that the generic language semantics for
  $\BBT$-automata correctly instantiates the usual language semantics
  for a number of known classes of machines/languages,
  including~regular, context-free, recursively-enumerable and various
  subclasses of context free languages (e.g.~deterministic and
  real-time ones). Moreover, our approach provides new generic
  techniques for studying the expressivity power of various
  machine-based models.
\end{abstract}


\terms{Theory}


\maketitle

\thispagestyle{empty}

\section{Introduction}
In recent decades much interest has been drawn to studying generic
abstraction devices that not only formally generalize various computation
models and tools, but also help to identify core principles and reasoning
patterns behind them. One example of this kind is given by the notion
of \emph{computational monad}~\cite{Moggi91b}, which made an impact
both on the theory of programming (as an organization tool for
denotational semantics~\cite{FioreMoggiEtAl02,PlotkinPower02}) and on
the practice (e.g.\ being implemented as a programming language
feature of Haskell~\cite{Haskell98} and
F\#~\cite{SymeGraniczEtAl07}). Another pivotal abstraction device is
given by the notion of \emph{coalgebra}, providing a uniform
syntax-independent framework for concurrency theory and observational
semantics of state based systems (see e.g.~\cite{Rutten00}).

\iffull In this paper, we combine the use of monads and coalgebras for
formalizing semantics and behaviors of systems to give a unified
(bialgebraic) perspective of classical automata theory as well as
of some less standard models such as weighted automata and valence
automata.

\else
In this paper, we use previous work on monads and coalgebras to give a combined (bialgebraic) perspective of the classical automata theory as well as of some less standard models such as weighted automata. This does not only provide a unifying framework to study various computational models but also suggests new perspectives for proving expressivity bounds for various kinds of machines in a generic way.
\fi

We base our framework on the notion of \emph{$\BBT$-automaton} whose
original definition goes back to~\cite{Jacobs06}. A $\BBT$-automaton
is a coalgebra of the form
\begin{displaymath}
\M:X\to B\times (TX)^{A},
\end{displaymath}
where $T$ is the functor part of a monad $\BBT$, which we understand
as a mathematical abstraction of a \emph{computational effect} (in the
sense of~\cite{Moggi91b}) happening in conjunction with state
transitions of the automaton, $A$ is the set of inputs, and $B$ is the
set of outputs which is required to be a $\BBT$-algebra. For example,
nondeterminism, viz.~the computational effect of nondeterministic
machines, is modelled by the finite-powerset monad
$\BBT = \PFin$, together with the $\PFin$-algebra $B=\{0,1\}$ for marking 
accepting $(1)$ and rejecting $(0)$ states. Analogously, we show that certain
(nondeterministic) extensions of the pushdown store form the
underlying effect of pushdown automata.

A crucial ingredient of our framework is the \emph{generalized
  powerset construction}~\cite{SilvaBonchiEtAl13}, which serves as a
coalgebraic counterpart of classical Rabin-Scott determinization
algorithm~\cite{RabinScott59} and allows us to provide a generic
\emph{(deterministic) semantics} of $\BBT$-automata. %
By instantiating the operational analysis of computational effects
from~\cite{PlotkinPower02} to our setting we axiomatize relevant
monads and algebras and thus arrive at syntactic fixpoint expressions,
which we dub~\emph{reactive expressions}, representing
$\BBT$-automata. Furthermore, we prove a Kleene-style theorem relating
$\BBT$-automata and the corresponding expressions, thus generalizing
previous work in~\cite{brs_lmcs,bbrs_ic}. This generic correspondence
instantiates to three large classes of machines actively studied in
the literature:
\begin{itemize}
\item state machines over various types of store, as classically
  studied in formal language theory~\cite{RozenbergSalomaa97}; here we
  elaborate in detail push-down stores and their combination with one
  another and with nondeterminism, as well as Turing tapes;
\item valence
  automata~\cite{RenderKambites09,Kambites09,Zetzsche2016c}, capturing
  nondeterministic computations over a store modelled by various
  classes of monoids;
  \item weighted automata~\cite{dkv,sakarovitch_2009}.
\end{itemize}
We also capture systems combining probability and
nondeterminism~\cite{Segala95,SegalaLynch95}, which do not fit any of
the above classes.

A unifying semantic domain of our framework is the set $B^{A^*}$ of
\emph{formal power series}, standardly used in weighted automata
theory (where $B$ is assumed to be a semiring). With $B$ being the
two-element set $\{0,1\}$, $B^{A^*}$ is isomorphic to the set of all formal
languages over $A$, which is the semantic domain for finite state
automata. In the case of \emph{stack $\BBT$-automata}, i.e.\
$\BBT$ models a pushdown store,~$B$ consists of certain predicates in
$2^{\Gamma^*}$, where $\Gamma$ denotes the stack alphabet. Hence
formal power series may be identified with certain functions
$\Gamma^* \to 2^{A^*}$, and our semantics assigns to a state of a
given $\BBT$-automaton the function which maps a word
$w \in \Gamma^*$ to the language recognized by the automaton with
initial stack content $w$. Analogous considerations apply to
$\BBT$-automata where~$\BBT$ models a Turing tape. Furthermore, note that
most textbooks (e.g.~\cite{HopcroftMotwani}) define a Turing machine
with a single tape both for performing computations and for
communicating the data. However, it is important in our approach to
delimit the \emph{reactive} and \emph{computational} parts of a
machine. Therefore we consider~\emph{online Turing
  machines}~\cite{Hennie66} that have a designated (one-way) input tape
alongside with the Turing tape. Essentially the same type of machines
(but subject to \emph{bisimulation semantics} instead of \emph{language semantics}) was 
recently studied under the name~\emph{reactive Turing machines}~\cite{BaetenLuttikEtAl11}.
The format of our general reactive expressions deviates from the
format of the familiar Kleene's regular expressions. This is
inevitable, for the latter use various features of the underlying
model that are not generally available, most notably nondeterministic
choice, but also the fact that~$B$ is precisely the two-element set
$\{0,1\}$. 
However, our syntax features precisely the operations
coming from an equational presentation of the computation monad $\BBT$. This
allows us to cover instances which are beyond the reach of expression
formats with ``hard-wired''  nondeterminism. Specifically, we elaborate the
case of deterministic machines over a pushdown store, recognizing
precisely \emph{real-time deterministic context-free languages}, which
are properly contained in the class of all context-free languages,
which in turn are recognized by the respective nondeterministic stack
$\BBT$-automata. 
Moreover, we show that our syntax can be simplified for monads
whose presentation features a finitary summation operation              
(generalizing nondeterministic choice), and under further expected assumptions, 
become convertible to the one familiar in weighted automata theory for         
defining \emph{rational} formal power series~\cite{dkv,sakarovitch_2009}.               

A considerable part of our technical development (especially
Section~\ref{sec:mon}) is devoted to characterizing monads for
realizable transitions of state machines. For example, the stack of a
pushdown automaton is standardly modelled by the set of finite
sequences $\Gamma^*$ over an alphabet $\Gamma$ of stack
symbols. However, not all transformations $\Gamma^*\to\Gamma^*$ are
realizable by such an automaton (they need not even be computable). We
characterize the relevant \emph{stack monad} of realizable stack
transformers in two complementary ways: as a submonad of the store
monad $TX = (X\times \Gamma^*)^{\Gamma^*}$ and as an algebraic theory
over primitive stack operations $push$ and $pop$. We then obtain an
analogous result for the \emph{(Turing) tape monad},
whose theory, in contrast to the stack theory, fails to be
finitely axiomatizable.

The main salient feature of our approach is that it allows one to
untie from the standard enumerative and diverse definitions of various
kinds of state machines and reason about them collectively in a
uniform way. We demonstrate this by providing some initial
constructions on $\BBT$-automata, specifically by \emph{tensoring} the
underlying monads for obtaining machines over combined effects, e.g.\
store and nondeterminism.
Another construction we present is a certain
\emph{continuations passing style (CPS) transformation} of a given
$\BBT$-automaton allowing us to define an extension of the canonical
coalgebraic semantics to the case of unobservable (aka \emph{silent})
transitions.  The latter semantics allows us to capture recursively
enumerable languages by (deterministic) $\BBT$-automata over the
Turing tape. This provides an answer to a long standing challenge of
giving a coalgebraic description for any Turing complete computation
model.

Using a reduction to previous work~\cite{BookGreibach70} on \emph{real-time machines}
we show that
$\BBT$-automata with nondeterminism and an arbitrary number of stacks
\emph{without} unobservable moves capture precisely the class
$\mathsf{NTIME}(n)$ of nondeterministic linear time languages. Based on this 
we argue that it seems unlikely to be able to capture languages beyond
$\mathsf{NTIME}(n)$ by any computationally feasible class of
$\BBT$-automata without unobservable moves. In fact, we conjecture
that this bound remains valid also for our tape $\BBT$-automata.  The
requirement to be real-time is an inherent feature of coalgebraic
models and is often regarded a desirable feature of
\emph{reactivity} or \emph{productivity} of computations.

Finally, we prove a coalgebraic version of one direction
of the classical \emph{Chomsky-Sch\"utzen\-berger theorem}
(Theorem~\ref{thm:chsh}). As an instance, this allows to conclude that for
every polycyclic monoid $M$ of rank at least 2, every context-free
language is recognized by a valence automaton over $M$; that
context-free languages are precisely the languages recognized by
valence automata over polycyclic monoids was proven in~\cite{RenderKambites09}.

\medskip\noindent
\textbf{Related work.} We build on previous work on coalgebraic modelling and
monad-based semantics. Most of the applications of coalgebra to automata and
formal languages however address rational models (e.g.\ rational streams,
regular languages) from which we note~\cite{Rutten03} (regular languages and
finite automata),~\cite{Jacobs06} (bialgebraic treatment of Kleene algebra and
regular expressions),~\cite{brs_lmcs,bbrs_ic,Milius10,BonsangueMiliusEtAl13}
(coalgebraic regular expressions).

More recently, some further generalizations were proposed. In recent
work~\cite{WinterBonsangueEtAl13} a coalgebraic model of context-free
grammars is given, and~\cite{BonsangueRuttenEtAl12} captures weighted
context-free grammars and algebraic formal power-series
coalgebraically, without however an analogous treatment of (weighted)
push-down automata. Winter~\shortcite{Winter14} devotes a chapter of his thesis to the
treatmeant of push-down automata (and weighted push-down systems),
including e.g.\ a bisimulation-based proof of the result that any
power series recognizable by a weighted pushdown system is also
recognizable by a weighted pushdown system with a single state, the
latter of which coincide with weighted grammars in Greibach normal
form. However, a final coalgebra based semantics of push-down systems,
like the one we present for stack $\BBT$-automata, is not presented in loc.~cit.
Finally, \cite{MiliusEA16} gives a unifying
account of various finite state behaviours, and in particular
characterizes the domain of finite state behaviours by a universal
property; applications include all known coalgebraic models of
rational behaviour, but also (weighted) context-free languages and
algebraic power-series and the languages recognized by
$\BBT$-automata.
Myers established a rather general form of a Kleene theorem for surjection
preserving functors on varieties~\cite{Myers13}, while we stick to a concrete
 functor $B\times(\argument)^A$. His Kleene Theorem is parametric in a given
presentation of the variety and the type functor by operations and equations;
but we do not derive our Kleene-type theorem from his general one.
The specific form of the functor we are using allows us to directly associate
$\BBT$-automata and the corresponding expressions with their semantics, which
are formal power series from $B^{A^*}$. Moreover, this enables us to give
a direct syntactic translation between the \emph{reactive expressions} in
Section~\ref{sec:react} and the more convenient \emph{additive expressions} in
Section~\ref{sec:ex} (see Proposition~\ref{prop:aexp}).

The notion of $\BBT$-automata appeared for the first time in \cite{Jacobs06}. In
addition, we will also use in our development two results from \cite{Jacobs06}
(these appeared also in Bartels' thesis~\cite{Bartels04} and Turi and Plotkin's
seminal paper~\cite{TuriPlotkin97}) stating that: (i) in the presence of a
distributive law $\BBT G \Rightarrow G\BBT$, the final $G$-coalgebra carries
a $\BBT$-algebra structure; (ii) there is a bijective correspondence between
$GT$-coalgebras (in $\Set$) and $\lambda$-bialgebras. \cite{Jacobs06} gives
a list of $\BBT$-automata examples, including non-deterministic automata and
semiring automata, but these are not treated in detail and, more importantly,
this list does not include machines with memory such as pushdown automata. We go
beyond \cite{Jacobs06} both in terms of examples, but more importantly, in that 
we provided a uniform expression syntax for a large class of automata, which 
include automata equipped with memory, for which we make use of algebraic 
presentations of monads.

Pattinson and Sch\"oder~\citeyear{PattinsonSchroder16} independently
investigated an axiomatization of the Turing tape equivalent to ours and showed
that the axioms precisely characterize the Turing tape as a \emph{final comodel}
of the corresponding algebraic theory. They proved a completeness theorem which
can be read as the fact that the induced monad injectively embeds into the store
monad with the Turing tape as the store. In contrast to the latter result in our
work we additionally characterize precisely that submonad by a collection of
conditions on the store transformers.

The present paper is based on our previous conference
publication~\cite{GoncharovMiliusEtAl14}.

\medskip\noindent
{\bf Electronic Appendix.} All omitted proofs as well as a full proof
of Proposition~\ref{prop:aexp} and additional proof details for
Proposition~\ref{prop:stens} may be found in the electronic appendix
accompanying this publication.

\section{Deterministic Moore Automata, Coalgebraicaly}\label{sec:pre}
In this section we recall the main definitions and existing results on
coalgebraic modelling of state machines that we need. This material, as well as
the material of the following sections, uses the language of category theory,
hence we assume readers to be familiar with basic notions. We use $\Set$ as the
main underlying category throughout. Further abstraction from $\Set$ to a more
general category, while possible (and often quite straightforward), will not be
pursued in this paper.

Our central notion are \emph{$F$-coalgebras}, where $F$ is an
endofunctor on $\Set$ called \emph{transition type}. An $F$-coalgebra is a pair
$(X,f:X\to FX)$ where $X$ is a set called the \emph{state space} and $f$ is a map
called \emph{transition structure}.
We shall often identify a coalgebra with its state space if no confusion arises.

Coalgebras of a fixed transition type $F$ form a category whose
morphisms are maps of the state spaces commuting with the transition
structure: a map $h\colon X\to Y$ is a \emph{(coalgebra) homomorphism}
from $(X, f\colon X\to FX)$ to $(Y,g\colon Y\to FY)$ if the square
below commutes:
\begin{equation*}
\begin{tikzcd}[column sep=normal, row sep=normal]
X \rar["f"]\dar["h"'] & FX \dar["Fh"]\\
Y \rar["g"] & FY
\end{tikzcd}
\end{equation*}
A final object of this category (if it exists)
plays a particularly important role and is called \emph{final
  coalgebra}. We denote the final $F$-coalgebra by
\[
  (\nu F,\iota \colon \nu F\to F\nu F),
\]
and write $\widehat f\colon  X \to \nu F$ for the unique homomorphism from $(X,f)$ to $(\nu F, \iota)$. 

Our core example is the standard formalization of Moore automata as coalgebras~\cite{Rutten00}. For the rest of the paper we fix a finite set $A$ of \emph{actions} and a set $B$ of \emph{outputs}. We call the functor $\Lfun =B\times(-)^A$ the \emph{language functor} (over $A$, $B$).
The coalgebras for $\Lfun$ are given by a set $X$ of states with a transition structure on $X$ given by maps
\begin{flalign*}
&&o:X\to B &&\text{and}&&  \partial_a: X\to X, && (a \in A)
\end{flalign*}
where the left-hand map, called the \emph{observation map}, yields outputs in $B$
(e.g.\ an acceptance predicate if $B=2$; here and elsewhere we identify $2$ with $\{0, 1\}$) and the right-hand maps, called \emph{$a$-derivatives}, are the next state functions indexed by input actions from $A$. Finite $\Lfun$-coalgebras are hence precisely classical Moore automata. It is straightforward to extend $a$-derivatives to $w$-derivatives with $w\in A^*$ by induction: $\partial_{\eps}(x)=x$; $\partial_{a w}(x)=\partial_a(\partial_w(x))$ where $\eps\in A^*$ is the empty word. 

The final $\Lfun$-coalgebra $\nu \Lfun$ always exists and is carried by the set of all \emph{formal power series} $B^{A^*}$. The transition structure on $B^{A^*}$ is 
given by 
\begin{flalign*}
&&o(\sigma) = \sigma(\eps) &&\text{and}&& \partial_a(\sigma) = \lambda w.\, \sigma(aw), && (a \in A)
\end{flalign*}
for every formal power series $\sigma: A^* \to B$. The unique
homomorphism from an $\Lfun$-coalgebra $X$ to the final one $B^{A^*}$
assigns to every state $x_0 \in X$ a formal power series that we
regard as the \emph{(language) semantics} of $X$ with $x_0$ as an
initial state. Specifically, if $B=2$ then finite $\Lfun$-coalgebras
are deterministic automata and $B^{A^*}\cong\PSet(A^*)$ is the set of
all formal languages over $A$ and the language semantics assigns to
every state of a given finite deterministic automaton the language
accepted by that state. The transition structure on $\PSet(A^*)$ is
given by the predicate $o$ distinguishing languages containing the
empty word and by the maps $\partial_a$ assigning to a language their
left derivatives:
\begin{flalign*}
&&o(L) = 1\iff\eps\in L &&\text{and}&&\partial_a(L) = \{w\mid aw\in L\} && (a \in A)
\end{flalign*}

\begin{definition}[Language semantics, Language equivalence]\label{defn:tsem}
  Given an $\Lfun$-coalgebra $(X,f)$, the \emph{language semantics} \sgnote{Perhaps, we should change the name, because formal power series are more general than languages. But to what?}\smnote{Perhaps \emph{final semantics} and \emph{behavioural equivalence}?} is
  given by
  \[
    \widehat f: X \to B^{A^*}
  \]
  For every $x \in X$, $\widehat f(x)$ is the formal power series
  recognized by $x$.

  \emph{Language equivalence} identifies exactly those $x$ and $y$ for
  which $\widehat f(x) = \widehat g(y)$ (for possibly distinct
  coalgebras $(X, f)$ and $(Y,g)$); this is denoted by $x\sim y$.
\end{definition}
We obtain the following characterization of language equivalence. 
\begin{proposition}\label{prop:test}
Given $x\in X$ and $y\in Y$ where $X$ and $Y$ are $\Lfun$-coalgebras, $x\sim y$ iff for any $w\in A^*$, $o(\partial_w(x))=o(\partial_w(y))$.
\end{proposition}
It is well-known that Moore automata, i.e.~\emph{finite} $\Lfun$-coalgebras, can be characterized in terms of formal power series occurring as their language semantics (see e.g.~\cite{Rutten03}).  
\begin{definition}[Regular power series]\label{defn:rat}
We call a formal power series $\sigma$ \emph{regular} if the set $\{\partial_w(\sigma)\mid w\in A^*\}$ is finite. 
\end{definition}  
The following result is a rephrasing of a classical result on regular
languages (see e.g.~\cite[Theorem~III.8.1]{Eilenberg74}). The proof for
formal power series is similar and left to the reader. 
\begin{proposition}
  \label{prop:rat}
  A formal power series is accepted by a Moore automaton if and
  only if it is regular. 
\end{proposition}
\begin{remark} 
  Formal power series are usually considered when $B$ is a semiring,
  in which case one usually also speaks of \emph{recognizable formal
    power series} as behaviours of finite weighted automata over $B$
  (see e.g.~\cite{DrosteKuichEtAl09}). Our notion of \emph{regular
    formal power series} (Definition~\ref{defn:rat}) generally
  disagrees with the latter one (unless $B$ is finite) and is in
  conceptual agreement with such notions as `regular events' and
  `regular trees'~\cite{GoguenThatcherEtAl77,Courcelle83}.

  Regular formal power series as the semantics of precisely the finite
  $\Lfun$-coalgebras are a special instance of a general coalgebraic
  phenomenon~\cite{JirAdamekMiliusEtAl06,Milius10}. Let $F$ be any
  finitary endofunctor on $\Set$. Define the set $\varrho F$ to be the
  union of images of all \emph{finite} $F$-coalgebras $(X,f:X\to FX)$
  under their respective unique homomorphisms $\widehat f:X\to\nu
  F$. Then $\varrho F$ is a subcoalgebra of $\nu F$ with an isomorphic
  transition structure map; $\varrho F$ is therefore called the
  \emph{rational fixpoint} of $F$. It is (up to isomorphism) uniquely
  determined by either of the two following universal properties:
  (1)~as an $F$-coalgebra it is the final locally finite coalgebra and
  (2)~as an $F$-algebra it is the initial iterative algebra. We refer
  to~\cite{JirAdamekMiliusEtAl06,Milius10} for details.
\end{remark}
The characteristic property of regular formal power series can be used as a definitional principle. In fact, given a regular power series $\sigma$ and assuming that $A=\{a_1,\ldots,a_n\}$, we can view $\{\sigma_1,\ldots,\sigma_k\}=\{\partial_w(\sigma)\mid w\in A^*\}$ as a formal solution of a system of recursive equations of the form
\begin{align}\label{eq:bdeq}
\sigma_i = a_1.\sigma_{i_1}\pitchfork\ldots\pitchfork a_{n}.\sigma_{i_n}\pitchfork c_i, \qquad i = 1, \ldots, k, 
\end{align}
where for all $i=1,\ldots, k$ and $j = 1,\ldots, n$ we have $\partial_{a_j}(\sigma_i)=\sigma_{i_j}$ and $\sigma_i(\eps)=c_i$. Here we introduce~$\pitchfork$ as a syntax to combine the information about the ``heads'' of regular formal series with its derivatives. Reading the $\sigma_1, \ldots, \sigma_k$ as recursion variables, the system~\eqref{eq:bdeq} uniquely determines the corresponding regular power series: for every $i$ it defines $\sigma_i(\eps)$ as $c_i$ and for $w=au$ it reduces calculation of $\sigma_i(w)$ to calculation of some $\sigma_j(u)$ -- this induction is obviously well-founded.

Any recursive equation system~\eqref{eq:bdeq} can be rewritten as a term using the fixpoint operator $\mu$. To do this, first write
\begin{align}\label{eq:sigmu}
\sigma_i = \mu\sigma_i.\,a_1.\sigma_{i_1}\pitchfork\ldots\pitchfork a_{n}.\sigma_{i_n}\pitchfork c_i
\end{align} 
where $\mu\sigma_i$ binds the occurrences of $\sigma_i$ in the
right-hand term. One can then successively eliminate all the variables $\sigma_i$ using the equations~\eqref{eq:sigmu} as assignments and thus obtain a syntactic description of the given regular power series as $\sigma = t$ where $t$ is a closed term given by the following grammar:
\begin{align}\label{eq:kleene}
\gamma \Coloneqq \mu x.\,a.\delta\pitchfork\ldots\pitchfork
  a.\delta\pitchfork b&&\delta \Coloneqq x\mid\gamma&&(a\in A, x\in X, b\in B)
\end{align}
Here $X$ refers to an infinite stock of recursion variables. The term
$t$ according to~\eqref{eq:kleene} is then nothing but a condensed
representation of the system~\eqref{eq:bdeq} and as such it uniquely
defines $\sigma$. Thus every regular formal power series yields a closed
term. Proposition~\ref{prop:bro} below together with
Poposition~\ref{prop:rat} then establish that closed expressions
according to~\eqref{eq:kleene} capture precisely regular formal power series;
this can be viewed as a coalgebraic reformulation of Kleene's
theorem. This view has been advanced recently (in a more
general form) in~\cite{brs_lmcs,bbrs_ic,Myers13} and is instrumental for our present work.

Admittedly, the expressions of the form~\eqref{eq:kleene} are still
quite close to Moore automata. However, for $\BBT$-automata (introduced in Section~\ref{sec:react}) we shall extended this syntax with
operations from an algebraic theory given by the monad $\BBT$
(Definition~\ref{def:react}) and show how to simplify that syntax in
the case where $\BBT$ is an additive monad
(Definition~\ref{defn:guard}); in the special case of weighted
automata, this yields a syntax that is equivalent to
the familiar rational expressions (Remark~\ref{rem:rat}).

Proposition~\ref{prop:rat} together with the presentation of
regular formal power series as expressions~\eqref{eq:kleene} suggest
that every expression gives rise to a finite $\Lfun$-coalgebra, whose 
state space consists of expressions. This is
indeed true and can be viewed as a coalgebraic counterpart of 
Brzozowski's classical theorem for regular
expressions~\cite{Brzozowski64}. Given
$e=\mu x.\,a_1.e_1\pitchfork\ldots a_n.e_n\pitchfork c$, let
\begin{equation}\label{eq:bro}\setlength{\belowdisplayskip}{-2ex}
o(e) = c\qquad\text{and}\qquad\partial_{a_i}(e) = e_i[e/x].
\end{equation}
\begin{proposition}\label{prop:bro}
Let $e$ be a closed expression~\eqref{eq:kleene}. Then the set
$\{\partial_{w}(e)\mid w\in A^*\}$\/ forms a finite $\Lfun$-coalgebra
under the transition structure defined by~\eqref{eq:bro}.  
\end{proposition}
\begin{proof}
We only have to show that $E=\{\partial_{w}(e)\mid w\in A^*\}$ is
finite. Let $S$ be the set of all closed expressions $u\rho$ where
$u$ is a subexpression of $e$ and $\rho$ is a substitution sending
free variables of $u$ to closed subexpressions of $e$. Then, $S$ is
closed under $a$-derivatives, for
\[
  \partial_{a_i}(u\rho)=u_i\rho[u/x]\qquad  \text{if\qquad $u=\mu
    x.\,a_1.u_1\pitchfork\ldots\pitchfork a_n.u_n\pitchfork c$},
\]
and for $u = x \in X$, we have
$\partial_{a_i}(u\rho)=\partial_{a_i}(\rho(x))$, which lies in $S$
by the previous case because $\rho(x)$ is a closed subexpression of
$e$, which must start with a $\mu$-operator. By definition, $e\in
S$, hence $E\subseteq S$. Since $S$ is finite, so is $E$.
\end{proof}
\begin{remark}
If $B=2$, then Proposition~\ref{prop:bro} is essentially equivalent to Brzozowski's theorem, for in that case the expressions~\eqref{eq:kleene} are equivalently convertible into the standard regular expressions; the proof of the latter conversion is similar to the one found in~\cite{Silva10}. The conversion from regular expressions to $\mu$-expressions deploys a determinization procedure, which is available for the underlying notion of automaton. We revisit the question of converting $\mu$-expressions into generalized regular expressions in a broader context in Section~\ref{sec:tensor}.  
\end{remark}
\begin{figure}[t]
\begin{minipage}{.5\textwidth}
\begin{eqnarray*}
\\
\left\{
\begin{aligned}
 q_0 =&~a.q_1\pitchfork b.q_2\pitchfork 1&&\hspace{8ex}\\
 q_1 =&~a.q_2\pitchfork b.q_0\pitchfork 2\\
 q_2 =&~a.q_0\pitchfork b.q_1\pitchfork 3
\end{aligned}
\right.
\end{eqnarray*} 
\end{minipage}%
\begin{minipage}{.5\textwidth}%
\begin{tikzpicture}[shorten >=1pt,
  node distance=2.7cm,
  on grid,
  >=stealth',
  baseline=(current bounding box.north),
  every state/.style={inner sep=0cm,ellipse,minimum height=2.3em}
]

\node[state] (q_0) {$q_0,$ {\small 1}};
\node[state] (q_1) [right=of q_0] {$q_1,$ {\small 2}};
\node[state] (q_2) [right=of q_1] {$q_2,$ {\small 3}};

\path[->] 
  (q_0) edge [out=-40, in=220, looseness=.6]       node [above] {$a$} (q_1)
		    edge [out=90, in=90, looseness=.45]   node [above] {$b$} (q_2)
	(q_1) edge [out=-40, in=220, looseness=.6]       node [above] {$a$} (q_2)
	(q_1) edge                                  node [above] {$b$} (q_0);

\path[->] 
	(q_2) edge [out=130, in=50, looseness=.3]  node [above] {$a$} (q_0)
	(q_2) edge                                 node [above] {$b$} (q_1);
\end{tikzpicture}
\end{minipage}
\caption{A Moore automaton over $A=\{a,b\}$, $B=3=\{1,2,3\}$ as a graph (right) and as the corresponding system of equations (left).}
\label{fig:moore_aut}
\end{figure}
We close this section with a small illustration of the presented material. 
\begin{example}
  Let $B=\{1,2,3\}$ and let $A=\{a,b\}$. Consider a Moore automaton
  over these data as depicted in Fig.~\ref{fig:moore_aut}. Besides the
  standard pictorial representation as a graph, we consider an
  equivalent representation as a system of recursive equations. Given
  $w\in A^*$ let $\sharp_a w$ and $\sharp_b w$ denote the number of
  occurrences of $a$ and $b$ in $w$, respectively. Then the power
  series $\sigma$ recognized by state $q_i$ is the one for which
\begin{align*}
\sigma(w)=(\sharp_a w + 2\cdot \sharp_b w + i)~\text{mod}~3+1.
\end{align*} 
After picking $q_0$ as the initial state we can fold the system of equations into a single fixpoint expression
\begin{align*}
q_0 =\mu x.\, a.\mu y.\,\bigl(a.\mu z.\,(a.x\pitchfork b.y\pitchfork 3)\pitchfork b.x\pitchfork 2\bigr)\pitchfork
             b.\mu z.\,\bigl(a.x\pitchfork b.\mu y.\,(a.z\pitchfork b.x\pitchfork 2)\pitchfork 3\bigr)\pitchfork 1.
\end{align*}
If we replace $1$ in $B$ with $\top$ and both $2$ and $3$ with $\bot$, then
we obtain a deterministic automaton in which $q_0$ is the only final
state. This state then accepts exactly those words $w\in A^*$ for
which $\sharp_a w + 2\cdot \sharp_b w$ is divisible by $3$.
\end{example}

\section[Monads and Sigma-theories]{Monads and $\Sigma$-theories}\label{sec:mon}
In the previous section we summarized a coalgebraic presentation
of deterministic Moore automata, essentially capturing regular
languages and regular formal power series. In order to capture bigger
language classes we introduce \emph{(finitary) monads} and
\emph{$\Sigma$-theories} as a critical ingredient of our
formalization; this is following and extending ideas in 
previous work~\cite{JacobsSilvaEtAl12,SilvaBonchiEtAl13}.
In this work we find it easiest to work with monads in the form of \emph{Kleisli triples}.
\begin{definition}[Kleisli triple]
  A Kleisli triple $(T,\eta,\argument^{\klstar})$ consists of an
  object assignment $T$ sending sets to sets, a set-indexed family of
  maps $\eta_X:X\to TX$ and an operator, called \emph{Kleisli
    lifting}, sending any map $f:X\to TY$ to $f^{\klstar}:TX\to TY$. These
  data are subject to the following axioms: \iffull
\begin{align*} 
\eta^{\klstar}=\id, && f^{\klstar}\cdot\eta=f, && (f^{\klstar} \cdot g)^{\klstar}=f^{\klstar}\cdot g^{\klstar}.
\end{align*}
\else
$\eta^{\klstar}=\id$, $f^{\klstar}\cdot\eta=f$ and $(f^{\klstar}\cdot g)^{\klstar}=f^{\klstar}\cdot g^{\klstar}$.
\fi
\end{definition}
It is well-known that the definition of a monad as a Kleisli triple is equivalent to the usual definition of a monad $\BBT$ as an endofunctor $T$ equipped with natural transformations $\eta:Id\to T$ (\emph{unit}) and $\mu:TT\to T$ (\emph{multiplication}) satisfying standard 
identities~\cite{MacLane98}.

A \emph{$\BBT$-algebra} over a set $X$ (called the \emph{carrier}) is a pair 
$(X,a:TX\to X)$ where $a$ (called the \emph{structure}) satisfies $a\eta_X = 
\id_X$ and $a\mu_X = (Ta) a$.
A morphism of $\BBT$-algebras from $(X, a)$ to $(Y, b)$ is a map 
$h: X \to Y$ between carriers, such that $h a = b (Th)$. 

The category of $\BBT$-algebras and their morphisms is called
\emph{Eilenberg-Moore category of\/ $\BBT$} and is denoted by
$\Set^{\BBT}$. Note that $(TX, \mu_X)$ is the \emph{free
  $\BBT$-algebra} on the set $X$; that means that for every map $f: X
\to Y$, where $Y$ is the carrier set of a $\BBT$-algebra $(Y,t)$, there
exists a unique $\BBT$-algebra morphism $f^\sharp: (TX, \mu_X) \to
(Y,t)$ extending $f$, i.e.~such that $f^\sharp \cdot \eta_X = f$.
For more background material on monads and $\BBT$-algebras see~\cite{MacLane98}.

We find it useful to consider monads not only as a technical tool, but also as a metaphor for a notion of computation as manifested by Moggi~\citeyear{Moggi91b}. We therefore rely on the syntax of Moggi's \emph{computational metalanguage} (aka, Haskell $\mathsf{do}$-notation): 
\begin{notation}[$\DO$-notation]
Given $p \in TX$, $q: X \to TY$, we use the following notation for $q^{\klstar}(p)$: 
\begin{align*}
  \letTerm{x\leteq p}{q(x)}.
\end{align*}
\end{notation}
Intuitively, the construction $\letTerm{x\leteq p}{q(x)}$ should be read as
follows: run the computation $p$; bind the result to $x$ and then run the computation $q(x)$ depending on $x$.
This becomes particularly suggestive when considering state-based
monads, for which one can form expressions like 
\begin{displaymath}
  \letTerm{x\leteq\mathit{get}(\mathit{l}_1)}{\mathit{set}(\mathit{l}_2,f(x))},
\end{displaymath}
meaning: get a
value under location $\mathit{l}_1$, apply $f$ to it and put the result under
$\mathit{l}_2$.
\begin{remark} Some comments regarding the $\DO$-notation are in order.
  \begin{enumerate}[wide]
  \item The interpretation of the $\DO$-notation in
    general requires that the corresponding monad is \emph{strong},
    i.e.\ equipped with a natural transformation
    $\tau_{X,Y}:X\times TY\to T(X\times Y)$ called \emph{strength} and
    satisfying a number of obvious coherence conditions, which are elided here because
    every monad on $\Set$ is strong via the following canonical strength~\cite{Kock72}:
    $\tau_{X,Y}(x,p) = T(\lambda y.\,\brks{x,y})\, (p)$.
    Strength is needed for propagating values along the
    $\mathsf{do}$-expressions. For example, the meaning~  of
    \[
      \letTerm{x\leteq p;y\leteq q(x)}{r(x,y)}\quad \text{for every $p\in TX$,
        $q: X\to TY$ and $r:X\times Y\to TZ$}
    \]
    is precisely $r^\klstar((\tau_{X,Y}\brks{\id_X,q})^\klstar(p))$ (which is
    $(\lambda x.\, (\lambda y.\,
    r(x,y))^\klstar(q(x)))^\klstar(p)$ in~$\Set$).

  \item Further standard notational conventions are as
    follows:
    \begin{center}
      \begin{tabular}{l|l|l}
        Notation & Meaning & Condition \\
        \hline
        $\letTerm{x\leteq p}{q}$
        &
        $\letTerm{x\leteq p}{(\lambda x.\,q)(x)}$
        & --
        \\
        $\letTerm{p}{q}$
        &
        $\letTerm{x\leteq p}{q}$
        &
        $x$ not a free variable in $q$;
        \\
        $\letTerm{{\brks{x,y}\leteq p}}{q(x,y)}$
        &
        $\letTerm{z\leteq p}{q(z)}$
        &
        for $p\in T(X_1\times X_2)$
        \\
        && and $q:X_1\times X_2\to TY$;
        \\
        $\letTerm{x_1 \leteq p_1;\ldots;x_n\leteq p_n}{q}$
        &
        $\letTerm{x_1\leteq p_1}{\ldots;\letTerm{x_n\leteq p_n}{q}}$ & --  
      \end{tabular}
    \end{center}
\item Moggi~\citeyear{Moggi91b} has indeed proved that the following axiomatization
  of $\DO$-expressions is sound complete for strong monads 
\begin{align*}
  \letTerm{x\leteq (\letTerm{y\leteq p}{q})}{r} &= \letTerm{y\leteq p;x\leteq q}{r}&&\text{($y$ not free in $r$)}\\ 
  \letTerm{x\leteq\eta_X(a)}{p} &= p[a/x]\\
  \letTerm{x\leteq p}{\eta_X(x)} &= p.
\end{align*}  
making the $\DO$-notation a fully fledged \emph{internal language} of strong monads.
\end{enumerate}
\end{remark}
A monad $\BBT$ is \emph{finitary} if the underlying functor $T$ is finitary, i.e., $T$ preserves filtered colimits. Informally, $T$ being finitary means that $T$ is determined by its action on finite sets. In addition, finitary monads admit a presentation in terms of (finitary) equational theories over an algebraic signature as we now outline.
\begin{definition}[$\Sigma$-theory]
\label{dfn:theory}
An \emph{algebraic signature} $\Sigma$ consists of operation
symbols~$f$, each of which comes together with its \emph{arity} $n$,
which is a nonnegative integer -- we denote this by $f\colon n\to
1$. Symbols of zero arity are also called
\emph{constants}. $\Sigma$-terms are constructed from the operations
in~$\Sigma$ and variables in the usual way. A \emph{$\Sigma$-theory}
is given by a set of $\Sigma$-term equations closed under inference of
the standard equational logic. We shall usually present an algebraic
theory $\CE$ by its signature $\Sigma$ together with a set of
\emph{axioms}; we then obtain $\CE$ as the \emph{deductive closure} of
the given set of axioms under standard equational reasoning.
\end{definition}
\iffull Given a $\Sigma$-theory $\CE$ we can form a monad $\BBT_\CE$
as follows: $T_\CE X$ is the set of equivalence classes of terms of
the theory over free variables from $X$ (in what follows we shall
refer to equivalences of terms always by terms representing them);
$\eta_X:X\to T_\CE X$ casts a variable to a term; given
$\rho:X\to T_\CE Y$ and $p \in T_\CE X$, $\rho^{\klstar}(p)$ is the
term $p\rho$ obtained by substituting the free variables in the term
$p$ according to the substitution $\rho$.

Conversely, we can pass from a finitary monad $\BBT$ to the
$\Sigma_\BBT$-theory $\CE_{\BBT}$, where $\Sigma_\BBT$ is the
signature that contains an operation symbol $f_a:n\to 1$ for each
element $a$ of $Tn$. Such an operation symbol can be interpreted as a
map
\[
\brks{t_1,\ldots,t_n}\mapsto (\lambda i.\,t_i)^{\klstar}(a) 
\]
from $(T X)^n$ to $TX$.  This yields a semantics of
$\Sigma_{\BBT}$-terms over $TX$ and we define $\CE_\BBT$ to be the
$\Sigma_{\BBT}$-theory given by all term equations valid over any
$TX$. Notably, $\BBT$-algebras are then exactly the models of the
$\Sigma$-theory $\CE_{\BBT}$.

While the passage from a monad to the $\Sigma_{\BBT}$-theory
$\CE_\BBT$, followed by the passage in the opposite direction yields
an identical transformation, the passage from a $\Sigma$-theory,
followed by the passage from monads to theories does not yield the
original $\Sigma$-theory, but instead produces its \emph{clone}, i.e.\
a theory, obtained from the original $\Sigma$-theory by recognizing
all $\Sigma$-terms as (possibly new) operation symbols.
This fundamental observation, going back to
Lawvere~\citeyear{Lawvere63}, allows us to consider $\Sigma$-theories as
presentations of finitary monads. It will be instrumental in our study
of syntactic presentations of generic automata, e.g.\ our Kleene Theorem
(Theorem~\ref{thm:kleene}).
\begin{definition}[Presentation of a monad]
  A $\Sigma$-theory $\CE$ is said to be a \emph{presentation} of the
  monad~$\BBT$ if $\BBT$ is naturally isomorphic to $\BBT_\CE$. We
  also say that $\CE$ \emph{generates} $\BBT$.
\end{definition}
While the $\Sigma_\BBT$-theory $\CE_\BBT$ yields a canonical
presentation of the monad $\BBT$ we shall subsequently be interested
in working out more compact presentations. In order to do this we will
consider semantics of $\Sigma$-terms and $\Sigma$-theories over monads
not necessarily of the form~$\BBT_{\CE}$. We will make free use of the
equivalence between \emph{$n$-ary algebraic operations} over a monad
$\BBT$ and the elements of $Tn$ (where we identify $n$ with the set
$\{1,\ldots,n\}$). This equivalence was presented by Plotkin and
Power~\citeyear{PlotkinPower03} (more generally as a duality between
algebraic operations $n\to m$ and Kleisli morphisms $m\to Tn$), and we
recall it below.

Let $\BBT$ be any monad, and recall that an \emph{$n$-ary algebraic
  operation over $\BBT$} is a natural transformation $\alpha: T^n \to
T$, where $T^n$ denotes the $n$-fold product $T\times\cdots\times
T$,\footnote{We will use exponents on $T$ only in this sense and
  \emph{not} to indicate $n$-fold composition of $T$ with itself.} such that for every $f: X \to TY$,
\begin{equation}\label{eq:alpha}
\begin{tikzcd}[column sep=normal, row sep=normal]
(TX)^n \dar["(f^\klstar)^n"'] \rar["\alpha_X"] & TX \dar["f^\klstar"]\\
(TY)^n \rar["\alpha_Y"] & TY
\end{tikzcd}
\end{equation}
Any element $a \in Tn$ yields $\alpha: T^n \to T$ by defining 
\[
\alpha_X(f)=f^\klstar(a)=\letTerm{x\leteq a}{f(x)}
\] 
for any $f: n \to TX$. And given an $n$-ary algebraic operation $\alpha: T^n \to T$ over $\BBT$ we obtain $\alpha_n(\eta_n) \in Tn$. It is not difficult to show that these two passages are mutually inverse.

The technical advantage of using elements of $Tn$ is that they are unconstrained whereas $n$-ary algebraic operations $\alpha: T^n \to T$ need to satisfy the above coherence condition~\eqref{eq:alpha}.
\begin{definition}\label{dfn:sem}
Let $\Sigma$ be a signature and let $\BBT$ be a (not necessarily finitary) monad. A \emph{semantics} of $\Sigma$ over $\BBT$ is an assignment $\tsem{-}_T$ sending any $f:n\to 1$ in $\Sigma$ to $\tsem{f}_T\in Tn$. For every $\Sigma$-term $t$ over a set of variables $X$ this determines $\tsem{t}_{TX}\in TX$ inductively as follows: 
\begin{itemize}
 \item $\tsem{x}_{TX}=\eta_X(x)$ for $x\in X$;
 \item $\tsem{f(t_1,\dots,t_n)}_{TX}=\letTerm{i\leteq\tsem{f}_T}{\tsem{t_i}_{TX}} 
   = \alpha_X(\tsem{t_1}_{TX}, \ldots, \tsem{t_n}_{TX})$, where $\alpha: T^n \to T$ is the $n$-ary algebraic operation over $\BBT$ corresponding to $\tsem{f}_T$. 
 \end{itemize}
 Now let $\CE$ be a $\Sigma$-theory. We call a semantics of $\Sigma$ over $\BBT$
\begin{itemize}
  \item\emph{sound} if for any equation $s=t$ from $\CE$ with free variables included in $X$, $\tsem{s}_{TX}=\tsem{t}_{TX}$;
  \item\emph{complete} if $s=t\in\CE$ whenever $\tsem{s}_{TX}=\tsem{t}_{TX}$ for some $X$ containing all the free variables of $s$ and $t$;
  \item\emph{expressive} if for every $p\in TX$ there is a $\Sigma$-term $t$ over $X$ such that $p=\tsem{t}_{TX}$.  
\end{itemize}
In the future we shall omit the subscripts of $\tsem{-}$ whenever $T$ or $TX$, respectively, are clear from the context.
If a semantics of $\CE$ over $\BBT$ is assumed, we simply call $\CE$
\emph{sound}, \emph{complete} and \emph{expressive} over $\BBT$ in the
corresponding cases. 
\end{definition}
\begin{remark}
  Note that if a $\Sigma$-theory is presented by a signature and
  axioms then it suffices to verify soundness for every
  axiom. Soundness of all equations in the closure $\CE$ of the set of
  axioms under inference of standard equational logic then follows
  easily by induction. 
\end{remark}
\begin{example}
  For every $\Sigma$-theory $\CE$ we have a \emph{canonical semantics}
  over the monad $\BBT_\CE$ given by setting
  $\tsem{f} = f(1, 2, \ldots, n) \in T_\CE n$.
\end{example}
The following theorem shows that the fact that a $\Sigma$-theory $\CE$ generates 
a monad $\BBT$ entails a canonical presentation of $\BBT$ in terms of $\CE$ up to isomorphism.
\begin{theorem}\label{thm:mon_thm_eq}
  Let $\CE$ be a $\Sigma$-theory and let\/ $\BBT$ be a finitary
  monad. Then $\CE$ generates $\BBT$ iff\/ there exists a sound,
  complete and expressive semantics of $\Sigma$ over\/ $\BBT$.
\end{theorem}
\begin{proof}
  As we outlined after Definition~\ref{dfn:theory}, from $\CE$ we can
  construct a finitary monad $\BBT_{\CE}$ such that $T_{\CE}X$
  consists of $\Sigma$-terms over $X$ modulo $\CE$ and equip it with
  the canonical semantics $\tsem{-}$. Essentially due to
  Lawvere~\citeyear{Lawvere63} this semantics is sound, complete and
  expressive. Thus if $\CE$ generates $\BBT$, i.e.~we have a natural
  isomorphism $\gamma: \BBT_\CE \to \BBT$, then we can define the
  semantics
  $\tsem{-}_\BBT = \gamma_n \cdot \tsem{-}$, and show by an easy
  induction that
  \begin{equation}\label{eq:cohsem}
    \tsem{-}_{T X}
    = \gamma_X \cdot \tsem{-}_{T_\CE X}
    \qquad
    \text{for every set $X$.}
  \end{equation}
  Soundness, completeness and expressivity now easily follow from the fact that $\gamma_X$ is
  bijective.

  Conversely, we have to show that for any sound, complete and
  expressive semantics $\tsem{-}_\BBT$ of $\CE$ over $\BBT$, the latter is
  isomorphic to $\BBT_{\CE}$ via some natural isomorphism $\gamma$.   
  Indeed,
  since any element of $T_{\CE}X$ is represented by a $\Sigma$-term
  $t$ we can define $\gamma_X:T_{\CE} X\to T X$ by sending $t$ to
  $\tsem{t}_{TX}$. It immediately follows by soundness that this
  definition is well-defined (i.e.\ independent of the concrete choice of
  $t$). Completeness and expressiveness of the given semantics imply
  injectivity and surjectivity, respectively, of $\gamma_X$. It is
  also easy to see by definition that $\gamma_X$ respects unit and
  Kleisli lifting, hence it extends to a monad isomorphism.
\end{proof}

\fi
\begin{example}[Monads, $\Sigma$-theories]
\label{ex:mon}
Standard examples of computationally relevant monads include (cf.~\cite{Moggi91b}) the following ones.
\begin{cenumerate}
  \item The \emph{finite and unbounded powerset monads} $\PFin$ and $\PSet$. For both monads the unit is the singleton map $\eta_X: x \mapsto\{x\}$ and the Kleisli-lifting extends a map $f: X \to \PSet Y$ to $f^\klstar: \PSet X \to \PSet Y$ taking direct images: $f^\klstar(M\subseteq X)  = \bigcup_{x \in M} f(x)$ (and similarly for $\PFin$). Only $\PFin$ is finitary and corresponds to the $\Sigma$-theory of join-semilattices with bottom over $\Sigma=\{\bot,\lor\}$, or equivalently to the theory of commutative idempotent monoids.
  \item The \emph{monoid action monad} for a monoid $(M, \cdot, 1)$ maps a set $X$ to $M \times X$. Its unit is formed by the maps $\eta_X: x \mapsto (1,x)$ and the Kleisli-lifting extends $f: X \to M \times Y$ to $f^\klstar: M \times X \to M \times Y$ with $f^\klstar(m,x) = (m\cdot n, y)$ where $(n,y) = f(x)$. The corresponding $\Sigma$-theory is the theory of $M$-actions, i.e., $\Sigma$ has a unary operation symbol $m \cdot (-)$ for every $m \in M$ with the usual axioms $m\cdot (n\cdot x) = (m\cdot n)\cdot x$ and $1\cdot x = x$.  
  \item The \emph{store monad} over a store $S$. The object assignment of this monad is $X\mapsto (X\times S)^S$ and the unit $\eta_X: X \to (X \times S)^S$ assigns $\eta_X(x) = \lambda s. \langle x, s \rangle$. Typically, $S$ is the set of maps $L\to V$ from locations $L$ to values $V$. A function $f: X \to (Y \times S)^S$ represents a computation that takes a value in $X$ and, depending on the current contents of the store $S$ returns a value in $Y$ and a new store content. The Kleisli lifting sends $f$ to $f^\klstar: (X \times S)^S \to (Y \times S)^S$ with
\[
  f^\klstar(h)
  = \bigl(S \xrightarrow{h} X \times S \xrightarrow{f \times S} (Y\times
  S)^S \times S \xrightarrow{\mathsf{ev}} Y \times S\bigr),
\]
where $\mathsf{ev}$ is the obvious evaluation map. 
As shown in~\cite{PowerShkaravska04}, if $V$ is finite then the
corresponding store monad can be presented by a $\Sigma$-theory for
$\Sigma=\{\lookup_l:|V|\to 1\}_{l\in L}\cup\{\update_{l,v}:1\to 1\}_{l\in L,v\in V}$.
  \item The \emph{continuation monad}. Given any set $R$, the assignment $X\mapsto R^{R^{X}}$ yields a monad under the following definitions: \iffull
\[
\eta_X(x)=\lambda f.\,f(x)\qquad\text{and}\qquad f^{\klstar}(k)=\lambda c.\,k(\lambda x.\, f(x)(c)).
\]
\else
$\eta(x)=\lambda f.\,f(x)$ and $f^{\klstar}(k)=\lambda c.\,k(\lambda x.\, f(x)(c))$. %
\fi%
This monad is known to be non-finitary, unless $R = 1$.
\end{cenumerate}
\end{example}
We will need the following technical lemma for monads on $\Set$ and specifically implications from it for submonads of the store monad.
\begin{lemma}\label{lem:subalg}
Let $\BBT'$ be a submonad of\/ $\BBT$ and let $\alpha:\BBT\to\BBP$ be a monad morphism. Then $\alpha$ restricted to $\BBT'$ induces a monad morphism $\alpha':\BBT'\to\BBP'$ such that 
\begin{equation}
\begin{tikzcd}[column sep=normal, row sep=normal]
  \BBT'\rar["\alpha'"] \dar[hook, "i"'] & \BBP'\dar[hook, "j"]  \\
  \BBT\rar["\alpha"] & \BBP
\end{tikzcd}
\end{equation}
\end{lemma} 
\begin{corollary}\label{cor:subalg}
Let $\BBT_S$ be the store monad over $S$ and let $\BBR_S$ be the corresponding reader monad (i.e.~$R_SX=X^S$). For any submonad $\BBT$ of $\BBT_S$, the monad morphism $\alpha$ sending any $f:S\to X\times S$ to $\pi_1 f:S\to X$ restricts to a submonad $\BBR$\/ of\/ $\BBR_S$.  
\end{corollary}
The following class of examples is especially relevant for the coalgebraic modelling.
\begin{definition}[Semimodule monad, Semimodule theory]\label{defn:srmon}
Given a semiring $R$, the semimodule monad $\BBT_R$ assigns to  a set $X$ the free left $R$-semimodule $\brks{X}_R$ over $X$. Explicitly, $\brks{X}_R$ consists of all formal linear combinations of the form
\begin{align}\label{eq:sr_lin}
r_1\cdot x_1+\ldots+r_n\cdot x_n && (r_i\in R, x_i\in X)
\end{align}
Equivalently, $\brks{X}_R$ consists of maps $f:X\to R$ with finite support (i.e.~$|\{x\in X\mid f(x)\neq 0\}|<\omega$).
The assignment $X\mapsto\brks{X}_R$ extends to a monad, which we call the \emph{(free) semimodule monad}: $\eta_X$ sends any $x\in X$ to $1\cdot x$ and $\theta^\klstar(p)$ applies the substitution $\theta:X\to\brks{Y}_R$ to $p\in\brks{X}_{R}$ and renormalizes the result as expected.

The semimodule monad corresponds to the $\Sigma$-theory of $R$-semimodules. Explicitly, we have a constant $\emptyset:0\to 1$, a binary operation $+:2\to 1$, and a unary operation $\bar r: 1 \to 1$ for each $r\in R$. The axioms presenting this theory are the laws of commutative monoids for $+$ and $\emptyset$, plus the following identities for the (left) semiring action of $R$:
\begin{align*}
\bar r(x+y) =&~ \bar {r}(x)+\bar {r} (y) & \bar r(x)+\bar s(x)=&~\overline {r+s}(x) & \bar r(\bar s(x)) =&~ \overline {r\cdot s}(x) &\\
\bar {r}(\emptyset) =&~ \emptyset & \bar {0}(x)=&~\emptyset & \bar {1}(x) =&~ x
\end{align*}
It can be shown by using these laws that any term can by normalized to
a term of the form $\bar r_1(x_1)+\ldots+\bar r_n(x_n)$, and the
latter represent precisely the element~\eqref{eq:sr_lin} of
$\brks{X}_R$. Thus, the above $\Sigma$-theory generates $\BBT_R$.

Some notable instances of $\BBT_R$ are the following:
\begin{citemize}
  \item If $R$ is the Boolean semiring $\{0,1\}$ then $\BBT_R$ is (isomorphic to) the finite powerset monad $\PFin$.
  \item If $R$ is the semiring of natural numbers then $\BBT_R$ is the \emph{multiset} monad: the elements of $\langle X\rangle_R$ are in bijective correspondence with finite multisets over $X$.
  \item If $R$ is the interval $[0,+\infty)$ then $\BBT_R$ is the monad of \emph{finite valuations} used for modelling probabilistic computations~\cite{VaraccaWinskel06}. \iffull Two other well-known monads of \emph{finite distributions} and \emph{finite subdistributions} serving the same purpose embed into $\BBT_R$: the formal sums~\eqref{eq:sr_lin} for them are requested to satisfy the additional constraints $r_1+\ldots+r_n=1$ and $r_1+\ldots+r_n\leq 1$, respectively.\fi
\end{citemize}
\end{definition}
\subsection{The Stack Monad} 
The following example shows how to model a push-down store, see~\cite{Goncharov13}.
\begin{definition}[Stack monad, Stack theory]\label{defn:stack_mon}
Given a finite set of stack symbols $\Gamma$, the \emph{stack monad (over $\Gamma$)} is the submonad $\BBT$ of the store monad $(\argument\times\Gamma^*)^{\Gamma^*}$ for which the elements $\brks{r,t}$ of $TX\subseteq (X\times\Gamma^*)^{\Gamma^*}$ satisfy the following restriction: there exists $k$ depending on $r,t$ such that for every $w\in\Gamma^k$ and $u\in\Gamma^*$, 
\iffull
\begin{align}\label{eq:stack_cond}
r(wu)=r(w)\qquad\text{and}\qquad t(wu)=t(w)u.
\end{align}
\else
$r(wu)=r(w)$ and $t(wu)=t(w)u$.
\fi
Intuitively, a map $f: X \to TY$ (cf.~Example~\ref{ex:mon}) computes an output 
value in $Y$ and a result stack based on the prefix of the input stack 
of size $k$, which does not depend on the content of the stack.

The \emph{stack signature} w.r.t.\ $\Gamma=\{\gamma_1,\ldots,\gamma_n\}$ consists of operations $pop:{n+1}\to 1$ and $push_i:1\to 1$, $1 \leq i \leq n$. The intuition here is as follows (in each case the arguments represent continuations, i.e.~computations that will be performed once the operation has completed its task, cf.~\cite{PlotkinPower02}):
\begin{itemize}
\item $pop(x_1,\ldots,x_n,y)$ proceeds with $y$ if the stack is empty;
  otherwise it removes the top element from it and proceeds with
  $x_i$, where $\gamma_i\in\Gamma$ is the removed stack element.
  \item $push_i(x)$ adds $\gamma_i\in\Gamma$ on top of the stack and proceeds with~$x$.  
\end{itemize}
The \emph{stack theory} is presented by these operations and the
axioms in Fig.~\ref{fig:stack-ax}. These axioms capture semantic
equivalences of terms considered as programs transforming the
underlying store. This implies that composition is to be read from
left to right, e.g.~the left-hand term of the first equation means
``push $\gamma_i$, then pop one symbol from the stack, then proceed
with with $y$ if the stack was empty or with $x_j$ if the popped
symbol was $\gamma_j$.
\begin{figure}[t]
\noindent
\begin{flalign*}
&\textbf{(push-pop)}&  push_i(pop(x_1,\ldots,x_n,y)) =\;& x_i&&  \\[.6ex]
&\textbf{(pop-push)}&  pop(push_1(x),\ldots,push_n(x),x) =\;&x&& \\[.6ex]
&\textbf{(pop-pop)}& pop(x_1,\ldots,x_n,pop(y_1,\ldots,y_n,z)) =\;&pop(x_1,\ldots,x_n,z)&&
\end{flalign*} 
\caption{Axioms for the stack monad ($i\in\{1,\ldots,n\}$).}
\label{fig:stack-ax}
\end{figure}
We connect the stack theory with the stack monad~$\BBT$ by the following semantics:
\begin{align*}
\tsem{pop}(\eps)=\brks{n+1,\eps},&&\tsem{pop}(\gamma_i w)=\brks{i,w}, && \tsem{push_i}(w) = \brks{1,\gamma_i w}
\end{align*} 
where $w\in\Gamma^*$, and $\eps$ denotes the empty stack.
\end{definition}
As claimed in~\cite{Goncharov13} the stack theory generates the stack
monad. We include a proof of this fact below. It relies on the
following auxiliary statement.
\begin{lemma}\label{lem:stack_sem}
The semantic identity 
$\tsem{pop(p_1,\ldots,p_n,p)}=\tsem{pop(q_1,\ldots,q_n,q)}$
with $p$ and $q$ not containing $pop$ implies 
$
\tsem{p_1}=\tsem{q_1}\comma\ldots\comma\tsem{p_n}=\tsem{q_n},\tsem{p}=\tsem{q}.
$   
\end{lemma}
\begin{theorem}\label{thm:stack_comp}
The stack theory generates the stack monad.  
\end{theorem}
\begin{proof}
We directly verify soundness, expressiveness and completeness in order. 
\begin{citemize} 
 \item[]\emph{Soundness} is straightforward to verify. Consider for example the left-hand side of the second axiom of the stack theory:
\begin{align*}
 \tsem{pop(push_1(x),\ldots,push_n(x),x)}
=\letTerm{i\leteq\tsem{pop}}{\ifTerm{\;(i<n+1)}{\tsem{push_i(x)}}{\tsem{x}}}.
\end{align*}
Using the definition of the store monad, and the semantic of $push$ and $pop$, 
\begin{align*}
\tsem{pop(push_1(x),\ldots,push_n(x),x)}(\eps)       =&\; \tsem{x}(\eps)\\
\tsem{pop(push_1(x),\ldots,push_n(x),x)}(\gamma_i w) = \tsem{push_i(x)}(w) =&\; \tsem{x}(\gamma_i w)
\end{align*}
which is in agreement with the right-hand side of the identity in question.
 
\item[]\emph{Expressiveness.} Let $\brks{r,t}\in TX$. By definition,
  there is $k$ such that for any $w\in\Gamma^k$ and any
  $u\in\Gamma^*$, \eqref{eq:stack_cond} is satisfied. Using these data
  we construct by induction over $k$ a $\Sigma$-term $p_k(r,t)$ over~$X$:
\begin{itemize}
 \item~ if $k=0$ then $p_k(r,t)=push_{{i_m}}(\ldots push_{{i_1}}(r(\eps))\ldots)$ where $t(\eps)=\gamma_{i_1}\ldots\gamma_{i_m}$;
 \item~ if $k>0$ let us define for any $1\leq i\leq n$,
   $\brks{r_i,t_i}\in TX$ by the following equations
   \[
     r_i(w)=r(\gamma_i w)
     \qquad\text{and}\qquad
     t_i(w)=t(\gamma_i w)\qquad\text{for every $w\in \Gamma^*$.}
   \]
   Then we put $p_k(r,t)=pop(p_{k-1}(r_1,t_1),\ldots,p_{k-1}(r_n,t_n),p_{0}(r,t))$.
 \end{itemize}
 We now prove that $\tsem{p_k(r,t)} = \brks{r,t}$ by
 induction over $k$. For the base case $k=0$, \eqref{eq:stack_cond}
 states that for all $w \in\Gamma^*$ we have $r(w) = r(\eps)$ and $t(w)
 = t(\eps)w$. Hence, by definition of~$\tsem{push_i}$, and by~\eqref{eq:stack_cond},
 \begin{flalign*}
 \tsem{p_0(r,t)}(w) = \tsem{push_{{i_m}}(\ldots push_{{i_1}}(r(\eps))\ldots)}(w)= \brks{r(\eps),\gamma_{{i_1}}\cdots\gamma_{{i_m}}w}  
 = \brks{r(w),t(w)}.
 \end{flalign*}
 For the induction step note first that we may apply the induction
 hypothesis with $r_i, t_i$ since this pair
 satisfies~\eqref{eq:stack_cond} for every $w\in\Gamma^{k-1}$. Thus we
 have
\begin{align*}
\tsem{p_k(r,t)}(\eps)
&= \tsem{pop(p_{k-1}(r_1,t_1),\ldots,p_{k-1}(r_n,t_n),p_{0}(r,t))}(\eps)
 =\tsem{p_0(r,t)}(\eps) = \brks{r,t}(\eps),
\intertext{and analogously, using the induction hypothesis,}
\tsem{p_k(r,t)}(\gamma_iw)
&= 
\tsem{p_{k-1}(r_i,t_i)}(w) = \brks{r_i,t_i}(w) = \brks{r,t}(\gamma_iw).
\end{align*}
 
\item[]\emph{Completeness.} We turn the stack axioms into a rewriting system by orienting each equation from left to right. This rewriting system is obviously strongly normalizing because each application of the rule decreases the term size. There are no nontrivial critical pairs and therefore using the standard argument from term rewriting any term has a unique normal form~\cite{Terese03}. From the structure of the rules we can see that any normal form $p$ either does not contain $pop$ or is of the form $pop(p_1,\ldots,p_n,p')$ where each $p_i$ is in a normal form and $p'$ does not contain $pop$.

By soundness, it remains to show that for any normal $p$ and $q$, $\tsem{p}=\tsem{q}$ implies $p=q\in\CE$. We proceed by induction over the total number of the $pop$ operators in $p$ and $q$.
\begin{cenumerate}
 \item If both $p$ and $q$ do not contain $pop$ they must be of the
   form $push_{{i_1}}(\ldots push_{{i_m}}(x)\ldots)$ and
   $push_{{j_1}}(\ldots push_{{j_l}}(y)\ldots)$, respectively. Then  $\tsem{p}(w)=\tsem{q}(w)$ amounts to $\brks{x,\gamma_{i_m}\ldots\gamma_{i_1} w}=\brks{y,\gamma_{j_l}\ldots\gamma_{i_1} w}$ and therefore $x=y$, $m=l$ and $i_1=j_1,\ldots,i_m=j_m$, i.e. $p$ is identical to $q$.
\item If $p=pop(p_1,\ldots,p_n,p')$ and $q$ does not contain
  $pop$,
  then we have 
  \[
    \tsem{p}=\tsem{q}=\tsem{pop(push_1(q),\ldots,push_n(q),q)}.
  \]
  By Lemma~\ref{lem:stack_sem}, $\tsem{p_1}=\tsem{push_1(q)}\comma\ldots\comma\tsem{p_n}=\tsem{push_n(q)}$, $\tsem{p'}=\tsem{q}$. Note that the terms $push_i(q)$ need not be normal, but they can be normalized and since normalization only decreases the number of the $pop$ operators the induction hypothesis applies to the result, and we have $\{p_1=push_1(q),\ldots,p_n=push_n(q),p'=q\}\subseteq\CE$. Hence, in $\CE$,
$p=pop(p_1,\ldots,p_n,p')=pop(push_1(q),\ldots,push_n(q),q)=q$.
\item If $q=pop(q_1,\ldots,q_n,q')$ and $p$ does not contain $pop$,
  then we proceed analogously to the previous case.
\item If $p = pop(p_1,\ldots,p_n,p')$ and $q=pop(q_1,\ldots,q_n,q')$,
  then $\tsem{p_i} = \tsem{q_i}$, $i=1, \ldots, n$, and $\tsem{p'} =
  \tsem{q'}$ by Lemma~\ref{lem:stack_sem}. By induction hypothesis, we
  have $\{p_1 = q_1, \ldots, p_n = q_n, p' = q'\} \subseteq
  \CE$. Hence, in $\CE$, 
  $p = pop(p_1, \ldots, p_n, p') = pop(q_1,\ldots, q_n, q') = q$.
  \qed
\end{cenumerate}
\end{citemize}
\noqed\end{proof}
\subsection{The Tape Monad}
We now introduce a monad and the corresponding theory underlying the tape of a Turing machine. The idea we use here is the same as in the case of the stack theory: we specify a submonad of a suitable store monad in such a way that only local transformations of the Turing tape are allowed.  

Let $\int$ be the set of integers. We will need the following notation: given two maps $\rho,\rho':\int\to\Gamma$ and a set $I\subseteq\int$ we write 
\begin{align}\label{eq:congI}
\rho\equiv\rho'\pmod{I}
\end{align}
if $\rho(i)=\rho'(i)$ for all $i\in I$. We use interval notation
to specify subsets of $\int$, e.g.
\[
  [i-k, i+k] = \{j \mid i-k \leq j \leq i +k\},
\]
and by $\overline I$ denote the complement of $I \subseteq \int$. 
\takeout{%
Specifically, given
$k\geq 0$, we write
\[
  \begin{array}{l@{\qquad}l}
    \sigma\equiv\sigma'\pmod{[i-k,i+k]}
    &
    \text{if $\sigma(j)=\sigma'(j)$ for $i-k\leq j\leq i+k$ and}
    \\
    \sigma\equiv\sigma'\pmod{\overline{[i-k,i+k]}}
    &
    \text{if $\sigma(j)=\sigma'(j)$ for $j<i-k$ and $j>i+k$.}
  \end{array}
\]
}%
Also, for any $\rho:\int\to\Gamma$ and any $i$, let $\rho_{+i}:\int\to\Gamma$ be
such that $\rho_{+i}(j)=\rho(i+j)$. The intuition here is that the
maps $\rho$ and $\rho'$ represent snapshots of a Turing tape being
filled with symbols from $\Gamma$ ($\Gamma$ may contain a special
symbol for a blank cell, but it does not play a role sofar). The
relation~\eqref{eq:congI} indicates that $\rho$ and $\rho'$ agree
on the positions indexed by~$I$. The tape $\rho_{+i}$ is obtained
from $\rho$ by reindexing the cells with the function
$\lambda x.\,x-i$. We also commonly use the notation
$\rho[k\mapsto\gamma_i]$ to refer to $\rho':\int\to\Gamma$
defined by $\rho'(k)=\gamma_i$ and $\rho'(l)=\rho(l)$ if
$l\neq k$. This generalizes to sequences of assignments
$k\mapsto\gamma_i$ in the obvious way.
\begin{definition}[Tape monad, Tape theory]\label{defn:tape}
Let $\Gamma$ be a finite set of tape symbols. The \emph{tape monad (over $\Gamma$)} is the submonad $\BBT$ of the store monad $(\argument\times\int\times\Gamma^\int )^{\int\times\Gamma^\int}$ for which $TX$ consists of exactly those maps 
\[
  p=\brks{r,z,t}:\int\times\Gamma^\int\to(X\times \int\times\Gamma^\int)
\] 
satisfying the following restriction: there exists a $k\geq 0$, which we call a \emph{locality parameter} of $p$, such that for any $i,j\in \int$ and $\rho,\rho':\int\to\Gamma$ if $\rho\equiv\rho'\pmod{[i-k,i+k]}$ then the conditions in Fig.~\ref{fig:tape-monad} are satisfied.
\begin{figure}[t]
\noindent
\begin{flushleft}
\textbf{Locality conditions:}
\end{flushleft}
\begin{flalign*}
\begin{gathered}
t(i,\rho')\equiv t(i,\rho) \pmod{[i-k,i+k]}\hspace{8ex}  t(i,\rho)\equiv\rho\pmod{\overline{[i-k,i+k]}} \\[1ex]
z(i,\rho') = z(i,\rho)\hspace{8ex} |z(i,\rho) - i|\leq k\hspace{8ex} r(i,\rho') = r(i,\rho)
\end{gathered}
\end{flalign*}
\medskip
\noindent
\begin{flushleft}
\textbf{Shift-invariance conditions:}
\end{flushleft}
\begin{flalign*}
\qquad t(i,\rho_{+j})=t(i+j,\rho)_{+j}&&z(i,\rho_{+j})=z(i+j,\rho)-j&& r(i,\rho_{+j})=r(i+j,\rho)\qquad
\end{flalign*}
\caption{Conditions of the tape monad, assuming $\rho\equiv\rho'\pmod{[i-k,i+k]}$.}
\label{fig:tape-monad}
\end{figure}

The \emph{tape signature} w.r.t.\ $\Gamma=\{\gamma_1,\ldots,\gamma_n\}$ consists of the operations $\tmread:{n\to 1}$, $\tmwrite_i:1\to 1$ ($1\leq i\leq n$),  $\tmmove_{k}:1\to 1$ ($k\in\{-1,1\}$), which we interpret over any $TX$ as follows:
\begin{align*}
\tsem{\tmread}(j,\rho) = \brks{i,j,\rho} \text{\quad if $\rho(j) = \gamma_i$}~\quad
\tsem{\tmwrite_i}(j,\rho) = \brks{1,j,\sigma[j\mapsto\gamma_i]}, ~\quad
  \tsem{\tmmove_k}(j,\rho) = \brks{1,j+k,\rho}.
\end{align*}
The \emph{tape theory} w.r.t.\ $\Gamma$ consist of all those equations $p=q$ in the tape signature, which are valid over every $TX$.
\end{definition}
We shall henceforth use $\tmmove_k(p)$ with arbitrary integer $k$ as an abbreviation for $p$ if $k=0$; 
$\tmmove$ nested $k$ times and applied to $p$ if $k>0$; and $\tmmove_{\mone}$ nested $-k$ times and applied to $p$ if $k<0$. 
It is easy to see that the semantic assignments remain intact under such extended use of~$\tmmove_k$.

It is not obvious that Definition~\ref{defn:tape} does indeed define a monad. To show this, we need the following auxiliary fact.
\begin{lemma}\label{lem:pm}
  Suppose, for some $\rho,\rho',\theta,\theta'\colon \int \to \Gamma$
  and $I \subseteq J \subseteq \int$ that
  \begin{align*}
    \rho \equiv \rho' \pmod{J}, && \theta\equiv \theta' \pmod{I},&&
    \rho \equiv \theta  \pmod{\overline I}, && \rho'\equiv \theta'\pmod{\overline I}.
  \end{align*}
  Then $\theta \equiv \theta' \pmod{J}$. 
\end{lemma}
\takeout{%
\begin{lemma}\label{lem:pm}
Suppose, for some $\sigma,\sigma',\theta,\theta'\colon \int\to\Gamma$, $m\geq k$, 
\begin{align*}
\sigma'\equiv\;&\sigma\pmod{[i-m,i+m]}& \theta'\equiv\;&\theta\:\,\pmod{[i-k,i+k]}\\
\theta \equiv\;&\sigma\pmod{\overline{[i-k,i+k]}}& \theta'\equiv\;&\sigma'\pmod{\overline{[i-k,i+k]}}
\end{align*}
Then $\theta'\equiv\theta\pmod{[i-m,i+m]}$. 
\end{lemma}
\begin{proof}
Let us depict the assumptions:
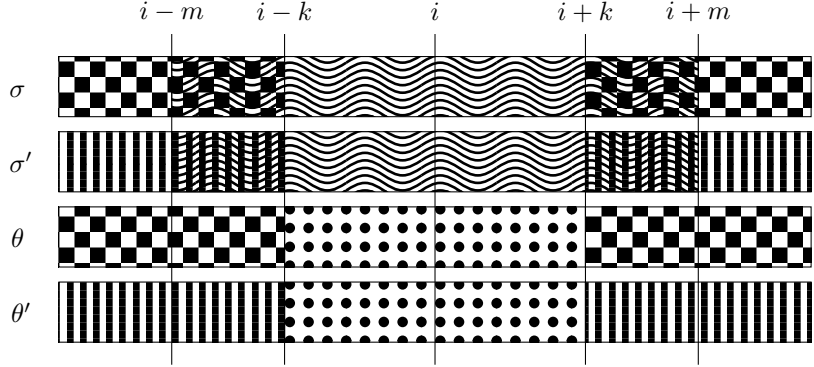
\begin{center}
\begin{tikzpicture}

\filldraw[pattern=vertical stripes] (0,0) rectangle (3,.8);
\filldraw[pattern=polka dot]        (3,0) rectangle (7,.8);
\filldraw[pattern=vertical stripes] (7,0) rectangle (10,.8);

\filldraw[pattern=checkerboard] (0,1) rectangle (3,1.8);
\filldraw[pattern=polka dot]    (3,1) rectangle (7,1.8);
\filldraw[pattern=checkerboard] (7,1) rectangle (10,1.8);

\filldraw[pattern=vertical stripes]   (0,2) rectangle (3,2.8);
\filldraw[pattern=wave]               (1.5,2) rectangle (8.5,2.8);
\filldraw[pattern=vertical stripes]   (7,2) rectangle (10,2.8);

\filldraw[pattern=checkerboard]   (0,3) rectangle (3,3.8);
\filldraw[pattern=wave]           (1.5,3) rectangle (8.5,3.8);
\filldraw[pattern=checkerboard]   (7,3) rectangle (10,3 .8);

\node at (-.5,0.4) (n) {$\theta'$};
\node at (-.5,1.4) (n) {$\theta\phantom{'}$};
\node at (-.5,2.4) (n) {$\sigma'$};
\node at (-.5,3.4) (n) {$\sigma\phantom{'}$};

\draw (5,-.3) to (5,4.1) ++(0,.3) node (n) {$i$};

\draw (1.5,-.3) to (1.5,4.1) ++(0,.3) node (n) {$i-m$};
\draw (3,-.3)   to (3,4.1) ++(0,.3)   node (n) {$i-k$};
\draw (7,-.3)   to (7,4.1) ++(0,.3)   node (n) {$i+k$};
\draw (8.5,-.3) to (8.5,4.1) ++(0,.3) node (n) {$i+m$};

\end{tikzpicture}
\end{center}
Observe that $\sigma'\equiv\sigma\pmod{[i-m,i+m]}$ in conjunction with $m\geq k$ imply that $\sigma'\equiv\sigma\pmod{[i-k,i+k]}$, as depicted above (wave pattern). 

We want to show that for all $j$ such that $|i-j|\leq m$ we have $\theta(j)=\theta'(j)$. We break this into two cases.
\begin{itemize}
\item $|i-j| \leq k$: $\theta'(j)=\theta(j)$ follows by the assumption $\theta\equiv\theta'\pmod{[i-k,i+k]}$ (polka dots in the picture).\\[-1ex]
\item $k< |j-i|\leq m$: $\theta(j)=\theta'(j)$ follows by the following calculation:
\begin{flalign*}
\hspace{1cm}&&\theta(j) &= \sigma(j)\hspace{1cm} & \text{chessboard pattern,~} \theta\equiv\sigma~\hspace{-2ex}  &\pmod{\overline{[i-k,i+k]}}\\
&&&= \sigma'(j)          & \text{wave pattern,~}       \sigma'\equiv\sigma~\hspace{-2ex} &\pmod{[i-m,i+m]}\\[-.1ex]
&&&= \theta'(j)          & \text{vertical stripes pattern,~} \theta'\equiv\sigma'\hspace{-2ex} &\pmod{\overline{[i-k,i+k]}}
\end{flalign*} 
\end{itemize}
We conclude that $\theta'\equiv\theta\pmod{[i-m,i+m]}$. 
\end{proof}}%
Now we can prove that Definition~\ref{defn:tape} correctly defines a monad.
\begin{theorem}
The conditions in Fig.~\ref{fig:tape-monad} identify a submonad $\BBT$ of the store monad over $\int\times\Gamma^\int$.
\end{theorem}
\begin{proof}
  We have to show that the unit and Kleisli lifting of the store monad
  restrict to $T$. First recall the definition of the monad structure
  of the store monad over $\int\times\Gamma^\int$: for any $x \in X$,
  $f: X \to TY$ and $p = \brks{r,z,t} \in TX$,
  \[
  \begin{array}{r@{\ }l@{\qquad}l@{\qquad}r@{\ }l}
    \eta_X(x): & \int\times \Gamma^\int \to X \times\int \times \Gamma^\int
    &\text{with} & \eta_X(x) (i,\rho) &= \brks{x,i,\rho},
    \\
    f^\klstar(p):& \int \times \Gamma^\int \to Y \times \int \times
    \Gamma^\int  &\text{with}  & f^\klstar (p)(i,\rho) &=
      f(r(i,\rho))(z(i,\rho),t(i,\rho)).
  \end{array}
\]
It is our task to prove that the maps $\eta_X(x)$ and $f^\klstar(p)$ lie in $TX$ and $TY$, respectively, i.e.~they satisfy the conditions in Fig.~\ref{fig:tape-monad}.
For $\eta_X(x)$ this clearly holds, for $\eta_X(x) = \brks{r,z,t}$,
where $z$ and $t$ are the left- and right-hand product projections and
$r$  is the constant map on $x \in X$. We proceed to prove this for $f^\klstar(p)$. 

Let $p\in TX$, let $f\colon X\to TY$ and for any $x\in X$ let $r,z,t,r_x,z_x,t_x$ be defined by 
\begin{align*}
p(i,\rho) =\brks{r(i,\rho),z(i,\rho),t(i,\rho)},&&
f(x)(i,\rho)= \brks{r_{x}(i,\rho),z_x(i,\rho),t_x(i,\rho)}.
\end{align*}

\begin{cenumerate}[label=(\arabic*)]
  \item We first show the locality conditions for $f^\klstar(p)$.  Let us fix a locality parameter $k_p$ of $\brks{r,z,t}$. For any $\rho$ and $\rho'$ such that $\rho\equiv\rho'\pmod{[i-k_p,i+k_p]}$ we have that $r(i,\rho') = r(i,\rho)$ by the locality condition for $r$, and hence $f(r(i,\rho')) = f(r(i,\rho))$. Let $k_f$ be a locality parameter of $f(r(i,\rho))$. Finally put $k=k_p+k_f$ and let us verify the locality conditions in Fig.~\ref{fig:tape-monad} for $f^\klstar(p)$ using $k$ as the corresponding locality parameter.
First we calculate using the above notation:
\begin{align*}
 f^\klstar(p)(i,\rho)=&~f(r(i,\rho))(z(i,\rho),t(i,\rho)) \\
  =&~ \brks{r_{r(i,\rho)}(z(i,\rho),t(i,\rho)),
z_{r(i,\rho)}(z(i,\rho),t(i,\rho)),
  t_{r(i,\rho)}(z(i,\rho),t(i,\rho))}\\
 =&~ \brks{r_{x}(j,\theta),z_x(j,\theta),t_x(j,\theta)},
\intertext{where $x=r(i,\rho)$, $j=z(i,\rho)$ and $\theta =t(i,\rho)$ will be fixed from now on. 
Similarly,}
 f^\klstar(p)(i,\rho')=&~f(r(i,\rho'))(z(i,\rho'),t(i,\rho')) \\
  =&~ \brks{r_{x'}(j',\theta'),z_{x'}(j',\theta'),t_{x'}(j',\theta')},
\end{align*}
where we also fix $x'=r(i,\rho')$, $j'=z(i,\rho')$ and $\theta' =t(i,\rho')$. 

\smnote[inline]{Do we want abbreviations for the intervals? It was
  suggested by the referee.}
Let us fix such $\rho,\rho'$ that
$\rho\equiv\rho'\pmod{[i-k,i+k]}$. Note that this implies that
\[
  \rho\equiv\rho'\pmod{[i-k_p,i+k_p]}
  \qquad\text{and}\qquad
  \rho\equiv\rho'\pmod{[i-k_f,i+k_f]},
\]
and therefore we can apply the locality conditions for both $p$ (with locality parameter $k_p$) and for $f(r(i,\rho)) = f(r(i,\rho'))$ (with locality parameter $k_f$). The former immediately implies that 
\begin{align}\label{eq:xj}
x = x' \qquad\text{and}\qquad j=j'.
\end{align}
On the other hand, using the locality condition for $f(r(i,\rho)) = f(r(i,\rho'))$ we obtain: 
\begin{align*}
  \theta' \equiv \theta\pmod{[i-k_f,i+k_f]},~~  
  \theta \equiv \rho \pmod{\overline{[i-k_f,i+k_f]}},~~ %
  \theta' \equiv \rho'\pmod{\overline{[i-k_f,i+k_f]}}.
\end{align*}
Hence, by Lemma~\ref{lem:pm}, we conclude
\begin{align}\label{eq:theta_cong}
\theta'\equiv\theta\pmod{[i-k,i+k]}.
\end{align}
Since $|i-j|\leq k_p$ and $k = k_p + k_f$, the interval $[i- k,i+ k]$
includes $[j- k_f,j+ k_f]$, hence~\eqref{eq:theta_cong} implies
\begin{align}\label{eq:sigt}
\theta'\equiv\theta\pmod{[j-k_f,j+k_f]}.
\end{align}
We proceed to show the locality conditions for $f^\klstar(p)$.
\begin{itemize}[wide]
 \item $t_x(j,\theta)\equiv t_{x'}(j',\theta')\pmod{[i- k,i+ k]}$.
By applying the locality conditions for $f(x)=f(r(i,\rho))$ to~\eqref{eq:sigt} we obtain
\begin{align}
t_x(j,\theta')\equiv&\; t_x(j, \theta)\pmod{[j- k_f,j+ k_f]},\label{eq:lcond1}\\
t_x(j, \theta)\equiv&\;\theta\pmod{\overline{[j- k_f,j+ k_f]}},\label{eq:lcond2}\\
t_x(j,\theta')\equiv&\;\theta'\pmod{\overline{[j- k_f,j+ k_f]}}.\label{eq:lcond3}
\end{align} 
Recall that $|i-j| \leq k_p$ and $k = k_p+k_f$. Then by combining~\eqref{eq:lcond2},~\eqref{eq:lcond3} with~\eqref{eq:theta_cong} we obtain 
\begin{align*}
t_x(j,\theta')\equiv t_x(j, \theta)\pmod{{[i-k,j-k_f)\cup(j+ k_f,i+ k]}}.
\end{align*}
In conjunction with~\eqref{eq:lcond1} and~\eqref{eq:xj} this yields
the desired result.
\item $t_x(j,\theta)\equiv\rho\pmod{\overline{[i- k,i+ k]}}$. Since
  $[i-k_f,i+k_f]$ is contained in $[i-k,i+k]$ this follows
  from~\eqref{eq:lcond2}.
\item $z_x(j,\theta') = z_{x'}(j',\theta')$. By applying the locality
  condition for $f(x)$ to the assumption~\eqref{eq:sigt} we obtain $z_x(j,\theta)= z_{x}(j,\theta')$ and then we are done by~\eqref{eq:xj}. 
\item $|z_x(j,\theta) - i| \leq k$. We estimate the left-hand side as follows:
\begin{align*}
|z_x(j,\theta) - i| 
=\;& |z_x(j,\theta) - z(i,\rho) + z(i,\rho) - i | \\ 
\leq\;&  |z_x(j,\theta) - z(i,\rho)| + |z(i,\rho) - i |\\
=\;&  |z_x(j,\theta) - j| + |z(i,\rho) - i |.  
\end{align*}
By applying the locality conditions for $f(x)$ and $p$ to the assumptions $\rho\equiv\rho'\pmod{[i-k_p,i+k_p]}$ and~\eqref{eq:sigt}, we see that the last sum is smaller or equal than $k_f+k_p=k$.
\item $r_x(j,\theta') = r_{x'}(j',\theta')$. Using the locality
  conditions for $f(x)$ with~\eqref{eq:sigt} as assumption we obtain $r_x(j,\theta)=r_x(j,\theta')$ and hence $r_x(j,\theta)=r_{x'}(j',\theta')$ by~\eqref{eq:xj}.
\end{itemize}

  \item We now prove the shift-invariance conditions for $f^\klstar(p)$. For this we need to compare $f^\klstar(p)(i,\rho_{+m})$ and $f^\klstar(p)(i+m, \rho)$ for any $i, m \in \int$. Similarly as before let us define
\begin{align*}
  x &= r(i,\rho_{+m}) & j &=z(i,\rho_{+m}) & \theta &=t(i,\rho_{+m})  \\
  x' &= r(i+m,\rho) & j' &=z(i+m,\rho) & \theta' &=t(i+m,\rho)  
\end{align*} 
so that
\begin{align*}
  f^\klstar(p)(i,\rho_{+m}) &= \brks{r_{x}(j,\theta),z_x(j,\theta),t_x(j,\theta)}, \\
  f^\klstar(p)(i+m,\rho) &= \brks{r_{x'}(j',\theta'),z_{x'}(j',\theta'),t_{x'}(j',\theta')}.
\end{align*}
Establishing the desired conditions now boils down to proving the following equations:
\begin{align*}
  r_x(j,\theta) = r_{x'}(j',\theta'), &&
  z_x(j,\theta) = z_{x'}(j',\theta') - m, &&
  t_x(j,\theta) = t_{x'}(j',\theta')_{+m}.
\end{align*}
The shift-invariance conditions of $p = \brks{r,z,t}$ state that 
\[
x = x', \qquad j+m = j',\qquad \theta = \theta_{+m}'.
\]
Using these equations and the shift-invariance conditions of $f(x)$ we obtain the desired equations:
\[
\begin{array}{l}
r_x(j,\theta) = r_{x'}(j, \theta) = r_{x'}(j,\theta_{+m}') = r_{x'}(j+m, \theta') = r_{x'}(j',\theta'),\\
z_x(j,\theta) = z_{x'}(j,\theta) = z_{x'}(j,\theta_{+m}') = z_{x'}(j+m,\theta') - m = z_{x'}(j',\theta') - m, \\
t_x(j,\theta) = t_{x'}(j,\theta) = t_{x'}(j,\theta_{+m}') = t_{x'}(j+m, \theta')_{+m} = t_{x'}(j', \theta')_{+m}.
\end{array}
\]
This completes the proof.\qed
\end{cenumerate}
\noqed\end{proof}
In contrast to the stack theory, the tape theory is so far defined
indirectly.  We present the corresponding {\em infinitary} axiomatization
for it in Fig.~\ref{fig:tape-ax}.  Like in the case of the stack
theory, these equations capture semantic equivalences of terms
considered as programs
transforming the underlying store. This implies that composition is to
be read from left to right, e.g.\ $\tmwrite_i(\tmwrite_j(x))$ means
``write~$\gamma_i$, then~$\gamma_j$, then proceed with $x$''.
\begin{figure}[t]
\noindent
\begin{flalign*}
&\textbf{(mv-l)}& \tmmove_{\mone}(\tmmove_{1}(x)) =\;& x 	&\kern-2ex\textbf{(rd-wr)}&& \tmread(\tmwrite_1(x),\ldots,\tmwrite_n(x)) =\;&x \\[.6ex]
&\textbf{(mv-r)}& \tmmove_{1}(\tmmove_{\mone}(x)) =\;& x 	&\kern-2ex\textbf{(wr-rd)}&&\tmwrite_i(\tmread(x_1,\ldots,x_n)) =\;&\tmwrite_i(x_i)\\[.6ex]
&\textbf{(wr-wr)}&\tmwrite_i(\tmwrite_j(x)) =\;&\tmwrite_j(x)
\end{flalign*} 
\vspace{-2.2ex}
\begin{flalign*}
&\textbf{(wr-mv)}\quad\tmwrite_i(\tmmove_k(\tmwrite_j(\tmmove_{\mkah}(x)))) =\tmmove_{k}(\tmwrite_j(\tmmove_{\mkah}(\tmwrite_i(x)))) && %
\end{flalign*} 
\caption{Axioms for the tape monad ($k\in\int\setminus\{0\}$, $i,j\in\{1,\ldots,n\}$).}
\label{fig:tape-ax}
\end{figure}
\begin{theorem}\label{thm:tape_compl}
The deductive closure of the axioms in Fig.~\ref{fig:tape-ax} generates the tape monad over $\Gamma=\{\gamma,\ldots,\gamma_n\}$.
\end{theorem}
Proving Theorem~\ref{thm:tape_compl} requires some preliminaries. Let us introduce the following auxiliary operations: $\tmwrite_{i,k}:1\to 1$, $\tmread_{k}:n\to 1$ with $k$ ranging over all integers and $i$ ranging from $1$ to $n$. These are just abbreviations for the following derived operations: 
\begin{align*}
\tmwrite_{i,k}(x) =&\;\tmmove_k(\tmwrite_i(\tmmove_{\mkah}(x)))\\
\tmread_{k}(x_1,\ldots,x_n) =&\;\tmmove_k(\tmread(\tmmove_{\mkah}(x_1),\ldots,\tmmove_{\mkah}(x_n)))
\end{align*} 
Note that $\tmwrite_{i,0}=\tmwrite_i$, $\tmread_{0}=\tmread$. Clearly,
we have 
\begin{align*}
\tsem{\tmwrite_{i,k}}_T(j,\rho)=&\;\brks{1,j,\rho[j+k\mapsto\gamma_i]},\\
\tsem{\tmread_{k}}_T(j,\rho[j+k\mapsto\gamma_i])=&\;\brks{i,j,\rho[j+k\mapsto\gamma_i]}.%
\end{align*}
It is easy to establish the following implications of the axioms in Fig.~\ref{fig:tape-ax}. %
\begin{lemma}\label{lem:tape_rule}
The following proof rule is sound w.r.t.\ the axioms in Fig.~\ref{fig:tape-ax}:
\[
  \vcenter{
  \infer{s=t}{\tmwrite_{1,k}(s)=\tmwrite_{1,k}(t)\quad\cdots\quad\tmwrite_{n,k}(s)=\tmwrite_{n,k}(t)}}
  \qquad\text{for every $k\in\int$.}
\]
\end{lemma}
\begin{lemma}\label{lem:tape_more}
The following equations are derivable from the ones in Fig.~\ref{fig:tape-ax}.
\begin{flalign}
&&\tmwrite_{i,k}(\tmwrite_{j,k}(x)) 		=&\; \tmwrite_{j,k}(x)\label{eq:tape_more1}\\
&&\tmwrite_{i,k}(\tmwrite_{j,{k'}}(x)) 		=&\; \tmwrite_{j,k'}(\tmwrite_{i,k}(x))&& (k\neq k')&&\label{eq:tape_more2}\\
&&\tmwrite_{i,k}(\tmread_{k}(r_1,\ldots,r_n)) 	=&\;\tmwrite_{i,k}(r_i)\label{eq:tape_more3}\\
&&\tmwrite_{i,k}(\tmread_{k'}(r_1,\ldots,r_n)) 	=&\;\tmread_{k'}(\tmwrite_{i,k}(r_1),\ldots,\tmwrite_{i,k}(r_n))&& (k\neq k')&&\label{eq:tape_more4}
\end{flalign}
\end{lemma}
\begin{proof}[Proof of Theorem~\ref{thm:tape_compl}]
By Theorem~\ref{thm:mon_thm_eq} it suffices to verify the following.
\begin{citemize}
 \item\textit{Soundness.} This is a routine calculation using $\tsem{-}$ from Definition~\ref{dfn:sem}.
 \item\textit{Expressiveness.} 
Recall that for any $p=\brks{r,z,t}:\int\times\Gamma^\int\to(X\times \int\times\Gamma^\int)$ in $TX$ there exists~$k$ for which the conditions in Fig.~\ref{fig:tape-monad} are satisfied. We claim that
\begin{equation}\label{eq:tape_expr}
\begin{aligned}
p = \letTerm{&\;x_{\mkah}\leteq\tsem{\tmread_{\mkah}}_T;\ldots;x_k\leteq\tsem{\tmread_{k}}_T;\\
&\;\tsem{\tmwrite_{t(0,p(x_{\mkah},\ldots,x_k))(\mkah),\mkah}}_T;\ldots;\tsem{\tmwrite_{t(0,p(x_{\mkah},\ldots,x_k))(k),k}}_T;\\
&\;\tsem{\tmmove_{z(0,p(x_{\mkah},\ldots,x_k))}}_T}{\tsem{r(0,h(x_{\mkah},\ldots,x_k))}}~~
\end{aligned}
\end{equation}
(slightly abusing the notation by writing $\tmwrite_{\gamma_i,j}$ in
lieu of $\tmwrite_{i,j}$) where $h:\Gamma^{2k+1}\to \Gamma^\int$ is
any map for which
$h(\gamma_{i_{\mkah}},\ldots,\gamma_{i_{k}})(j)=\gamma_{i_{j}}$ whenever
${-k\leq j\leq k}$. Intuitively, the constructed program works as
follows: in the first step it reads values from the interval $[-k,k]$
on the tape relative to the current head position, and stores the
obtained results in $x_{\mkah},\ldots,x_k$; in the step round it
updated the tape according to $t$; in the third step, it moves the
head according to $z$; and in the final fourth step, it returns the
result from $X$ computed by $r$.

Once we prove~\eqref{eq:tape_expr} we are done with the proof of      
expressiveness; indeed, recall from Definition~\ref{dfn:sem} that for any $f:   
m \to 1$ and any family of terms $t_1,\ldots t_m$, 
\begin{displaymath}
\tsem{x} = \eta_X(x)\qquad  \tsem{f(t_1,\ldots,t_m)}  
= \letTerm{i\leteq\tsem{f}_T}{\tsem{t_i}}. 
\end{displaymath}
Then by straightforward induction, the right-hand side of~\eqref{eq:tape_expr} is $\tsem{t}$   
for some term $t$.                                                              

Now we prove~\eqref{eq:tape_expr}. Let $j\in\int$ and let
$\rho:\int\to\Gamma$. Applying the right-hand side
of~\eqref{eq:tape_expr} to $\brks{j,\rho}$ and using the semantics
of $\tmread$ we obtain
\begin{align*}
\bigl(\letTerm{\tsem{\tmwrite_{t(0,\theta)(-k),-k}}_T;\ldots;\tsem{\tmwrite_{t(0,\theta)(-k),-k}}_T;\tsem{\tmmove_{z(0,\theta)}}_T}{\tsem{r(0,\theta)}}\bigr)\brks{j,\rho}
\end{align*}
for some $\theta:\int\to\Gamma$ such that
\begin{equation}\label{eq:mod}
  \rho_{+j} \equiv \theta \mod{[-k,k]}.
\end{equation}
Using the semantics of $\tmwrite$ we further reduce the right-hand side of~\eqref{eq:tape_expr} to
\begin{align*}
\bigl(\letTerm{\tsem{\tmmove_{z(0,\theta)}}_T}{\tsem{r(0,\theta)}}\bigr)\brks{j,\rho[j-k\mapsto t(0,\theta)(-k),\ldots,j+k\mapsto t(0,\theta)(k)]}
\end{align*}
and the latter is equal to
\begin{align*}
\brks{r(0,\theta), j+z(0,\theta),\rho[j-k\mapsto t(0,\theta)(-k),\ldots,j+k\mapsto t(0,\theta)(k)]}.
\end{align*}
It remains to show that this is equal to
$p(j,\rho)=\brks{r(j,\rho),z(j,\rho),t(j,\rho)}$. Consider the
first component: using shift-invariance and the locality condition
with assumption~\eqref{eq:mod} for $r$ we have
$
r(j,\rho) = r(0,\rho_{+j}) = r(0,\theta)
$.

Analogously, for the second component, $z(j,\rho)=j+z(0,\rho_{+j})=j+z(0,\theta)$. 

Finally, consider the third component. We shall prove separately that
\begin{align*}
t(j,\rho)\equiv \rho[j-k\mapsto t(0,\theta)(-k),\ldots,j+k\mapsto t(0,\theta)(k)]\mod{[j-k,j+k]}\\
t(j,\rho)\equiv \rho[j-k\mapsto t(0,\theta)(-k),\ldots,j+k\mapsto t(0,\theta)(k)]\mod{\overline{[j-k,j+k]}}
\end{align*}
The first congruence is equivalent to 
\begin{align*}
t(j,\rho)_{+j}\equiv\rho_{+j}[-k\mapsto t(0,\theta)(-k),\ldots,k\mapsto t(0,\theta)(k)] \equiv t(0,\theta)\mod{[-k,k]}.
\end{align*}
The last congruence clearly holds: by shift-invariance for $t$, $t(j,\rho)_{+j}=t(0,\rho_{+j})$ and by the first locality condition for $t$, $t(0,\rho_{+j})\equiv t(0,\theta)\mod{[-k,k]}$ (using~\eqref{eq:mod}). 

Let us check the second congruence. It is equivalent to $t(j,\rho)_{+j}\equiv\rho_{+j} \mod{\overline{[-k,k]}}$. Like before $t(j,\rho)_{+j}=t(0,\rho_{+j})$ and using the second locality condition for $t$, we have $t(0,\rho_{+j})\equiv \rho_{+j}\mod{\overline{[-k,k]}}$ which completes the proof of expressiveness. 

 \item\textit{Completeness.} Let $s$ and $t$ be tape theory terms such that $\tsem{s} = \tsem{t}$. We have to show $s = t$. Consider first the case where both $s$ and $t$ are \emph{normal}, that means that $s$ and $t$ are composed from $\tmread_k$, $\tmwrite_{i,k}$, $\tmmove_k$ and variables in such a way that (i) no $\tmread$ occurs under $\tmwrite$; (ii) $\tmmove$ is only applied to variables.
We proceed by induction over the total number of $\tmread$ operations in $\brks{s,t}$.

In the base case neither $s$ nor $t$ contains $\tmread$ and hence
\begin{align*}
s=\tmwrite_{k_1,i_1}(\,\ldots(\tmwrite_{k_m,i_m}(\tmmove_k(x)))\ldots),&&
t=\tmwrite_{l_1,j_1}(\,\ldots(\tmwrite_{l_w,j_w}(\tmmove_l(y)))\ldots)
\end{align*} 
with suitable indices and variables $x,y$. W.l.o.g.\ we can assume
that the sequences $k_1,\ldots,k_m$ and $l_1,\ldots,l_w$ are
increasing and nonrepetetive -- otherwise we can rearrange and
possibly remove some of the $\tmwrite$ operators by
Lemma~\ref{lem:tape_more}. Then we argue that $s$ must be provably
equal to $t$. We have for all $v$ and $\rho$ that
$\tsem{s}(v,\rho)=\brks{x,v+k,\rho_1}$ and
$\tsem{t}(v,\rho)=\brks{y,v+l,\rho_2}$ for some
$x,y, k, l, \rho_1$ and $\rho_2$. By hypothesis we have $k=l$,
$x=y$ and $\rho_1 = \rho_2$. Moreover,
$\rho_1=\rho[k_1\mapsto\gamma_{i_1},\ldots,k_m\mapsto\gamma_{i_m}]$
and
$\rho_2=\rho[l_1\mapsto\gamma_{j_1},\ldots,l_w\mapsto\gamma_{j_w}]$. Since
these are equal for all $\rho$ it follows that the sequences
$\brks{k_1,i_1},\ldots,\brks{k_m,i_m}$ and
$\brks{l_1,j_1},\ldots,\brks{l_w,j_w}$ must be equal, too.

For the induction step, let $s=\tmread_k(r_1,\ldots,r_n)$. We then apply $\tmwrite_{i,k}$ to $s$ and $t$ for every~$i$. Note that 
any term~$\tmwrite_{i,k}(s)$ can be brought to a normal form $s_i$ using~\eqref{eq:tape_more3} and~\eqref{eq:tape_more4} as rewrite rules: 
\begin{align}
\tmwrite_{i,k}(\tmread_{k}(r_1,\ldots,r_n)) \to\;&\tmwrite_{i,k}(r_i)\label{eq:wr_red_rewr1}\\
\tmwrite_{i,k}(\tmread_{k'}(r_1,\ldots,r_n)) \to\;&\tmread_{k'}(\tmwrite_{i,k}(r_1),\ldots,\tmwrite_{i,k}(r_n))\label{eq:wr_red_rewr2}
\end{align}
where $k'\neq k$. Analogously, any $\tmwrite_{i,k}(t)$ reduces to a normal form $t_i$. Since $\tsem{s} = \tsem{t}$, for every $i$, 
\[
\tsem{s_i} = \tsem{\tmwrite_{i,k}(s)} = \tsem{\tmwrite_{i,k}(t)} = \tsem{t_i}.
\]
Now notice that $s_i$ has at least one $\tmread$ operator less than
$s$; thus, the total number of $\tmread$ operators in
$\langle s_i, t_i\rangle$ is lower than that of $\langle s,
t\rangle$. Hence, by induction hypothesis, the identities
\[
\tmwrite_{i,k}(s) = s_i = t_i = \tmwrite_{i,k}(t)
\]
belong to the tape theory for every $i$. By Lemma~\ref{lem:tape_rule}, $s= t$ is a provable identity as desired.

The remaining case $t=\tmread_k(t_1,\ldots,t_n)$ is symmetric to the previous one.

In order to complete the proof it remains to show how an arbitrary tape theory term $t$ can be reduced to a normal form satisfying the above conditions (i) and (ii) in such a way that the reductions are sound w.r.t.\ the identities in Fig.~\ref{fig:tape-ax}. Given $t$ we ensure first (ii) and then (i) as follows.

Ad~(ii). We exhaustively apply the reductions
\begin{align*}
\tmmove_k(\tmmove_{l}(s))\to&\;\tmmove_{k+l}(s)\\
\tmmove_k(\tmwrite_{i,l}(s))\to&\;\tmwrite_{i,k+l}(\tmmove_{k}(s))\\
\tmmove_k(\tmread_{l}(s_1,\ldots,s_n))\to&\;\tmread_{k+l}(\tmmove_k(s_1),\ldots,\tmmove_k(s_n))
\end{align*}
which are easily seen to be sound by~\textbf{(mv-l)} and~\textbf{(mv-r)}.

Ad~(i). Then we exhaustively apply~\eqref{eq:wr_red_rewr1} and~\eqref{eq:wr_red_rewr2}.\qed
\end{citemize} 
\noqed\end{proof}
It can now be readily shown that any axiomatization of the tape monad is necessarily
infinitary.
\begin{theorem}\label{thm:nfa}
The tape theory over $\Gamma$ is not finitely axiomatizable, unless ${|\Gamma|\leq 1}$.
\end{theorem} 
\begin{proof}
If $|\Gamma|=0$ then the axiom scheme~\textbf{(wr-mv)} disappears instantly, and 
if $|\Gamma|=1$ then it is entailed by the axioms~\mbox{\textbf{(mv-l)}},~\mbox{\textbf{(mv-r)}}
and the identity $\tmwrite(x)=x$ (we omit the index $1$ at $\tmwrite$); this identity is derived
as follows using~\textbf{(rd-wr)} and~\textbf{(wr-wr)}:
\[
  x=\tmread(\tmwrite(x))=\tmread(\tmwrite(\tmwrite(x))) = \tmwrite(x).
\]
Note that for $|\Gamma|=0$, the monad becomes trivial ($TX=\emptyset$) and 
for $|\Gamma|=1$, $TX = X\times\int$. 

Let us assume henceforth that $|\Gamma|\geq 2$.
Given a finite set of identities $\CA$ belonging to the tape theory, we prove the claim by 
constructing a model $M$ of $\CA$ which does not satisfy all instances of~\textbf{(wr-mv)}.
Let $m$ be greater than the total number of instances of operations
$\tmmove_{\mone}$ and $\tmmove_1$ in any equation from~$\CA$. Our
model $M$ is carried by the set of all endomaps on a tape of length
$m$, i.e.~all endomaps on the set
$\int_m\times\Gamma^{\int_m}\to\int_m\times\Gamma^{\int_m}$, where
$\int_m=\{0,\ldots,m-1\}$ is the finite ring of integers modulo
$m$. We interpret the operations of the tape theory on $M$ (here we
overload our previous notation $\tsem{-}_{TX}$ and write $\tsem{-}_M$
for this interpretation) as follows:
\begin{align*}
  \tsem{\tmread}_M(p_1,\ldots,p_n)(z,\rho)
  &= p_{i}(z,\rho),\qquad\qquad \text{where $\rho(z)=\gamma_i$},\\
  \tsem{\tmwrite_i}_M(p)(z,\rho)
  &= p(z,\rho[z\mapsto\gamma_i])\\
  \tsem{\tmmove_k}_M(p)(z,\rho) &= p(z+_mk,\rho)
\end{align*}
where $i$ ranges from $1$ to $n=|\Gamma|$ and $+_m$ denotes addition modulo $m$. %
By additionally defining $\tsem{x}_M = \id$ for every variable
$x$, we extend
$\tsem{-}_{M}$ to terms over the tape signature. The inductive clauses
for
$\tsem{p}_M(z,\rho)$ are the same as for
$\tsem{p}_{\scriptscriptstyle
  T1}(z,\rho)$, except that the tuples returned by the latter
interpretation are extended to the left with an additional component
constantly equal~$1$, and now
$\tsem{p}_M(z,\rho)$ may call on addition modulo
$m$ for sufficiently large
$z$ and sufficiently many operations $\tmmove_k$ in
$p$. Specifically, this means that for any equation $p=q$ in
$\CA$, any $z\in \int$ and any $\theta:\int_m\to\Gamma$,
\begin{align*}
  \tsem{p}_{M}(0,\theta) = \brks{z',\theta'} &\text{\qquad iff \qquad} \tsem{p}_{T1}(0,\theta_*) = \brks{1, z,\theta'_*}
\end{align*}
for $\theta_*,\theta_*':\int\to\Gamma$ defined as follows:
\[
  \theta_*(i) = \theta (i \mathbin{\mathsf{mod}} m)
  \qquad\text{and}\qquad
  \theta_*'(i) = \theta'(i \mathbin{\mathsf{mod}} m)
  \qquad\text{for every $i \in \int$.}
\]
An analogous identity
holds for~$q$ and therefore
\begin{equation}\label{eq:rho0}
  \tsem{p}_{M}(0,\theta) = \tsem{q}_{M}(0,\theta)
  \qquad\text{for every $\theta: \int_m \to \Gamma$.}
\end{equation}
Now note that, for any $z \in \int_m$, $\theta\in \Gamma^{\int_m}$, in
order to compute $\tsem{p}_M$ on $(z,\theta)$ one can first perform a
cyclic left-shift on the model, then apply $\tsem{p}_M$ with $0$ it its
first argument and then shift the result back to the right. More
precisely, let $\theta_{+z}(i) = \theta(i +_m z)$ for every
$i, z \in \int$ and $\theta: \int \to \Gamma$ (in analogy the same
notation $\rho_{+z}$ we previously used for
$\rho: \int \to \Gamma$). Then we have
\[
  \tsem{p}_M(z,\theta) = \brks{z' +_m z, \theta_{-z}'},
  \qquad
  \text{where $\brks{z',\theta'} = \tsem{p}_M\brks{0,\theta_{+z}}$.}
\]
This can be shown by a straightforward induction over the term $p$.

Hence, from~\eqref{eq:rho0}, we obtain that
$\tsem{p}_M(z,\theta)=\tsem{q}_M(z,\theta)$ for every
$\theta:\int_m\to\Gamma$ and $z\in\int_m$. We have thus shown that $\CA$ is valid over $M$. 

Now, if we take $k=m$ in~\textbf{(wr-mv)} we obtain that for $i \neq j$ (such a pair of indices exists for $|\Gamma|\geq 2$, by assumption):
\begin{align*}
\tsem{\tmwrite_i(\tmmove_m(\tmwrite_j(\tmmove_{\mmm}(x))))}_{M}(0,\rho)
=\;&\brks{0,\rho[0\mapsto \gamma_j]}\neq\\
\tsem{\tmmove_m(\tmwrite_j(\tmmove_{\mmm}(\tmwrite_i(x))))}_{M}(0,\rho) =\;&\brks{0,\rho[0\mapsto \gamma_i]}
\end{align*}
This concludes the proof.
\end{proof}

\section{\texorpdfstring{Reactive $\BBT$-algebras and $\BBT$-automata}{Reactive $T$-algebras and $T$-automata}}\label{sec:react}
\noindent As in Section~\ref{sec:pre} we fix a finite set of actions $A$. We first consider $\BBT$-algebras which are equipped with a transition structure similar to that of Moore automata but which, in addition, preserves the algebraic structure. Such a transition structure extends a $\BBT$-algebra with dynamic behaviour (making it into a coalgebra) and hence we call such structures \emph{reactive} $\BBT$-algebras.
\begin{definition}[Reactive $\BBT$-algebra]
  Let $B$ and $X$ be $\BBT$\dash algebras. Then $X$ is a
  \emph{reactive $\BBT$\dash algebra} if~$X$ is an coalgebra for
  $\Lfun = B \times (-)^A$ (cf.~Definition~\ref{defn:tsem}) for which
  $\partial_a:X\to X$ and $o:X\to B$ are $\BBT$-algebra morphisms.
\end{definition}
\iffull
\begin{remark}\label{rem:react}
  The definition of a reactive $\BBT$-algebra is an instance of a more
  general construction~\cite{Bartels04} (the main idea goes back to
  Turi and Plotkin~\citeyear{TuriPlotkin97}). Any endofunctor
  $F:\Set\to\Set$ equipped with a distributive law
  $\delta:\BBT F\to F\BBT$ is known to lift to the Eilenberg-Moore
  category $\Set^{\BBT}$. Under $F=\Lfun$ there is a standard
  distributive law, given by
\begin{align*}
T(B \times X^A)~\xrightarrow{\brks{T\pi_0, T\pi_1}} 
    ~TB \times T(X^A) ~\xrightarrow{\alpha\times \brks{T\mathsf{ev}_a}_{a \in
        A}}  
    B \times (TX)^A
\end{align*}
where $\pi_0, \pi_1$ denote the product projections, $\alpha:TB\to B$ is the $\BBT$-algebra structure on $B$,
$\mathsf{ev}_a: X^A \to X$ is the obvious evaluation at $a \in A$,
and we regard $(TX)^A$ as the $|A|$-fold power of $TX$.
A reactive $\BBT$-algebra is then simply a coalgebra in $\Set^{\BBT}$
for the lifting of $\Lfun$. Putting it yet differently, a reactive
$\BBT$-algebra is a \emph{$\delta$\dash bialgebra} for the above
distributive law $\delta$~\cite{Jacobs06} (see also~\cite{Klin11}). 
\end{remark}

\noindent
\else
Observe that the functor $LX = B \times X^A$ lifts to $\Set^\BBT$; in fact, each $LX$ can be equipped with the pointwise $\BBT$-algebra structure. Thus, reactive $\BBT$-algebras are simply coalgebras for this lifting of $L$ to $\Set^\BBT$.

\fi%
Given a $\BBT$-algebra $B$, the set of all formal power series $B^{A^*}$ (which is the carrier of the final $\Lfun$-coalgebra in $\Set$) can also be viewed as a reactive $\BBT$-algebra with a pointwise $\BBT$-algebra structure. The morphisms $\partial_a$ and $o$ are easily seen to be $\BBT$-algebra morphisms.
Since every reactive $\BBT$-algebra is an $\Lfun$\dash coalgebra,
reactive $\BBT$\dash algebras inherit the general coalgebraic theory
from Section~\ref{sec:pre}. In particular, we use for reactive
$\BBT$\dash algebras the same notions of language semantics and
language equivalence as for $\Lfun$-coalgebras (see
Definition~\ref{defn:tsem}).
\begin{definition}[$\BBT$-automaton, cf.~\cite{Jacobs06}]\label{defn:taut}
Suppose, \iffull\/$\BBT$ is finitary and \fi $B$ is finitely generated, i.e.\ there is a finite set $B_0$ of \emph{generators} and a surjection $TB_0\to B$ underlying a $\BBT$-algebra morphism. 
A \emph{$\BBT$-automaton} $\M$ is given by a triple of maps 
\begin{align}\label{eq:aut}\tag{$\bigstar$}
o^\M:X\to B,&&  t^\M:A\times X\to TX,&&  
\algstruc:TB\to B,
\end{align}
where $\algstruc$ is a $\BBT$-algebra and $X$ is finite. %
The first two maps in~\eqref{eq:aut} can be aggregated into a coalgebra transition structure, which we write as 
\[
\M:X\to B\times (TX)^{A}
\]
slightly abusing the notation.%
\end{definition}
\iffull
\begin{remark}\label{rem:fincoref}
  We require the monad $\BBT$ in~\eqref{eq:aut} to be finitary
  in order to be able to represent $\BBT$-automata using finite
  syntax. For technical reasons, it is sometimes convenient to drop
  this restriction (e.g.~in Section~\ref{sec:cps} where $\BBT$ is the
  continuation monad). This is not in conflict with
  Definition~\ref{defn:taut}, since we apply~$\BBT$ to finite sets
  only, and therefore, in lieu of\/ $\BBT$, we can use its finitary
  coreflection $\BBT_{\omega}$ whose object part is defined by
  $T_{\omega} X = \bigcup_{Y\subseteq X,|Y|<\omega} T Y$.
\end{remark}
\fi

\noindent
A simple nontrivial example of a $\BBT$-automaton is given with the \emph{nondeterministic finite state machines (NFSM)} by taking $B=\{0,1\}$, $\BBT=\PFin$ and $\algstruc(s\subseteq\{0,1\})=1$ iff $1\in s$.

In order to introduce the language semantics of a $\BBT$-automaton we will first convert it into a reactive $\BBT$-algebra, and the language semantics of the latter is settled by Definition~\ref{defn:tsem}. This conversion is called the \emph{generalized powerset construction}~\cite{SilvaBonchiEtAl13}, as it generalizes the classical Rabin-Scott NFSM determinization~\cite{RabinScott59} and amounts to the following.
Observe that $\Lfun TX$ is a $\BBT$-algebra, since $TX$ is the free $\BBT$-algebra on $X$ and $\Lfun$ lifts to $\Set^{\BBT}$\iffull\ (see Remark~\ref{rem:react})\fi. Hence, given a $\BBT$-automaton $\M:X\to B\times (TX)^A$ there exists a unique $\BBT$-algebra morphism 
\begin{align*}
\M^\sharp:TX\to B\times (TX)^A
\end{align*}
such that $\M^\sharp \cdot \eta_X = \M$; explicitly, $\M^\sharp(p) = (B \times \mu_X^A) \cdot \delta_{TX} \cdot T\M$ where
$\delta$ is the distributive law from Remark~\ref{rem:react}.
This $\M^\sharp$ is a reactive $\BBT$-algebra on $TX$.
\begin{definition}
  Given a $\BBT$-automaton~$\M: X \to B \times (TX)^A$, its
  \emph{language semantics} assigns to every state $x \in X$ the
  formal power series
\begin{align*}
\sem{x}_{\M} =\widehat\M^\sharp(\eta_X(x)): A^* \to B,%
\end{align*}
where $\widehat\M^\sharp$ is the unique $L$-coalgebra morphism from $(TX,\M^\sharp)$ to the final coalgebra $(B^{A^*},\iota)$. This can be summarized in the following diagram
\begin{equation}\label{diag:hatm}
\begin{tikzcd}[column sep=large, row sep=normal] %
X\rar["\eta_X"]\dar["\M"']\ar[rr, controls={++(60:1) and ++(110:.3)}, "\sem{-}_\M"] & TX\ar[dl, "\M^\sharp"']\rar["\widehat\M^\sharp"] & B^{A^*}\dar["\iota"] \\
B\times (TX)^A\ar[rr, "B\times(\widehat \M^\sharp)^A"] && B\times (B^{A^*})^A
\end{tikzcd}
\end{equation}

\end{definition}

\begin{remark}\label{rem:raaut}
  \begin{enumerate}%
  \item Due to the 1-1-correspondence of $\M$ and $\M^\sharp$ given by
    freeness of $TX$, $\BBT$-automata bijectively correspond to
    reactive $\BBT$-algebras whose carrier is a free algebra on a
    finite set; but we find it useful to retain the distinction.
  \item The term language semantics comes from the fact that for
    $\BBT = \PFin$ and $B = \{0,1\}$, our language semantics of\/
    $\BBT$-automata is precisely the classical language semantics of
    NFSM; $\sem{x}_{\M}$ is the formal language accepted by the NFSM
    given by $\M$ with initial state $x$.

    More generally, for any semiring $R$, take $B=R$ and the
    semimodule monad $\BBT_R$. Then $\BBT$-automata are precisely weighted
    automata with weights in $R$, and for every state $x$ the
    formal power-series $\sem{x}_\M: A^* \to R$ is the weighted
    language accepted by the weighted automaton given by~$\M$.

    However, for other monads $\BBT$ and algebras $B$ elements in
    $B^{A^*}$ may look very different than formal languages, e.g.~for
    the stack $\BBT$-automata we will discuss in
    Section~\ref{sec:stackT}.
  \item Note that $\BBT$-automata for the identity monad are precisely
    the same as Moore automata, and the above definition of their
    language semantics coincides with Definition~\ref{defn:tsem}.
  \end{enumerate}
\end{remark}
Note that the generalized powerset construction does not reduce a
$\BBT$-automaton to a Moore automaton over $TX$ as $TX$ need not be
finite. However, when this is the case, e.g.~for $\BBT=\PFin$, the
semantics of a $\BBT$-automaton falls within regular power series,
which is precisely the reason why the languages recognized by
deterministic and nondeterministic FSM coincide. Surprisingly, all
$\BBT$\dash automata with a finite $B$ have the same property:
\begin{proposition}\label{prop:fin_b}
For every $\BBT$-automaton~\eqref{eq:aut} with finite $B$ and $x\in X$, $\sem{x}_{\M}:A^*\to B$ is regular.
\end{proposition}
We will present the proof of this proposition after Corollary~\ref{cor:aut_cont}. 

We are now ready to introduce fixpoint expressions for $\BBT$-automata similar to~\eqref{eq:kleene}.
\begin{definition}[Reactive expressions]\label{def:react}
Let $\Sigma$ be an algebraic signature and let $B_0$ be a finite set. \emph{Reactive expressions} w.r.t.\ these data are closed terms $\delta$ defined according to the following grammar:
\begin{flalign*}%
&&\delta \Coloneqq&~ x\mid\gamma\mid f(\delta,\ldots,\delta)&(x\in X,f\in\Sigma) \\*
&&\gamma \Coloneqq&~ \mu x.\,a_1.\delta\pitchfork\ldots\pitchfork a_{n}.\delta\pitchfork \beta\quad&(x\in X)\\*
&&\beta \Coloneqq &~ b \mid f(\beta, \ldots, \beta) &(b \in B_0, f \in \Sigma)
\end{flalign*}
where we assume $A=\{a_1,\ldots,a_n\}$ and an infinite collection of variables $X$. 
\iffull Free and bound variables here are defined in the standard way. We do not distinguish expressions equivalent under $\alpha$\dash conversion (i.e.~renaming of bound variables).\fi
\end{definition}
\begin{notation}\label{not:terms}
\begin{enumerate}[wide] \item Let $t$ be a $\Sigma$-term over $\{1,\ldots,n\}$ (i.e.\ the numbers $1,\ldots,n$ are identified as variables) and let $t_1, \ldots, t_n$ be any $\Sigma$-terms. Then we write $t(t_1,\ldots,t_n)$ for $t[t_1/1,\ldots,t_n/n]$.
\item For every $\Sigma$-algebra $A$ (so, in particular for every
  $\BBT$-algebra, where $\Sigma$ is part of a presentation of $\BBT$) we write $f^A: A^n \to A$ for the operation
  associated to $f: n \to 1$ from $\Sigma$. We also write
  $t^A: A^n \to A$ for the map evaluating the $\Sigma$-term $t$ over
  $\{1, \ldots, n\}$ in $A$.
  \item Finally, we shall sometimes call $\Sigma$-terms over a set $X$ of variables simply $\Sigma$-terms.
  \end{enumerate}
\end{notation}
Observe that a reactive expression can be uniquely represented in the form $t(e_1,\ldots,e_n)$ where $e_1,\ldots,e_n$ are reactive expressions starting with $\mu$.

Let $\BBT$ be a finitary monad, generated by an algebraic theory $\CE$ over the signature $\Sigma$ and let $B$ be a finitely generated $\BBT$\dash algebra over a finite set of generators $B_0$ (witnessed by the surjective $\BBT$-algebra morphism $h: TB_0 \to B$). Let us denote by $\Exp{\Sigma}{B_0}$ the set of all reactive expressions over $\Sigma$ and $B_0$. We aim to define a reactive $\BBT$-algebra structure on a suitable quotient of $\Exp{\Sigma}{B_0}$. 
\takeout{ %
First, we interpret the operations from $\Sigma$ on $\Exp{\Sigma}{B_0}$ thus making it a $\Sigma$-algebra by putting
\begin{align*}
\tsem{f}(e_1,\ldots, e_k) = \mu x.\,\bigl(a_1.f(e_1^1,\ldots, e_1^k)&\pitchfork\ldots\pitchfork\\ a_{n}.f(e_n^1,\ldots, e_n^k)&\pitchfork f(b_1,\ldots, b_k)\bigr)
\end{align*}
where $e_i=\mu x.\,\bigl(a_1.e_1^i\pitchfork\ldots\pitchfork a_{n}.e_n^i\pitchfork b_i\bigr)$. This extends in the standard way to $\Sigma$-terms: for any $\Sigma$-term $t$ in $k$ variables we have the map $\tsem{t}: (\Exp{\Sigma}{B_0})^k \to \Exp{\Sigma}{B_0}$.}%
First, notice that $\Exp{\Sigma}{B_0}$ is obviously a $\Sigma$-algebra. Then we introduce an $\Lfun$-transition structure on $\Exp{\Sigma}{B_0}$ as follows: notice that expressions $b$ from the $\beta$-clause in Definition~\ref{def:react} are just $\Sigma$-terms on the generators from $B_0$. %
Recall also that $B$ is a surjective image of $TB_0$ and let $b^B$ be the image of $b\in B_0$ under
\begin{equation*}
\begin{tikzcd}[column sep=large, row sep=normal]
B_0 \rar["\eta_{B_0}"] & TB_0 \rar[twoheadrightarrow, "h"] & B.
\end{tikzcd}
\end{equation*}
This extends to arbitrary $\Sigma$-terms over $B_0$ by putting
$
(t(b_1,\ldots,b_k))^B=t^B(b_1^B,\ldots,b_k^B)
$.
Then let us define
\begin{gather}\label{eq:partial}
\begin{split}
o(f(e_1, \cdots, e_n)) =&~f^B(o(e_1), \ldots, o(e_n)),\\
\partial_{a_i}(f(e_1, \cdots, e_n)) =&~f(\partial_{a_i}(e_1), \ldots, \partial_{a_i}(e_n)), \\
o(\mu x.\,(a_1.e_1\pitchfork\ldots\pitchfork a_{n}.e_n\pitchfork b)) =&~b^B,\\
\!\!\partial_{a_i}(\mu x.\,(a_1.e_1\pitchfork\ldots\pitchfork a_{n}.e_n\pitchfork b)) = &~e_i[\mu x.\,(a_1.e_1\pitchfork\ldots\pitchfork a_{n}.e_n\pitchfork b)/x].
\end{split}
\end{gather}
This defines an $L$-transition structure
$s: \Exp{\Sigma}{B_0} \to B \times (\Exp{\Sigma}{B_0})^A$ and so
$\widehat s: \Exp{\Sigma}{B_0} \to B^{A^*}$ provides language
semantics to expressions and a language equivalence relation $\sim$ on
them according to Definition~\ref{defn:tsem}.

\begin{notation}
We overload notation and write $\sem{e}$ (i.e.~$\sem{-}$ with no subscripts) 
for the formal power series $\widehat s(e)$ denoted by the expression $e$.
\end{notation}
Note that the first two equations in~\eqref{eq:partial} above imply
that the $L$-transition structure $s$ is a $\Sigma$-algebra
homomorphism.
\begin{remark}
  Recall that the category of\/ $\Sigma$-algebras (and its full subcategory of all $\BBT$-algebras) has image factorizations. That means that every $\Sigma$-algebra morphism $f: A \to B$ can be factorized as a surjective $\Sigma$-algebra morphism $e: A \epito C$ followed by an injective one $m: C \monoto B$. This factorization system has the usual \emph{diagonalization} property: given a commutative square $m \cdot f = g \cdot e$ with $m$ injective and $e$ surjective we have a unique diagonal $d$ with $m \cdot d = g$ and $d \cdot e = f$. See e.g.~Ad\'amek, Herrlich and Strecker~\citeyear{AdamekHerrlichEtAl90} for basics on factorization systems. 
\end{remark}
\begin{theorem}\label{thm:expr}
The quotient $\Exp{\Sigma}{B_0}/\mathord{\sim}$ is a reactive $\BBT$-algebra whose $\Lfun$-coalgebra part is inherited from $\Exp{\Sigma}{B_0}$ and whose $\BBT$-algebra part is a quotient of the $\Sigma$-algebra structure on~$\Exp{\Sigma}{B_0}$.
\end{theorem}
\begin{proof}
Recall first that $\BBT$-algebras, being the variety of $\Sigma$-algebras 
satisfying the equations in~$\CE$, form a full subcategory of
the category of $\Sigma$-algebras. 
We have seen that $\Exp{\Sigma}{B_0}$ is a coalgebra for the lifting
of $L$ to the category of $\Sigma$-algebras and that the final
coalgebra for the lifting is $B^{A^*}$ (its $L$-transition structure
is a $\Sigma$-algebra morphism since it is a $\BBT$-algebra
morphism). Thus, the language semantics map $\sem{-}: \Exp{\Sigma}{B_0}
\to B^{A^*}$ is a $\Sigma$-algebra morphism. The quotient
$\Exp{\Sigma}{B_0}/\mathord{\sim}$ is obtained by taking its
factorization into a surjective followed by an injective
$\Sigma$-algebra morphism: 
\begin{equation*}
\begin{tikzcd}[column sep=large, row sep=normal]
\Exp{\Sigma}{B_0} \rar[twoheadrightarrow, "q"] & 
\Exp{\Sigma}{B_0}/\mathord{\sim} \rar[hook, "m"] & B^{A^*}
\end{tikzcd}
\end{equation*}
Since (the lifting of) $L$ preserves monos we obtain an $L$-transition
structure on the quotient by diagonalization:
\begin{equation*}
\begin{tikzcd}[column sep=large, row sep=normal]
\Exp{\Sigma}{B_0} 
  \rar["\brks{o, \partial}"] 
  \dar[twoheadrightarrow, "q"'] &
L(\Exp{\Sigma}{B_0})\dar["Lq"]\\
\Exp{\Sigma}{B_0}/\mathord{\sim} 
  \rar[r, dashed] 
  \dar[d, "m"'] &
L(\Exp{\Sigma}{B_0}/\mathord{\sim})\dar[hook, "Lm"]\\
B^{A^*} \rar["\brks{o, \partial}"] & L(B^{A^*})               
\end{tikzcd}
\end{equation*}
More explicitly, the $\Sigma$-algebra structure on
$\Exp{\Sigma}{B_0}/\mathord{\sim}$ is given for any operation $f: k
\to 1$ in $\Sigma$ by 
\begin{align*}
f^{\Exp{\Sigma}{B_0}/\mathord{\sim}}([t_1]_{\sim},\ldots,[t_k]_{\sim}) = [f(t_1,\ldots,t_k)]_{\sim}.
\end{align*}
And the $L$-transition structure on $\Exp{\Sigma}{B_0}/\mathord{\sim}$
is given by
\[
o([t]_\sim) = o(t)
\quad\text{and}\quad
\partial_a([t]_\sim) = [\partial_a(t)]_\sim.
\]
Now since $B^{A^*}$ is a $\BBT$-algebra and
$\Exp{\Sigma}{B_0}/\mathord{\sim}$ is its sub-$\Sigma$-algebra, $\Exp{\Sigma}{B_0}/\mathord{\sim}$ is a sub-$\BBT$-algebra of~$B^{A^*}$ (since
varieties are closed under subalgebras). Similarly,
$L(\Exp{\Sigma}{B_0}/\mathord{\sim})$ is a sub-$\BBT$-algebra of
$L(B^{A^*})$. It then follows that the $L$-transition structure on
$\Exp{\Sigma}{B_0}/\mathord{\sim}$ is a $\BBT$-algebra morphism as
 a restriction of the $L$-transition structure on~$B^{A^*}$.
\end{proof}
The following theorem is the main result of this section -- it is a variant 
of the celebrated Kleene theorem for regular languages. Like its classical counterpart
our theorem enables conversions from $\BBT$-automata to expressions and vice versa. 
\begin{theorem}[Kleene theorem]\label{thm:kleene}
  For any reactive expression $e\in\Exp{\Sigma}{B_0}$ there is a
  corresponding $\BBT$-automaton~\eqref{eq:aut} and a state $x\in X$
  such that $\sem{e}=\sem{x}_{\M}$. Conversely, for every $\BBT$\dash
  automaton~\eqref{eq:aut} and state $x\in X$ there is an expression
  $e\in\Exp{\Sigma}{B_0}$ such that $\sem{e}=\sem{x}_{\M}$.
\end{theorem}
\begin{proof}
($\Rightarrow$)~\emph{From expressions to $\BBT$-automata.}
Let $e\in\Exp{\Sigma}{B_0}$ and let us construct the corresponding $\BBT$-automaton. 
Recall that the grammar generating reactive expressions has $\gamma$- and $\delta$-clauses and let us call a not necessarily closed expression a \emph{$\gamma$-expression} if it matches the $\gamma$-clause.

We assume w.l.o.g.\ that distinct $\mu$-operators bind distinct
variables in $e$; this can be ensured by $\alpha$-conversion.
Let $X=\{x_1,\ldots,x_m\}$ be the set of variables occurring in~$e$. For $i = 1, \ldots, m$, let
\begin{align*}
t_i=\mu x_i.\,a_1.t_1^i\theta_1^i\pitchfork\ldots\pitchfork a_n.t_n^i\theta_n^i\pitchfork b_i
\end{align*}
be the uniquely determined subexpression of $e$ with each $t_j^i$ being
$\Sigma$-terms (i.e.\ not containing $\mu$) and each $t_j^i\theta_j^i$ being the
maximal proper $\delta$-subexpression of $t_i$ (cf.~Example~\ref{ex:expr}
further below); consequently, the $t_j^i$ are
obtained from the maximal proper $\delta$-subexpressions of $e$ by replacing
topmost occurrences of $t_k$ (i.e.\ topmost subexpressions starting with~$\mu
x_k$) with $x_k$, and $\theta_j^i$ being the derived substitution sending every
$x_k$ introduced in this way to $t_k$. Note that the $\theta_j^i$ need not be
total on $X$ and note that the $t_i$, the $\theta_j^i$ and the $t_j^i$ are
uniquely determined by~$e$. Without loss of generality we may assume
that $t_1 = e$. 

Starting with the triple 
\begin{equation}\label{eq:rule}
  \{\;\},~[\;],~\{x_1\doteq t_1\},
\end{equation}
where $\{\;\}$ denotes the emptyset and $[\;]$ the empty substitution,
we successively produce further triples of the form $I,~\theta,~S$, such that 
$I\subseteq\{x_1,\ldots,x_m\}$, $\theta$ is a substitution sending variables from 
$I$ to closed $\gamma$-expressions, and $S$ is a set of formal equations of the form 
$x_i\doteq t_i$ such that all the free variable of each $t_i$ are in $I$.
A successor of such triple is (nondeterministically) produced by the rule
\begin{align*}
\vcenter{
\infer{
  I\cup\{x_k\},~\theta[t_k\theta/x_k],~S\cup \{x_i\doteq t_i\mid\theta_j^k(x_i) = t_i \}
}{
  I,~\theta,~S\cup \{x_k\doteq t_k\}
}} \quad (\{x_k\doteq t_k\}\notin S)
\end{align*}  
This procedure of successively applying the above rule eventually
terminates with $S=\{\;\}$, for each
step reduces the number of $\mu$-operators that occur in the
terms on the right-hand side of equations in $S$. Note that the above
rule maintains the assumptions imposed on the triples $I,~\theta,~S$.
Hence we obtain a triple
$\{x_1,\ldots,x_m\},~\rho,~\{\,\}$ where the substitution $\rho$ sends each $x_i$ to a
closed $\gamma$-expression, which we denote by $e_i$, i.e.,
$\rho=[e_1/x_1,\ldots,e_m/x_m]$, or equivalently $e_i = x_i\rho$
for $i = 1, \ldots, m$.  %

We assume henceforth the representation
\begin{align}\label{eq:e_i_def}
e_i=\mu x_i.\,a_1.e_1^i\pitchfork\ldots\pitchfork a_n.e_n^i\pitchfork b_i.
\end{align}
Observe that 
\begin{equation}\label{eq:ek}
e_i=t_i\rho,
\end{equation}
which can be seen by induction as follows: if $t_i$ was handled at the first iteration of the above procedure then
\[
  e_i=x_i\rho=x_i[t_i/x_i]=t_i=t_i\rho;
\]
otherwise, by induction, we have
\[
  e_i=x_i\rho=x_i\theta[t_i\theta/x_i]=x_i[t_i/x_i]\theta=t_i\theta=t_i\rho,
\]
where the middle equation holds by the properties of substitution.

Using the above definition of $t_i$, we obtain 
\begin{align*}
e_i
=&~\mu x_i.\,a_1.t_1^i\theta_1^i\rho_{-i}\pitchfork\ldots\pitchfork a_n.t_n^i\theta_n^i\rho_{-i}\pitchfork b_i
\end{align*}
where $\rho_{-i}$ agrees with $\rho$ except that it leaves $x_i$ unchanged. By comparing it with~\eqref{eq:e_i_def}, we obtain for any $i,j$ that $t_j	^i\theta_j^i\rho_{-i}=e_j^i$ and therefore $e_j^i[e_i/x_i]=t_j^i\theta_j^i\rho$. Recall that for any $k = 1, \ldots, n$, $\theta_j^i$ sends $x_k$ to $t_k$ and $\rho$ sends $t_k$ to $e_k$, see~\eqref{eq:ek}. Therefore the composite substitution $\theta^i_j\rho$ sends each $x_k$ to $e_k$, i.e., we have $\theta^i_j\rho = \rho$, whence
\[
  t_j^i\theta_j^i\rho=t_j^i\rho. 
\]
We have thus obtained
\begin{align*}
e_j^i[e_i/x_i] = t_j^i\rho = t_j^i[e_1/x_1,\ldots,e_m/x_m].
\end{align*}
This allows us to restate the definitions for $o$ and $\partial$ as follows:
\begin{align*}
o(t(e_1,\ldots,e_m)) =&~ t^B(b_1^B,\ldots,b_m^B),\\
\partial_{a_j}(t(e_1,\ldots,e_m)) =&~ t(t_j^1[e_1/x_1,\ldots,e_m/x_m],\ldots,t_j^m[e_1/x_1,\ldots,e_m/x_m]), 
\end{align*}
for any $\Sigma$-term $t$ over $\{1,\ldots,m\}$. Let $\rho_{a_j}=[t_j^1/x_1,\ldots,t_j^m/x_m]$ and inductively define $\rho_{\eps}=\id$, $\rho_{a_jw}=\rho_{a_j}\rho_w$. By induction we obtain
\begin{equation}\label{eq:opit}
\begin{aligned}
o(\partial_w(t(e_1,\ldots,e_m)))=r^B(b_1^B,\ldots,b_m^B)
\text{\quad where\quad} t(x_1\rho_w,\ldots,x_m\rho_w)=r(x_1,\ldots,x_m)
\end{aligned}
\end{equation}
Suppose that $e=s(e_1,\ldots,e_m)$ with a $\Sigma$-term $s$ and 
let $\wave X=\{x,x_1,\ldots,x_m\}$.
We turn $T\wave X$ into a reactive $\BBT$-algebra. Recall that every
element of $T\wave X$ can be written as
$[t(x,x_1, \ldots, x_m)]_\equiv$, where $t$ is a $\Sigma$-term and
$[p]_\equiv$ denotes the equivalence class of the $\Sigma$-term $p$ in
$T \wave X$. Now let
\begin{align*}
o([t(x,x_1,\ldots,x_m)]_\equiv) =&~ t^B(s^B(b_1^B,\ldots,b_m^B), b_1^B,\ldots,b_m^B),\\
\partial_{a_j}([t(x,x_1,\ldots,x_m)]_\equiv) =&~ [t(s(t_j^1,\ldots,t_j^m), t_j^1,\ldots,t_j^m)]_\equiv.
\end{align*}
It is not difficult to see that the $\BBT$-algebra and the
$\Lfun$-transition structures interact properly, i.e.~$o$ and the
$\partial_{a}$ are $\BBT$-algebra morphisms; in fact, $o = \alpha\cdot 
Tf$, where $\alpha: TB \to B$ is the $\BBT$-algebra on $B$ and the map
$f: \wave X\to B$ is defined by $f(x_i) = b_i$, $i = 1, \ldots, m$,
$f(x)=s^B(b_1^B,\ldots,b_n^B)$; and $\partial_{a_j} = g_j^\klstar:
T\wave X \to T\wave X$ 
where $g_j:\wave X\to T\wave X$ is the map defined by $g_j(x_i) =
[t^i_j]_\equiv$ for $i = 1, \ldots, m$ and $g_j(x) =
[s(t^1_j,\ldots,t^m_j)]_\equiv$. Note that the induced language semantics
identifies $x$ and $s(x_1,\ldots,x_m)$, i.e.~we have
$[x]_\equiv \sim  [s(x_1,\ldots,
  x_n)]_\equiv$ (cf.~Definition~\ref{defn:tsem}).

For a $\Sigma$-term $t$, by definition $\partial_{a_j}([t]_\equiv) = [t\rho_{a_j}]_\equiv$, so an easy induction shows that
\begin{align*}
o(\partial_w([t(x_1,\ldots,x_m)]_\equiv))= r^B(b_1^B,\ldots,b_m^B) 
\text{\quad where\quad} t(x_1\rho_w,\ldots,x_n\rho_w)=r(x_1,\ldots,x_m).
\end{align*}
By comparing this to~\eqref{eq:opit} we obtain by Proposition~\ref{prop:test}, that 
\[
\sem{t(e_1,\ldots, e_n)} \sim [t(x_1,\ldots, x_n)]_\equiv
\]
Thus, specializing to $t = s$ we obtain
\[
  e = s(e_1,\ldots,e_m) \sim [s(x_1,\ldots,x_m)]_\equiv \sim [x]_\equiv.
\]
By Remark~\ref{rem:raaut}(1), the constructed reactive $\BBT$-algebra is equivalent to a $\BBT$-automaton $\M$ for which we then clearly have $\sem{e} = \sem{x}_\M$. 

($\Leftarrow$)~\emph{From $\BBT$-automata to expressions.} Suppose, we are given a $\BBT$-automaton~\eqref{eq:aut}. The generalized powerset construction yields a reactive $\BBT$-algebra over $TX$ for which
\begin{align}
  \label{eq:tij}
    o([t(x_1,\ldots,x_m)]_{\equiv}) = t^B(b_1^B,\ldots,b_m^B),&&
    \partial_{a_j}([t(x_1,\ldots,x_m)]_{\equiv}) = [t(t_j^1,\ldots,t_j^m)]_{\equiv},
\end{align}
where $b_i^B=o^{\M}(x_i)\in B$, $t_j^i$ is a term representing $t^{\M}(a_j,x_i)\in TX$ and $[t]_{\equiv}$ denotes the equivalence class of the $\Sigma$-term $t$ in $TX$. %
 We successively build expressions $u_m,\ldots,u_1$ such that all free variables of each $u_i$ with $i>1$ are in $\{x_{1},\ldots,x_{i-1}\}$ and $u_1$ is closed. Let
\begin{align*}
u_m =&~ \mu x_m.\,(a_1.t_1^m\pitchfork\ldots\pitchfork a_n.t_n^m\pitchfork b_m)\quad~\,\\
u_{i} =&~ \mu x_i.\,(a_1.t_1^i[u_{i+1}/x_{i+1},\ldots,
 u_m/x_m]\pitchfork\cdots\pitchfork a_n.t_n^i[u_{i+1}/x_{i+1},\ldots, u_m/x_m]\pitchfork b_i)
\end{align*}
for all $i = m-1, \ldots, 1$, and let $e_1=u_1$. We now apply the same construction to $e_1$ that we applied to $e$ in the first part of the proof. Note that the $\Sigma$-terms $t^i_j$ in the construction are precisely the $t^i_j$ from~\eqref{eq:tij} that we used to define the expressions $u_i$. Now the construction yields further expressions $e_i$, $i = 2, \ldots, m$ and, for every $i$, expressions $e^i_1, \ldots, e^i_n$ satisfying the identities 
\begin{align*}
e_i=&~\mu x_i.\,(a_1.e_1^i\pitchfork\ldots\pitchfork a_n.e_n^i\pitchfork b_i),\\
e_j^i[e_i/x_i] =&~t_j^i[e_1/x_1,\ldots,e_n/x_n].
\end{align*}
By the same argument as in the first part of the proof we obtain~\eqref{eq:opit}. Moreover, for the original reactive $\BBT$-algebra, also
\begin{align*}
o(\partial_w([t(x_1,\ldots,x_n)]_{\equiv}))=r^B(b_1^B,\ldots,b_n^B)
\text{\quad where\quad} t(x_1\rho_w,\ldots,x_n\rho_w)=r(x_1,\ldots,x_m),
\end{align*}
and therefore we are done by Proposition~\ref{prop:test}.
\end{proof}
\begin{example}\label{ex:expr}
Fig.~\ref{fig:aut} depicts a simple instance of the general
correspondence established by Theorem~\ref{thm:kleene} in the
particular standard case of NFSM. For the expression for $q_0$, the
subexpressions~$t_i$, $t_j^i$ and the substitutions $\theta_j^i$ are
as follows (we omit empty substitutions $\theta_j^i$):
\begin{align*}
  t_1 &= \mu x.\left(a.x\pitchfork b.\mu y.\left(a.\mathbf{\varnothing}\pitchfork b.(x\mathbf{+}\mu z.\left(a.x\pitchfork b.\mathbf{\varnothing}\pitchfork\top\right))\pitchfork\bot\right)\pitchfork\bot\right)\\
  t_2 &= \mu y.\left(a.\mathbf{\varnothing}\pitchfork b.(x\mathbf{+}\mu z.\left(a.x\pitchfork b.\mathbf{\varnothing}\pitchfork\top\right))\pitchfork\bot\right)\\
  t_3 &= \mu z.\left(a.x\pitchfork b.\mathbf{\varnothing}\pitchfork\top\right)
\end{align*}
\[
  \begin{array}{r@{\ }c@{\ }l@{\qquad\qquad}r@{\ }c@{\ }l@{\qquad}l}
    t_1^1 & = & x & t_2^1 & = & y & \theta_2^1 = [t_2/y] \\
    t_1^2 & = & \emptyset & t_2^2 & = & x + z & \theta_2^2 = [t_3/z]
    \\
    t_1^3 & = & x & t_2^3 & = & \emptyset 
  \end{array}
\]
Furthermore, the triples obtained by successively applying the
rule~\eqref{eq:rule} are as follows:
\[
  \begin{array}{c|c|c}
    I & \rho & S \\
    \hline
    \rule[11pt]{0mm}{0mm}
    \{\;\} & [\;] & \{ x \doteq t_1\} \\
    \{x\} & [t_1/x] & \{ y \doteq t_2\} \\
    \{x,y\} & [t_1/x, t_2[t_1/x]/y] & \{z \doteq t_3\} \\
    \{x,y,z\} & [t_1/x, t_2[t_1/x]/y, t_3[t_1/x, t_2[t_1/x]/y]/z] &
    \{\;\}
  \end{array}
\]
\end{example}
\begin{figure}[t!]
\begin{minipage}{.5\textwidth}
\begin{eqnarray*}
\\
\left\{
\begin{aligned}
 q_0 =&~a.q_0\pitchfork b.q_1\pitchfork \bot&&\hspace{8ex}\\
 q_1 =&~a.\mathbf{\varnothing}\pitchfork b.(q_0\mathbf{+}q_2)\pitchfork \bot\\
 q_2 =&~a.q_0\pitchfork b.\mathbf{\varnothing}\pitchfork\top
\end{aligned}
\right.
\end{eqnarray*}
\end{minipage}
\begin{minipage}{.45\textwidth}
\begin{tikzpicture}[shorten >=1pt,node distance=2cm,on grid,>=stealth',baseline=(current bounding box.north), every state/.style={inner sep=0cm}]
\node[state] (q_0) {$q_0$};
\node[state] (q_1) [right=of q_0] {$q_1$};
\node[state,accepting](q_2) [right=of q_1] {$q_2$};
\path[->] (q_0) edge [bend left] node [above] {$b$} (q_1)
	edge [loop left] node {$a$} ()
	(q_1) edge node [above] {$b$} (q_2)
	(q_1) edge [bend left] node [above] {$b$} (q_0);
\path[->] (q_2) edge [out=80, in=100, looseness=.5] node [above] {$a$} (q_0);
\end{tikzpicture}
\end{minipage}

\vspace{2ex}
$q_0 = \mu x.\left(a.x\pitchfork b.\mu y.\left(a.\mathbf{\varnothing}\pitchfork b.(x\mathbf{+}\mu z.\left(a.x\pitchfork b.\mathbf{\varnothing}\pitchfork\top\right))\pitchfork\bot\right)\pitchfork\bot\right)$
\caption{A $\PFin$-automaton over $A=\{a,b\}$, $B=\{\True,\False\}$ as
  a system of recursive definitions (left); as a nondeterministic FSM
  (right); and as a reactive expression (bottom).}
\label{fig:aut}
\end{figure}

\section{\texorpdfstring{$\BBT$-automata: Examples}{$T$-automata: Examples}}
\label{sec:ex}
As indicated in the previous section, a nondeterministic finite state machines (NFSM) is a specific case of a\/ $\BBT$-automaton under $B=2$ and $\BBT=\PFin$. More generally, we have the following definition.
\begin{definition}[Weighted $\BBT$-automata]\label{defn:waut}
A \emph{weighted $\BBT$\dash automaton} is a $\BBT$-automaton~\eqref{eq:aut} with $\BBT$ being the semimodule monad for the semiring $R$ (see Definition~\ref{defn:srmon}).
\end{definition}
\noindent
Let $\BBT$ be the semimodule monad for the semiring $R$. Besides the case ${R=B=2}$, where we obtain NFSMs, we also obtain $R$-weighted automata~\cite{DrosteKuichEtAl09} under $B=R$ (here $B$ is the free $\BBT$-algebra generated by $\{1\}$).

Weighted $\BBT$-automata can be further generalized as follows.
We call a monad \emph{additive} (cf.~\cite{CoumansJacobs13}) if the corresponding $\Sigma$-theory supports operations 
\[
+:2\to 1\qquad\text{and}\qquad\emptyset:0\to 1
\]
subject to the axioms of commutative monoids. We call a $\BBT$\dash
automaton \emph{additive} if $\BBT$ is additive. Semimodule monads
$\BBT_R$ are additive, of course. Besides the finite powerset monad
$\BBT = \PFin$, which is the semimodule monad for the Boolean semiring
$\{0,1\}$, a simple example is the bag monad $\BBT_\nat$, where $\nat$
is the usual semiring of natural numbers. This monad assigns to every
set $X$ the finite multisets on $X$ (i.e.~the free commutative monoid
on $X$).

\begin{example}[Probabilistic automata]\label{ex:alter}
Rabin's probabilistic automata~\cite{Rabin63} can be modelled as
weighted $\BBT$-automata over the semiring $[0,\infty)$ with the
standard arithmetic operations.
\takeout{%
Alternatively, as we mentioned in Definition~\ref{defn:srmon},
finitary probability distributions form a monad, which allows for a
slightly finer formalization: We take the finitary distribution monad
as $\BBT$ and $B=T2\cong [0,1]$. The $\Sigma$-theory of $\BBT$ is
defined over the signature of binary operations $+_p:2\to 1$ with
$p\in[0,1]$. The expression $x+_p\, y$ denotes probabilistic choice
between $x$ and~$y$: the left branch is chosen with probability $p$
and the right branch is chosen with probability $1-p$. The axioms of
this theory can be found e.g.\ in~\cite{Jacobs10}.}%

In fact, a Rabin automaton is precisely a
$\BBT$-automaton~\eqref{eq:aut} with a fixed initial state $x_0$. Then
given a \emph{cut-point} $\lambda\in[0,1)$, the set
$\{w\in A^*\mid \sem{x_0}_{\M}(w)>\lambda\}$
is precisely the language accepted by the Rabin automaton with cut-point $\lambda$ in the standard sense~\cite{Rabin63}.
\end{example}
We now give one example of an additive $\BBT$\dash automaton, which is not a weighted $\BBT$-automaton.
\begin{example}\label{ex:segal}
  (Simple) Segala systems~\cite{Segala95,SegalaLynch95} are systems
  combining probability and nondeterminism and are essentially
  coalgebras of transition type $\PSet(\CD\times A)\cong(\PSet\CD)^A$
  where $\CD$ is the probability distribution functor. Unfortunately,
  $\PSet\CD$ is not a
  monad~\cite[Theorem~25]{DahlqvistNeves17}. However, the combination
  of probability and nondeterminism can be modelled by a monad $\BBT$
  whose functorial part is the composition $CM$ of two functors given
  as follows: for every $X$, $M X$ consists of the finite valuations
  over $X$ (cf.~Definition~\ref{defn:srmon}); for any semimodule $U$,
  $C(U)$ consists of all subsets of~$U$, which are nonempty and
  \emph{convex}.  Convexity of a set $S$ here means that a convex
  combination $p_1\cdot\xi_1+\ldots+p_n\cdot\xi_n$, i.e.~where
  $\sum_i p_i=1$, belongs to~$S$ whenever $\xi_i\in S$ for
  every~$i$. $\BBT$-automata for $\BBT = CM$ are automata with
  combined probabilistic and nondeterministic branching. Taking
  $B=CM 1 = C[0,\infty)$, the set of all nonempty convex subsets of
  $[0,\infty)$, the semantics of a state of a $\BBT$-automaton is a
  formal power-series $A^* \to C[0,\infty)$. We leave the task of
  working out the relationship to Segala systems and their semantics
  for further work.
\end{example} 

We will now show that additive $\BBT$-automata allow for a more relaxed syntax of reactive expressions. As before we fix a finite set $A = \{a_1, \ldots, a_n\}$ of actions.
\begin{definition}[Guardedness, Additive expressions]\label{defn:guard}
Let $\Sigma$ be the signature of the algebraic theory of the additive monad $\BBT$, and let $B_0$ be a finite set.
We call an expression $e$ defined by the grammar
\begin{flalign}\label{eq:addexp}
&&\gamma \Coloneqq b\mid x\mid\mu x.\,\gamma\mid a.\gamma\mid f(\gamma,\ldots,\gamma)  && (a\in A,b\in B_0,f\in\Sigma)&&
\end{flalign}
\emph{guarded in $x$} if one of the following inductive clauses apply:
\begin{itemize}
 \item\emph{(induction base)} $e\in B_0$, $e$ is a variable distinct
   from $x$, $e=a.e'$, or $e=\mu x.\,e'$ for some expression $e'$;
 \item\emph{(induction step)} $e=f(e_1,\ldots,e_n)$ for some $e_1,\ldots,e_n$ guarded in $x$, or $e=\mu y.\,e'$ where $x\neq y$ and $e'$ guarded in $x$.
\end{itemize}
An expression generated by~\eqref{eq:addexp} is an \emph{open additive
  reactive expression} if for every of its subexpression $\mu x.\,e$, 
 $e$ is guarded in $x$. \emph{Additive reactive expressions} are those open ones in which all variables are bound. We denote by 
$\AExp{\Sigma}{B_0}$ the set of additive reactive expression over $\Sigma$, $B_0$ and by $\AExp{\Sigma}{B_0}^O$ the corresponding 
set of open additive expressions.
\end{definition}

\begin{proposition}\label{prop:aexp}
  Let $\BBT$ be an additive monad and let $B$ be a $\BBT$-algebra
  generated by the finite set $B_0$.  Given a reactive expression we
  obtain an additive reactive expression by replacing recursively each
  $\pitchfork$ with $+$. Conversely, one can also transform any
  additive reactive expression to a reactive expression, and both
  transformation are mutually inverse modulo the semantic
  equivalence~$\sim$.
\end{proposition}
\begin{proof}[Sketch of Proof]
  \begin{enumerate}
  \item Let $\Sigma$ be the signature of the $\Sigma$-theory of\/ $\BBT$.  
First, we observe that $\AExp{\Sigma}{B_0}$ clearly carries a $\Sigma$-algebra structure. Moreover, it also carries an $L$-transition structure.
In order to define it we first define an auxiliary normalization
function $\norm$ on (not necessarily closed) additive expressions as follows:
{
\begin{flalign*}
\quad\norm(f(e_1, \ldots, e_n)) = f(\norm(e_1), \ldots, \norm(e_n))\qquad (f\neq +)&&
\norm(p+q) = p\qquad (\norm(q)=\emptyset)\quad\\[.5ex]
\quad\norm(p+q) = \norm(p)+\norm(q)\qquad (\norm(p)\neq\emptyset,~\norm(q)\neq\emptyset)&&
\norm(p+q) = q\qquad (\norm(p)=\emptyset)\quad\\[-4.5ex]
\end{flalign*}
\begin{gather*}
\norm(\mu x.e) = \mu x.\, \norm(e)\qquad
\norm(a.e) = a. \norm(e)\qquad
\norm(p) = p\qquad (\text{$p$ a variable or $p\in B_0$}) 
\end{gather*}
}
Then we inductively define the $L$-transition structure on $\AExp{\Sigma}{B_0}$:
\begin{flalign*}
o(b) =&~b^B& 			o(\mu x.\, e) =&~o(e[\mu x.\, e/x])& o(a_i.e) =&~{\emptyset}^B\\
\partial_{a_i}(b) =&~\emptyset& \partial_{a_i}(\mu x.\, e) =&~\partial_{a_i}(e[\mu x.\, e/x])&
\partial_{a_i}(a_i.e) =&~\norm(e),~\partial_{a_i}(a_j.e)=\emptyset && (i\neq j)
\end{flalign*}\vspace{-4.5ex}
\begin{flalign*}
o(f(e_1, \ldots, e_n)) = f^B(o(e_1),\ldots, o(e_n))&&
\partial_{a_i}(f(e_1, \ldots, e_n)) = \norm(f(\partial_{a_i}(e_1),\ldots, \partial_{a_i}(e_n)))
\end{flalign*}

\item By Definition~\ref{defn:tsem}, the above $L$-coalgebra structure on
$\AExp{\Sigma}{B_0}$ induces a language semantics; again we
write $\sem{e}$ for the formal power series denoted by
$e \in \AExp{\Sigma}{B_0}$. We need to show that this semantics agrees
with the semantics of $\Exp{\Sigma}{B_0}$, that is
$\sem{e}=\sem{\otr(e)}$ for $e\in\Exp{\Sigma}{B_0}$ and
$\otr\colon \Exp{\Sigma}{B_0} \to \AExp{\Sigma}{B_0}$ defined inductively
as follows:
\begin{align*}
\otr(f(e_1, \ldots, e_n)) =&~\norm(f(\otr(e_1), \ldots, \otr(e_n))),& \otr(x) =&~ x,\\
\otr(\mu x.\,a_1.e_1\pitchfork\ldots\pitchfork a_{n}.e_n\pitchfork s)
=&~\mu x.\,\norm(a_1.\otr(e_1)+\ldots+ a_{n}.\otr(e_n)+ \otr(s)),& \otr(b) =&~ b.
\end{align*}
Note that $s$ in the bottom left equation is an arbitrary term in the theory of\/
$\BBT$ according to the $\beta$-clause of the grammar in
Definition~\ref{def:react}. In fact, the above assignments define
$\otr$ on expressions containing free variables and according to the
$\gamma$ and $\beta$-clauses of Definition~\ref{def:react}. The
verification of $\sem{e}=\sem{\otr(e)}$ may be found in the electronic appendix.

\item In order to prove the desired converse in the statement of the
proposition, we define a translation map
$\tr\colon \AExp{\Sigma}{B_0} \to \Exp{\Sigma}{B_0}$. To that end
we first define an auxiliary map $\bar o$ on every expression
according to~\eqref{eq:addexp} that is guarded in each of its
variables; $\bar o$ works similarly as $o$ but without interpreting
$\emptyset$, $f$ and $b$ in $B$, whence delivering a term in the
theory of $\BBT$ according to the $\beta$-clause of
Definition~\ref{def:react}:
\[
\begin{array}{r@{~}c@{~}l@{\qquad}r@{~}c@{~}l}
  \bar o(b) &=& b & \bar o(\mu x.\,e) &=& \bar o(e[\mu x\,e./x]) \\
  \bar o(a.e) &=& \emptyset & \bar o(f(e_1, \ldots, e_n)) &=& f(\bar o(e_1), \ldots, \bar o(e_n))
\end{array}
\]
Then $\bar o(e)$ is well-defined by guardedness of $e$. Similarly, we
define auxiliary maps $a^{\mone}$ completely similarly as
$\partial_{a}$; however, $a^{\mone}$ can be applied to expressions $e$
containing free variables but which are still guarded in each of their
variables. That means we do not (need to) define $a^{\mone}$ on variables~$x$. Now we define $\tr$ (on not necessarily closed expressions) as follows:
\begin{align*}
\tr(x) =&~x,\\
\tr(b) =&~\mu x.\,a_1.\emptyset \pitchfork\ldots\pitchfork a_{n}.\emptyset \pitchfork b, \\
\tr(a_i.e) =&~\mu x.\,a_1.\emptyset \pitchfork\ldots\pitchfork a_i .\tr(e) \pitchfork\ldots\pitchfork a_{n}.\emptyset \pitchfork \emptyset, \\
\tr(f(e_1, \ldots, e_n)) =&~f(\tr(e_1), \ldots, \tr(e_n)),\\
\tr(\mu x.\,e) =&~\mu x.\,a_1.\tr(a_1^{\mone}(e))\pitchfork\ldots\pitchfork a_{n}.\tr(a_n^{\mone}(e))\pitchfork \bar o(\mu x.\,e).
\end{align*}
It is then a matter of routine verification that
$\sem{e}=\sem{\tr(e)}$ for every  $e\in\AExp{\Sigma}{B_0}$. Some
details are in the electronic appendix.
\qedhere
\end{enumerate}
\end{proof}
\sgnote{I added new text here.}
\begin{remark}\label{rem:rat}
We note that for weighted automata additive expressions can be equivalently converted 
to the familiar rational expressions from weighted automata
theory. Suppose that $B_0=\{1\}$, so
$B = R$. Then we can define a composition operation 
$\bullet:\AExp{\Sigma}{B_0}^O\times \AExp{\Sigma}{B_0}^O\to
\AExp{\Sigma}{B_0}^O$ inductively~by: 
\begin{gather*}
x\bullet t = x\qquad\qquad 1\bullet t = t\qquad\qquad (\mu x.\,e)\bullet t = \mu x.\,(e\bullet t)\\[1.5ex]
(a.\,e)\bullet t = a.\,(e\bullet t) \qquad\qquad f(e_1,\ldots,e_n)\bullet t = f(e_1\bullet t,\ldots,e_n\bullet t)
\end{gather*}
According to this definition we have
$a.e = a.(1 \bullet e) = (a.1)\bullet e$, i.e.~every
expression $a.e$ can be expressed using $\bullet$ and expressions
$a.1$ only. The signature $\Sigma$ of the semimodule theory consists
of one binary operation symbol $+$ and unary
operation symbols, one for every $r \in R$, denoted $r \cdot -$ .
Thus, writing simply $a$ for $a.1$ and $r$ for
$r \cdot 1$, an alternative syntax for additive reactive
expressions can be defined by the following grammar:
\begin{flalign}\label{eq:rational-exp}
  &&
  \gamma \Coloneqq  x\mid \mu x.\,\gamma \mid a \mid r\mid\gamma+\gamma\mid\gamma\bullet\gamma && (a\in A,r\in R)&&
\end{flalign}
Guardedness becomes somewhat more complicated to formulate: $t$ is
guarded in $x$ if $x$ is contained in a subterm $t_l \bullet t_r$  of
$t$ in the right-hand subterm $t_r$, where the left-hand subterm $t_l$ contains some
letter $a \in A$. Again, we consider expressions in which in every
subexpression $\mu x.\, e$, $e$ is guarded in $x$ and where all
variables are bound. The syntax can be restricted further by requiring
that in every expression $\mu x.\, t$, $t$ is of the form $1 + e
\bullet x$, where $e$ is closed. Indeed, using the above sound equations, associativity of
$\bullet$, and the following distributive laws
\[
  (s+t)\bullet e = s\bullet e + t\bullet e,
  \qquad\qquad
  e\bullet (s+u) = e\bullet s + e\bullet u,
\]
this can be shown by induction over the number of $\mu$-operators as follows. Let
$\mu x.\,t$ be an expression with~$t$ satisfying the induction
hypothesis. Then $t$ can be brought to the form $q+ e\bullet x$ with~$q$ not
containing~$x$. It easily follows that $\mu x.\, t = \mu x.\, (q + e \bullet x)
(\mu x.\, (1 + e\bullet x))\bullet q$.
The usual notation for
$\mu x.\, (1 + e\bullet x)$ is \emph{Kleene star} $e^*$. Hence, by
replacing $\mu x.\,\gamma$ with $\gamma^*$ in the grammar~\eqref{eq:rational-exp}
we thus arrive at the grammar of rational expressions as in
the \emph{Kleene-Sch\"utzenberger theorem} (see e.g.~\cite{dkv}).
\end{remark}

\subsection{\texorpdfstring{Stack $\BBT$-automata}{Stack $T$-automata}}
\label{sec:stackT}
Here and in later sections we turn our attention to a different kind of examples of
$\BBT$-automata, where~$\BBT$ is related to the store monad. A
prominent instance are $\BBT$-automata where $\BBT$ is the stack monad
(Definition~\ref{defn:stack_mon}), which model finite state machines
manipulating a push-down store.
\begin{definition}[Stack $\BBT$-automaton]\label{defn:stack}
A \emph{stack $\BBT$-automaton} is a $\BBT$-automaton~\eqref{eq:aut} for which
\iffull
\begin{itemize}
  \item $\BBT$ is the stack monad over $\Gamma$;
  \item $B$ is the set of predicates over ${\Gamma^*}$ consisting of all those $p\in 2^{\Gamma^*}$ for each of which there exists a $k$ such that $p(w u)=p(w)$ whenever $|w|\geq k$; 
  \item $\algstruc:TB\to B$ is given by evaluation; it restricts the morphism
\begin{align*}
(2^{\Gamma^*} \times \Gamma^*)^{\Gamma^*} \xrightarrow{\quad\mathsf{ev}^{\Gamma^*}~} 2^{\Gamma^*},
\end{align*}
where $\mathsf{ev}: 2^{\Gamma^*} \times \Gamma^* \to 2$ is the evaluation morphism: 
\[
  \algstruc(r,t)(s) = r(s)(t(s)).
\] 
\end{itemize}
\else
$\BBT$ is the stack monad over $\Gamma$; $B$ is the set of predicates over ${\Gamma^*}$ consisting of all those $p\in 2^{\Gamma^*}$ for each of which there is $k$ such that $p(w u)=p(w)$ whenever $|w|\geq k$; $\algstruc:TB\to B$ is given by evaluation; it restricts the morphism $\mathsf{ev}^{\Gamma^*}: (2^{\Gamma^*} \times \Gamma^*)^{\Gamma^*} \to 2^{\Gamma^*}$, where $\mathsf{ev}: 2^{\Gamma^*} \times \Gamma^* \to 2$ is the evaluation morphism: $\algstruc(r,t)(s) = r(s)(t(s))$. 
\fi
\end{definition}
\noindent 
Intuitively, $o^\M: X \to B \subseteq 2^{\Gamma^*}$ %
models the acceptance condition by final states and stack contents, that is, we can consider $w\in A^*$ to be jointly accepted by a stack $\BBT$-automaton $\M$, an initial state $x_0$, an initial stack symbol $\gamma_0$, a finite set of final states $F$ and a set of final stack configurations $S$ if $\sem{x_0}_{\M}(w)(\gamma_0)=1$ where $o^\M(x)(s)=1$ iff $x\in F$, $s\in S$.  As $B$ obeys constraints analogous to those of $TX$, scanning an unbounded portion of the stack by $o^\M$ is disallowed\iffull; the role of the algebraic structure $\algstruc$ is roughly to trace acceptance conditions backwards along the transition structure~$t^\M$.

In terms of $\Sigma$-theories, $B$ is finitely generated over the set of generators $B_0=\{0,1\}$ and as such is a quotient of $T2$ under additional laws: $push_i(0) = 0$ and $push_i(1)=1$\fi.
The formal argument showing that $B$ is indeed an algebra for the stack monad is as follows.
By Corollary~\ref{cor:subalg}, the stack monad, being a submonad of the store monad over~$\Gamma^*$, induces a submonad $\BBP$ of the reader monad over~$\Gamma^*$. For this monad $\BBP$ we have that $PX$ consists of those $r:\Gamma^{*}\to X$ for each of which there exists $k$ such that for every $w\in\Gamma^*$ and $u\in\Gamma^*$, $r(wu)=r(w)$ whenever $|w| \geq k$. In particular, this makes $B = P 2$ a $\BBP$-algebra and hence a $\BBT$-algebra.

The expected fact that stack $\BBT$-automata can be used as a replacement for deterministic push-down automata without silent transitions (viz \emph{deterministic real-time push-down automata}) is justified by the following result. 
\begin{theorem}\label{thm:stack}
Let $\M$ be a stack $\BBT$-automaton. Given $x_0\in X$ and $\gamma_0\in\Gamma$, 
\iffull
\begin{align}\label{eq:dcfl}
\left\{w\in A^*\mid\sem{x_0}_{\M}(w)(\gamma_0)=1\right\}
\end{align}
\else
$\left\{w\in A^*\mid\sem{x_0}_{\M}(w)(\gamma_0)=1\right\}$
\fi
is a real-time deterministic context-free language. Conversely, for any real-time deterministic context-free language $\CL\subseteq A^*$ there exist a stack $\BBT$-automaton~\eqref{eq:aut}, an $x_0 \in X$, and a $\gamma_0\in \Gamma$ such that $\CL$ is the language in~\eqref{eq:dcfl}.
\end{theorem}

\noindent
As we shall see in Theorem~\ref{thm:bound}, one can obtain an
analogous characterization of ordinary context-free languages
(essentially because for nondeterministic push-down automata the
restriction of being real-time is omissible).

For the proof of Theorem~\ref{thm:stack} we need an explicit description of the action of the language semantics map $\sem{-}_\M$ defined in Diagram~\eqref{diag:hatm} in terms of the given data of the $\BBT$-automaton $\M$.
\begin{lemma}\label{lem:ind}
  Given any $\BBT$-automaton~\eqref{eq:aut}, $x\in X$ and $w\in A^*$ then 
\begin{align}\label{eq:sem_ind}
\sem{x}_\M(\eps)= o^\M(x),&&
\sem{x}_\M(a u) = \algstruc\left(\letTerm{y\leteq t^\M(a,x)}{\eta_X\sem{y}_\M(u)}\right).
\end{align}
\end{lemma}
Before we proceed with the proof of Theorem~\ref{thm:stack}, let us recall that a deterministic pushdown automaton (dpda) $M$ is determined by a transition function
\begin{align}\label{eq:dpda}
\delta:Q\times (A+\{\eps\})\times\Delta\to Q\times\Delta^* + \{\bot\},  
\end{align}
an initial stack symbol $\blk\in\Delta$, an initial state $q_0\in Q$
and a set of final states ${F\subseteq Q}$. Here $Q$ is a finite set
of all states, $A$ is a finite alphabet of actions and $\Delta$ is a
finite alphabet of stack symbols. The transition function $\delta$ is
subject to the following restrictions: for every $x\in Q$,
$\gamma\in\Delta$ (exclusively) either $\delta(x,\eps,\gamma)\neq\bot$
or $\delta(x,a,\gamma)\neq\bot$ for all $a\in A$.
Automaton configurations and transitions over them are defined in the standard way.

A word $w$ is recognized by $M$ if there is a chain of transitions over automaton configurations that starts at $\brks{x_0,\blk}$, consumes $w$, and finishes at some $\brks{x_n, s_n}$ with $x_n\in F$. A dpda $M$ is called \emph{real-time} if $\delta(x,\eps,\gamma)=\bot$ for every $x\in Q$, $\gamma\in\Delta$ and it is called \emph{quasi-real-time} if there is $n$ such that the following chain of transition is not admissible for any $x_1\in Q$, $s_1\in\Delta^*$ and $m>n$:
\begin{displaymath}
\brks{x_1,s_1}\xrightarrow{~~\eps~~}\brks{x_2,s_2}\xrightarrow{~~\eps~~}~\cdots~\xrightarrow{~~\eps~~}\brks{x_m,s_m}
\end{displaymath}
We will make use of the fact that the classes of languages recognized by real-time dpda and quasi-real-time dpda coincide~\cite{HarrisonHavel72}.
\begin{proof}[Proof of Theorem~\ref{thm:stack}]
Given~\eqref{eq:aut} over a stack monad and a finite $X$, let us
construct a quasi-real-time dpda $M$ as follows. For any $x\in X$ and
$a\in A$ let $n_{x,a}$ be the smallest $n\geq 1$ such that
$t^\M(x,a):\Gamma^*\to X\times\Gamma^*$ sends any $s u$ with
$s,u\in\Gamma^*$, $|s|=n$ to $\brks{y,s' u}$ where
$\brks{y,s'}=t^\M(x,a)(s)$. Analogously, let $n_x$ be the smallest
$n\geq 1$ such that $o^{\M}(x):\Gamma^*\to 2$ returns equal results on
words agreeing on the first $n$ letters. Note that the numbers
$n_{x,a}$ and $n_x$ exist by the definition of the stack monad.
Let $m=\max\{n_x,\max_a n_{a,x}\}$. As the state space of $M$ we take
\begin{displaymath}
 Q=\bigl\{\brks{x,s \blk^k}\mid x\in X,s\in\Gamma^*,|s|\leq m-k\bigr\}.
\end{displaymath}
Let $\Delta=\Gamma+\{\blk\}$. Then we define the transition function $\delta$ as follows: 
\begin{enumerate}[label=(\roman*)]
  \item $\delta(\brks{x,s},\eps,\gamma)=\brks{\brks{x,s\gamma},\eps}$ if $\gamma\neq\blk$ and $|s|<m$;
  \item $\delta(\brks{x,s},\eps,\blk)=\brks{\brks{x,s\blk},\blk}$ if $|s|<m$;
  \item $\delta(\brks{x,s \blk^k},a,\gamma)=\brks{\brks{y,\eps},s' \gamma}$ if $a\neq\eps$, $s\in\Gamma^{m-k}$ and $\brks{y,s'}=t^{\M}(x,a)(s)$.
\end{enumerate}
Finally, let 
\begin{align*}
F=\{\brks{x,s\blk^k}\in Q\mid o^\M(x)(s)=1, s\in\Gamma^{m-k}\}
\end{align*}
be the set of accepting states of~$M$. The intuitive motivation for $M$ comes from the need to save portions of the stack in the state. This is needed to model the behaviour of $\M$, which unlike a standard pda can read several symbols from the stack at once and not just the top one. For technical reasons it is convenient to assume that we always can transfer $m$ symbols from the stack to the state. We ensure this by allowing the completion of the second component of the state with an appropriate number of symbols $\blk$ added from the right if the stack happens to be shorter than~$m$.   

Our goal is to show that for any $w\in A^*$, $\sem{x_0}_{\M}(w)(\gamma_0)=1$ iff $w$ is accepted by $M$ with $\brks{x_0,\gamma_0}$ as the initial state. To that end we prove a (clearly) more general statement: for any $w\in A^*$, $x\in X$ and $s\in \Gamma^*$, $\sem{x}_{\M}(w)(s)=1$ iff there is a chain of transitions $C$ over configurations of $M$ corresponding to $w$, starting at $\brks{\brks{x,\eps},s\blk}$ and finishing in an accepting state. We proceed by  induction over the length of $w$.
\begin{citemize}
\item Let $w=\eps$. Then by Lemma~\ref{lem:ind}
  $\sem{x}_{\M}(w)(s)=o^\M(x)(s)=o^\M(x)(s')$ where $s'$ is the prefix
  of length $\min\{|s|, n_x\}$ of $s$. Therefore,
  $\sem{x}_{\M}(w)(s)=1$ iff $\brks{x,s'\blk^k}\in Q$ belongs to $F$
  with $k=m-|s'|$. On the other hand, by~(i)--(ii), every chain $C$ of
  transitions corresponding to $w=\eps$ and starting at
  $\brks{\brks{x,\eps},s\blk}$ must be a prefix of the following
  chain:
\begin{displaymath}
\brks{\brks{x,\eps},s\blk}\xrightarrow{\eps}\cdots\xrightarrow{\eps}\brks{\brks{x,s'\blk^k},u\blk}
\end{displaymath}
where $s = s'u$ and $k=m-|s'|$. Clearly, $C$ leads to an accepting configuration iff $\brks{x,s'\blk^k}$ is an accepting state.
 \item Let $w=a u$. Then by Lemma~\ref{lem:ind},
\begin{align*}
  \sem{x}_{\M}(w)(s)
  &= \algstruc\left(\letTerm{y\leteq t^\M(x,a)}{\eta_X\sem{y}_\M(u)}\right)(s)\\
  &= \sem{y}_\M(u)(s') \qquad \text{where $\brks{y,s'}=t^\M(x,a)(s)$}.
\end{align*}
The latter is equal to $1$ iff $\sem{y}_\M(u)(s')=1$ where $\brks{y,s'}=t^\M(x,a)(s)$. By the induction hypothesis $\sem{y}_\M(u)(s')=1$ iff there is a chain of transitions $C$ corresponding to $u$, starting at $\brks{\brks{y,\eps},s'\blk}$ and finishing in an accepting state. We shall show that there is a chain of transitions $C'$ starting in $\brks{\brks{x,\eps},s\blk}$ and finishing in an accepting state. There are two cases: (1) if $|s| < m$ then we obtain $C'$ by prepending $C$ with
\begin{displaymath}
\brks{\brks{x,\eps},s\blk}\xrightarrow{\eps}\cdots\xrightarrow{\eps}\brks{\brks{x,s\blk^k},\blk}\xrightarrow{a} \brks{\brks{y,\eps},s'\blk},
\end{displaymath}
where $k = m - |s|$; (2) if $|s| \geq m$ let $s = s''  w$ with $|s''| = m$ and let $t^\M(x,a)(s'') = (\hat y, \hat s)$. Then since $t^\M(x,a)(s''  u) = (\hat y, \hat s  u)$ holds by the properties of $t^\M(x,a): \Gamma^* \to X \times \Gamma^*$, we know that $\hat y = y$ and $\hat s  u = s'$. So we obtain $C'$ by prepending $C$ with 
\begin{displaymath}
\brks{\brks{x,\eps},s\blk}\xrightarrow{\eps}\cdots\xrightarrow{\eps}\brks{\brks{x,s''},u\blk}\xrightarrow{a} \brks{\brks{\hat y,\eps},\hat s  u \blk} = \brks{\brks{y,\eps},s'\blk}.
\end{displaymath}
Conversely, given a chain of transitions $C'$ for $w$ from $\brks{\brks{x,\eps},\mbox{$s\blk$}}$ and leading to a final state, then it must be a chain $C$ starting at $\brks{\brks{y,\eps},s'\blk}$ prepended by one of the above two prefixes (depending on $|s|$). This completes the induction and the proof of the first part of the theorem.
\end{citemize}
In order to show the second part of the claim, suppose we are given a
real-time deterministic pda $M$ with a transition
function~\eqref{eq:dpda}, an initial state $q_0\in Q$, a set of
accepting states $F\subseteq Q$ and an initial stack symbol
$\blk$. Let us define a $\BBT$-automaton~\eqref{eq:aut} with $X=Q+\{\bot\}$ and $\BBT$ being the stack monad over $\Delta$ as follows: for every $q\in X$, $s\in\Delta^*$, $a\in A$, $o^\M(q)(s)=1$ iff $q\in F$ and 
\begin{align*}
  t^\M(q,a)(\eps) &= t^\M(\bot,a)(\gamma s) = \brks{\bot,\eps}\\
  t^\M(q,a)(\gamma s)
  &= \brks{q',s's} \qquad \text{where $\brks{q',s'}=\delta(q,a,\gamma)$}.
\end{align*}
Let us show by induction over the length of $w\in A^*$ that for every $q\in Q$, $s\in\Delta^*$ an accepting configuration is reachable from $\brks{q,s}$ by $w$ iff $\sem{q}_{\M}(w)(s)=1$.
\begin{citemize}
  \item Let $w=\eps$. Then $\brks{q,s}$ is accepting iff $q\in F$ iff $o^\M(q)(s)=1$. By Lemma~\ref{lem:ind}, the latter is equivalent to $\sem{q}_{\M}(w)(s)=1$.
  \item Let $w=a u$. Then an accepting configuration is reachable from $\brks{q,s}$ iff $\brks{q,s}\xrightarrow{a}\brks{q',s'}$ for some $\brks{q',s'}$ from which an accepting configuration is reachable by $u$. By induction hypothesis and by definition of $t^\M$, an equivalent formulation is as follows: $\sem{q'}_{\M}(u)(s')=1$ where $\brks{q',s'}=t^\M(q,a)(s)$. On the other hand, by Lemma~\ref{lem:ind},
\begin{align*}
  \sem{q}_{\M}(w)(s)
  &= \algstruc\left(\letTerm{q'\leteq t^\M(q,a)}{\eta_X\sem{q'}_\M(u)}\right)(s)\\
  &= \sem{q'}_{\M}(u)(s') \qquad \text{where $\brks{q',s'}=t^\M(q,a)(s)$},
\end{align*}
i.e.~also $\sem{q}_{\M}(w)(s) =1$ iff $\sem{q'}_{\M}(u)(s')=1$ where $\brks{q',s'}=t^\M(q,a)(s)$.
\end{citemize}
As a result, the language recognized by $M$ is equal
to~\eqref{eq:dcfl} under $x_0=q_0$ and $\gamma_0=\blk$.
\end{proof}

\section{Monad Tensors for Combining Store and Nondeterminism}\label{sec:tensor}
Tensor products of monads (resp.\ algebraic theories) have been
introduced by Freyd~\citeyear{Freyd66} in the context of universal
algebra. Later, computational relevance of this operation has been
demonstrated by Hyland et al.~\citeyear{HylandLevyEtAl07}. Here, we
use tensors of monads as a tool for studying $\BBT$\dash automata,
where $\BBT$ combines (several kinds of) store with nondeterminism.
\begin{definition}[Tensor]
\label{defn:tensor}
Let $\CE_1$ and $\CE_2$ be two algebraic theories. Then the tensor
product $\CE=\CE_1\tensor\CE_2$ is the algebraic theory, whose
equations are obtained by joining the equations of $\CE_1$ and~$\CE_2$
and adding for every $f:n\to 1$ of $\CE_1$ and every $g:m\to 1$ of $\CE_2$ the following axiom
\[
f(g(x_1^1,\ldots, x_m^1),\ldots,g(x_1^n,\ldots, x_m^n))= g(f(x_1^1,\ldots, x_1^n),\ldots,f(x_m^1,\ldots, x_m^n))
\]
called the \emph{tensor laws}.  Given two finitary monads $\BBT_1$ and
$\BBT_2$, their tensor product $\BBT_1 \tensor \BBT_2$ arises from the
algebraic theory $\CE_{\BBT_1}\tensor\CE_{\BBT_2}$. Note that the
embedding of terms and equations of $\CE_i$, $i = 1,2$, into
$\CE_1 \tensor \CE_2$ gives rise to monad morphisms
$\BBT_i \to \BBT_1 \tensor \BBT_2$ called \emph{tensor injections}.
\end{definition}
\noindent
Intuitively, the tensor product of two monads captures a
noninterfering combination of the corresponding computational
effects. In the present work we shall use two kinds of tensor
products: tensors with \emph{submonads of the store monad} (see
Example~\ref{ex:mon}) and tensors with \emph{semimodule monads}
(see Definition~\ref{defn:srmon}). This allows us to combine
nondeterminism with one or several stores. %

It has been shown in~\cite{HylandLevyEtAl07} that tensoring with the store monad is equivalent to the application of the store monad transformer sending any monad $\BBT$ to the \emph{store monad transform} $\BBT_S$ whose functorial part is given by $T_SX=T(X\times S)^S$. Here we establish a similar result for stacks (Defnition~\ref{defn:stack_mon}).
\begin{proposition}\label{prop:stens}
Let\/ $\BBS$ be the stack monad over $\Gamma$. Then for any finitary monad~$\BBT$,\/ $\BBS\tensor\BBT$ is the submonad $\BBR$
of the store monad transform of\/ $\BBT$ with $\Gamma^*$\/ as the store, identified by the 
following condition: 
$p:\Gamma^*\to T(X\times\Gamma^*)$ is in $R X$ if 
\begin{align}\label{eq:stack-trans}
  \exists k\in\nat.\,\forall s \in \Gamma^k.\,\forall u \in \Gamma^*.\, p(su)=\letTerm{\brks{x,s'}\leteq p(s)}{\eta_{X\times\Gamma^*}\brks{x,s'u}}.
\end{align}
\end{proposition}
\begin{proof}[Sketch of Proof]
We will rely on the following equivalent reformulation of~\eqref{eq:stack-trans}:  
\begin{align}\label{eq:stack-trans-ext}
  \exists k\in\nat.\,\forall s,u\in \Gamma^*.\, |s|\geq k\impl p(su)=\letTerm{\brks{x,s'}\leteq p(s)}{\eta_{X\times\Gamma^*}\brks{x,s'u}}.
\end{align}
For the proof that~\eqref{eq:stack-trans}
and~\eqref{eq:stack-trans-ext} are indeed equivalent and that the
former indentifies a submonad of the previously mentioned store monad
tranform see the electronic appendix.

Let us refer to the stack theory over $\Gamma=\{\gamma_1,\ldots,\gamma_n\}$ as $\mathcal{E}$ and to the theory corresponding to the monad $\BBT$ as $\CT$. 
We define a semantics of the theory
$\CE\tensor\CT$ over $\BBR$ as follows:
\begin{align*}
\tsem{pop}_{\BBR}(\eps) =&\;\eta_{N\times\Gamma^*}\brks{n+1,\eps},&\tsem{pop}_{\BBR}(\gamma_i w)=&\;\eta_{N\times\Gamma^*}\brks{i,w},\\
\tsem{push_i}_{\BBR}(w) =&\;\eta_{\{1\}\times\Gamma^*}\brks{1,\gamma_i w},& \tsem{f}_{\BBR}(w) =&\;\letTerm{j\leteq\tsem{f}_{\BBT}}{\eta_{M\times\Gamma^*}\brks{j,w}},
\end{align*} 
where $\eta$ denotes the unit of the monad
$\BBT$, $N = \{1, \ldots, n+1\}$, $i$ ranges from $1$ to $n$,
$f$ ranges over the operations of $\CT$, and $M = \{1,\ldots, m\}$,
where $m$ is the arity of $f$.
We proceed to verify the properties prescribed by
Theorem~\ref{thm:mon_thm_eq}. This verification is analogous to the
proof of Theorem~\ref{thm:stack_comp} and hence we discuss only the
specific features of the case at hand.
\begin{citemize}
 \item[]\textit{Soundness.} We have to verify soundness of (i) the stack
   theory, (ii) the equations from $\CT$, and (iii) the tensor
   laws. Soundness of (i) is verified exactly as in
   Theorem~\ref{thm:stack_comp}. Soundness of~(ii) immediately follows
   from soundness of $\CT$ over~$\BBT$. Finally, soundness of~(iii) is
   verified directly for the $push$ and for $pop$ operations. For
   $push_i$ we have for every set $X$
\begin{align*}
\tsem{push_i&(f(x_1,\ldots,x_k))}_{RX}(w) \\
&= (\letTerm{\tsem{push_i}_{\BBR};j\leteq\tsem{f}_{\BBR}}{\tsem{x_j}_{RX}})(w)\\
&= \letTerm{\brks{x,u}\leteq\eta_{1\times\Gamma^*}\brks{1,\gamma_iw};\brks{j,v}\leteq\tsem{f}_{\BBR}(u)}{\tsem{x_j}_{RX}(v)}\\
&= \letTerm{\brks{j,v}\leteq\tsem{f}_{\BBR}(\gamma_iw)}{\tsem{x_j}_{RX}(v)}\\
&= \letTerm{j\leteq\tsem{f}_{\BBT}; \brks{x,v} \leteq\eta_{M\times\Gamma^*}\brks{j,\gamma_iw}}{\tsem{x_j}_{RX}(v)}\\
&= \letTerm{j\leteq\tsem{f}_{\BBT}}{\tsem{x_j}_{RX}(\gamma_iw)}
\intertext{and}
\tsem{f(push_i&(x_1),\ldots,push_i(x_k))}_{RX}(w) \\
&= (\letTerm{j \leteq \tsem{f}_{\BBR}; \tsem{push_i}_\BBR}{\tsem{x_j}_{RX})(w)}\\
&= \letTerm{\brks{j,v} \leteq \tsem{f}_{\BBR}(w); \brks{x,u}\leteq \tsem{push_i}_\BBR(v)}{\tsem{x_j}_{RX}(u)} \\
&= \letTerm{j \leteq \tsem{f}_{\BBT}; (y,v) \leteq \eta_{M\times\Gamma^*}\brks{j,w}; \brks{x,u}\leteq \tsem{push_i}_\BBR(v)}{\tsem{x_j}_{RX}(u)} \\
&= \letTerm{j \leteq \tsem{f}_{\BBT}; \brks{x,u}\leteq \tsem{push_i}_\BBR(w)}{\tsem{x_j}_{RX}(u)} \\
&= \letTerm{j \leteq \tsem{f}_{\BBT}; \brks{x,u}\leteq \eta_{1\times\Gamma^*}\brks{1,\gamma_iw}}{\tsem{x_j}_{RX}(u)} \\
&= \letTerm{j \leteq \tsem{f}_{\BBT}}{\tsem{x_j}_{RX}(\gamma_i w)}
\end{align*} 
We leave the verification for $pop$ to the reader.\smnote{OK, I've written up this computation notes are in the svn.}
 \item[]\textit{Expressiveness.} Let $p:\Gamma^*\to T(X\times\Gamma^*)$ be an element of $RX$ under some parameter $k$. We construct a term $p_k$ over $X$ such that $\tsem{p_k}=p$ by induction over $k$ adapting the construction from Theorem~\ref{thm:stack_comp}.
\begin{itemize}[wide]
  \item~Let $k=0$ and note that $p(\eps)\in T(X\times\Gamma^*)$. Let $q$ be a term over $X\times\Gamma^*$ for which $\tsem{q}_{T(X \times \Gamma^*)}=p(\eps)$ and let $p_0$ be obtained from $q$ by replacing any $\brks{x,\gamma_{i_m}\ldots\gamma_{i_1}}\in X\times\Gamma^*$ by the term $push_{i_1}(\cdots(push_{i_m}(x))\cdots)$.
  \item~If $k>0$ then we build $p_k$ from $p_{k-1}$ in the same way as in Theorem~\ref{thm:stack_comp}.
\end{itemize}
 \item[]\textit{Completeness.} Suppose we are given $s$ and $t$ such that
$\tsem{s}_{RX}=\tsem{t}_{RX}$. Let us normalize both $s$ and $t$ using the
equations of the stack theory oriented from left to right and additionally the 
rules:
\begin{align}
push_i(f(x_1,\ldots,x_m))\to&\; f(push_i(x_1),\ldots,push_i(x_m))\\
pop(x_1,\ldots,x_n,f(y_1,\ldots,&\;pop(z_1,\ldots,z_n,z),\ldots, y_m))\notag\\
\to&\; pop(x_1,\ldots,x_n,f(y_1,\ldots,z,\ldots, y_m))\label{eq:pop-pop-sigma}
\end{align} 
where $f(1,\ldots,m)$ is an $m$-ary term in the signature of $\CT$. 
Note that the obtained system is strongly normalizing because every rule either 
decreases the height of the term, or keeps it the same, but propagates the 
$push$ operator downwards. Except for the last rule, by definition, the respective
equations belong to $\CE\tensor\CT$. The equation corresponding to the last rule 
also belongs to $\CE\tensor\CT$, which can be shown by {\bf (pop-push)}, {\bf (pop-pop)}, 
and by tensor laws (see the electronic appendix for details).
It suffices to prove that $s = t \in \CE \tensor \CT$ for normal $s$
and $t$, which we do by induction over the number of
operations distinct from $push$ occurring both in $s$ and in $t$. Let
$f$ and $g$ be terms (possibly a single variable) in the signature of $\CT$ such
that $s=f(s_1,\ldots,s_m)$, $t=g(t_1,\ldots,t_l)$ and such that each of
the $s_1,\ldots,s_m,t_1,\ldots,t_l$ is either a variable or has an
operation of the stack theory at the~top.
\begin{itemize}[wide] 
\item~ If none of the terms $s_1,\ldots,s_m, t_1,\ldots,t_l$ contains
   the $pop$ operation on top, then by normality each of these terms
   must be an application of a sequence of the $push$ operations to a
   variable. Hence, by the definition of our semantics we obtain
   \begin{align*}
     \tsem{s}_{RX}(\eps)
     &= \tsem{f(\brks{x_1,w_1}, \ldots, \brks{x_m,w_m})}_{T(X \times \Gamma^*)}, 
     \\
     \tsem{t}_{RX}(\eps)
     & = \tsem{g(\brks{y_1,u_1},\ldots,\brks{y_l,u_l})}_{T(X \times \Gamma^*)},
   \end{align*}
   where $\brks{x_i,w_i} = \tsem{s_i}_{RX}(\eps)$ and $\brks{y_j,u_j} =
   \tsem{t_j}_{RX}(\eps)$. It follows that 
   \begin{align*}
     f(\brks{x_1,w_1},\ldots,\brks{x_m,w_m})=g(\brks{y_1,u_1},\ldots,\brks{y_l,u_l})
   \end{align*}
   is provable in $\CT$, and the desired proof of $s = t \in \CE \tensor \CT$ can be obtained from that proof
   by substituting every $\brks{x_i,w_i}$ by $s_i$ and every
   $\brks{y_j,u_j}$ by~$t_j$.

 \item Otherwise, suppose that $s_{j}=pop(\ldots,s')$ for some
   $j \in \{1, \ldots, m\}$.  Using equational reasoning (see the
   electronic appendix for details), we obtain that the following
   equations belong to $\CE \tensor \CT$:
\begin{equation}\label{eq:some-eq}
  \begin{array}{r@{\,}c@{\,}l}
  s & = & pop(push_1(s),\ldots, push_n(s), f(s_1,\ldots,s',\ldots,s_m)),
  \\
  t &= & pop(push_1(t),\ldots, push_n(t), t),
\end{array}
\end{equation}
where $s'$ occurs on the $j$-th position.
By (possibly repeated) application of the
rule~\eqref{eq:pop-pop-sigma}, we may replace
$f(s_1,\ldots,s',\ldots,s_m)$ and $t$ in the right-hand arguments by
terms $\tilde s$ and $\tilde t$, respectively, that do not contain
$pop$. Thus we obtain
\begin{align*}
  s = pop(push_1(s),\ldots, push_n(s), \tilde s), &&
  t = pop(push_1(t),\ldots, push_n(t), \tilde t),
\end{align*}
whence by soundness and since $\tsem{s}_{RX} = \tsem{t}_{RX}$ 
we have
\[
  \tsem{pop(push_1(s),\ldots, push_n(s), \tilde s)}_{RX}
  =
  \tsem{pop(push_1(t),\ldots, push_n(t), \tilde t)}_{RX}.
\]
Therefore, by Lemma~\ref{lem:stack_sem} (which is easily seen to be valid
over~$\BBR$), we obtain
\[
\tsem{push_1(s)}_{RX}=\tsem{push_1(t)}_{RX},\quad\ldots,\quad\tsem{push_n(s)}_{RX}=\tsem{push_n(t)}_{RX},\quad
\tsem{\tilde s}_{RX} = \tsem{\tilde t}_{RX}.
\]
Note that each $push_i(s)$ can be renormalized, and since $s_j$ has
the $pop$ operation on top, by \textbf{(push-pop)} the total number of
operations distinct from $push$ decreases at least by one. Hence,
using the induction hypothesis, we obtain
$push_i(s)=push_i(t)\in\CE\tensor\CT$ for every $i$. Analogously,
$f(s_1, \ldots, s', \ldots, s_m)$ has one $pop$ operator less that
$s$. Moreover, rewriting the former and $t$, respectively, in their
contexts in~\eqref{eq:some-eq} by the rule~\eqref{eq:pop-pop-sigma} might
only reduce the number of $pop$ operators further, while the number of
all other operators remains unchanged. Therefore, we obtain
$\tilde s = \tilde t \in \CE \tensor \CT$, and by standard equational
reasoning we have
\begin{align*}
s =pop(push_1(s),\ldots, push_n(s), \tilde s)
  =pop(push_1(t),\ldots, push_n(t), \tilde t)
  = t,
\end{align*}
i.e., we obtain that $s=t\in\CE\tensor\CT$ as desired.
\qed
\end{itemize}
\end{citemize}
\noqed\end{proof}
Using Proposition~\ref{prop:stens}, one can combine two stacks by computing the tensor square of the stack monad. The resulting monad $\BBT$ is a submonad of the store monad for $S = \Gamma^* \times \Gamma^*$, whence elements of $TX$ are certain maps of the form
$\brks{r,t_1,t_2}:\Gamma^*\times\Gamma^*\to X\times \Gamma^*\times\Gamma^*$. 
This allows one to define $\BBT$\dash stack automata over two and more stacks analogously to the one-stack case from Definition~\ref{defn:stack}. Before we do this formally in Definition~\ref{defn:mstack} we briefly discuss tensors with semimodule monads.

\begin{proposition}[\cite{Freyd66}]\label{prop:freyd}
The tensor product of any finitary monad with a semimodule monad is again a semimodule monad. %
\end{proposition}
\begin{remark}
Proposition~\ref{prop:freyd} in conjunction with
Proposition~\ref{prop:stens} provides two perspectives on machines with
memory and nondeterminism. On the one hand, e.g.\ we can regard the tensor 
product of $\PFin$ with the stack monad to model
(nondeterministic) push-down automata. As Proposition~\ref{prop:stens} indicates, this monad embeds into the monad with functorial part $TX=\PFin(X\times\Gamma^*)^{\Gamma^*}$. On the other hand, by Proposition~\ref{prop:freyd}, this tensor product is equivalent to a semimodule monad. 
\iffull
A rough intuition about this change of perspective can be gained from the isomorphism $\PSet(X\times\Gamma^*)^{\Gamma^*}\cong\PSet(\Gamma^*\times\Gamma^*)^X$ relating ``nondeterministic'' stateful computations over $X$ and values over $X$ weighted in the semiring $\PSet(\Gamma^*\times\Gamma^*)$.
\fi
\end{remark}
\begin{definition}[Multi-stack nondeterministic $\BBT$-automaton]\label{defn:mstack}
A \emph{multi-stack non\-det\-er\-mi\-nis\-tic $\BBT$-automaton} is a $\BBT$\dash automaton~\eqref{eq:aut} for which
\iffull
\begin{itemize}
  \item $\BBT$ is the tensor of $m$ copies of the stack monad with $\PFin$;
  \item $B$ is the set of $m$-ary predicates over ${\Gamma^*}$ consisting of all those $p\in 2^{\Gamma^*\times\cdots\times\Gamma^*}$ for each of which there is a $k$ such that for every $w_i \in \Gamma^k$ and $u_i \in \Gamma^*$, $i = 1, \ldots, m$, we have $p(w_1 u_1,\ldots,w_m u_m)=p(w_1,\ldots,w_m)$;
  \item %
for every $s\in(\Gamma^*)^m$, $f: (\Gamma^*)^m \to \PFin(B \times (\Gamma^*)^m)\in TB$
\begin{displaymath}
\algstruc(f)(s) = 1 \text{\qquad iff\qquad}  \exists s'\in (\Gamma^*)^m.~\exists p\in B.~(p,s')\in f(s)\land p(s').
\end{displaymath}
\end{itemize}
\else
$\BBT$ is the tensor of $m$ copies of the stack monad and $\PFin$; $B$ is the set of $m$-ary predicates over ${\Gamma^*}$ consisting of all those $p\in 2^{\Gamma^*\times\cdots\times\Gamma^*}$ for each of which there is a $k$ such that if for any $i$, $|w_i|\geq k$ then $p(w_1 u_1,\ldots,w_m u_m)=p(w_1\ldots,w_m)$; and for every $s\in(\Gamma^*)^m$, $f: (\Gamma^*)^m \to \PFin(B \times (\Gamma^*)^m)\in TB$ we have $\algstruc(f)(s)$ iff $\exists s'\in (\Gamma^*)^m.\,\exists p\in B.\,\ f(s)(p,s')\land p(s')$.
\fi
\end{definition}
To see that $B$ in Definition~\ref{defn:mstack} is indeed a $\BBT$-algebra, let us deduce the following corollary of Lemma~\ref{lem:subalg}.
\begin{corollary}
  Let $\BBT_S$ be the nondeterministic store monad over $S$
  (i.\,e.~$TX=\PFin(X\times S)^S$) and let~$\BBR_S$ be the
  nondeterministic reader monad over $S$ (i.e.~$R_SX=\PFin(X)^S$). For
  every submonad~$\BBT$ of\/ $\BBT_S$, the monad morphism $\alpha$
  sending any $f:S\to\PFin(X\times S)$ to
  $\PFin(\pi_1) \cdot f:S\to \PFin(X)$ restricts to a submonad of
  $\BBR_S$.
\end{corollary}
\begin{proof}
Recall that by Proposition~\ref{prop:stens}, the tensor of $\PFin$ with $m$ copies of the stack monad over $\Gamma^*$ is the submonad $\BBT$ of the nondeterministic store monad over $(\Gamma^*)^m$ identified by the following condition: $f:(\Gamma^*)^m\to\PFin(X\times(\Gamma^*)^m)\in TX$ iff there exists a $k$ such that whenever $|u_1|\geq k, \ldots, |u_m|\geq k$ then
\begin{align*}
f(u_1w_1,\ldots,u_mw_m)=\{ \brks{x,u_1'w_1,\ldots,u_m'w_m} \mid \brks{x,u_1',\ldots,u_m'}\in f(u_1,\ldots,u_m) \}.
\end{align*}
This induces a submonad $\BBR$ of the nondeterministic reader monad
over $(\Gamma^*)^m$ identified by the following condition: $f:(\Gamma^*)^m\to\PFin(X)\in TX$ iff there exists a $k$ such that whenever $|u_1|\geq k\comma \ldots\comma {|u_m|\geq k}$ then
$f(u_1w_1,\ldots,u_mw_m)= f(u_1,\ldots,u_m)$.
The $\BBT$-algebra used in Definition~\ref{defn:mstack} is thus obtained by taking $X=1$.
\end{proof}

We are now ready to prove the following result.
\begin{theorem}\label{thm:bound}
For any $m$ let $\CL_m$ be the following class of all languages 
\begin{align}\label{eq:bound}
\left\{w\in A^*\mid\sem{x_0}_{\M}(w)(\gamma_0,\ldots,\gamma_0)= \top\right\}
\end{align}
with $\M$ ranging over nondeterministic multistack $\BBT$\dash automata with $m$ stacks, $x_0$ ranging over the state space of $\M$ and $\gamma_0$ ranging over $\Gamma$. Then
\iffull 
\begin{enumerate}
 \item[(1)] $\CL_1$ is the class of context-free languages; 
 \item[(2)] for all $m>2$, $\CL_m$ is the class of nondeterministic linear time languages $\mathsf{NTIME}(n)$;
 \item[(3)] $\CL_2$ sits properly between $\CL_1$ and $\CL_3$.
\end{enumerate}  
\else
$\CL_1$ contains exactly context-free languages; for all $m>2$, $\CL_m$ contains exactly nondeterministic linear time languages, i.e.\ $\CL_m=\mathsf{NTIME}(n)$; and $\CL_2$ sits properly between $\CL_1$ and $\CL_3$.
\fi
\end{theorem}
\begin{proof}
The proof is completely analogous to the proof of Theorem~\ref{thm:stack}. We outline the main distinctions.
In lieu of quasi-real-time deterministic pda we use nondeterministic push-down quasi-real-time (NPDQRT) machines (see~\cite{BookGreibach70}). The transition function $\delta$ of such a machine has type 
\begin{align}\label{eq:NPDQRT}
\delta:Q\times (A+\{\eps\})\times\Delta^m\to \PFin(Q\times(\Delta^*)^m).  
\end{align}
This function is subject to the condition of being \emph{quasi-real-time}, i.e.\ there is a global bound on the lengths of $\eps$-transition chains over machine configurations.

Two acceptance conditions for NPDQRT are possible: (i) by final states and (ii) by the empty stack. It is a standard exercise to make sure that a language accepted by empty storage can be accepted by final states. In fact, the construction for ordinary PDAs (see e.g.~\cite{HopcroftMotwaniEtAl01}) also works for NPDQRT:  for a given PDA $P$ one forms a PDA $P'$ with a fresh initial stack symbol $\gamma_0'$ and a new inital state that pushes the original initial stack symbol on all stacks and then proceeds to the initial state of $P$. In addition, $P'$ has one final state that can be reached from all states by an (internal) $\eps$-transition if the stack content is $(\gamma_0', \ldots,\gamma_0')$ (which corresponds to configurations of $P$ with all stacks empty). Clearly, this construction preserves quasi real-timeness. 
Conversely, for every $m$, (i) can be modelled by~(ii), i.e.~a
language accepted by final states can be accepted by the empty stack: for
$m=1$, we obtain standard push-down automata for which the equivalence
of (i) and (ii) is well-known~\cite{RozenbergSalomaa97}; for $m=2$
this is shown in~\cite{GinsburgHarrison68}; for any $m>2$,
by~\cite{BookGreibach70}, the languages recognized under (ii) are
exactly $\mathsf{NTIME}(n)$ and since for quasi-real-time machines the
depths of all stacks is linearly bounded, these can be purged in
linear time once a final state is reached.

As shown in~\cite{Li85}, the class of languages recognized by NPDQRT with $m=2$ 
is properly between context-free and $\mathsf{NTIME}(n)$. 
It remains to show that for every $m$ the languages recognized by nondeterministic multistack $\BBT$-automata with $m$ stacks are the same as the languages recognized by NPDQRT with $m$ stacks with the acceptance condition chosen at pleasure.

As in Theorem~\ref{thm:stack}, given a nondeterministic multistack $\BBT$-automaton $\M$ with $m$ stacks we identify a global bound $n$ for the depths of the stack prefixes accessed at one step and then model $\M$ by an NPDQRT $M$ over the state space
\begin{displaymath}
 Q=\bigl\{\brks{x,s_1 \blk^{k_1},\ldots,s_m\blk^{k_m}}\mid x\in X,s_i\in\Gamma^*,|s_i|\leq n-k_i\bigr\}.
\end{displaymath}
The stack alphabet $\Delta$ is $\Gamma+\{\blk\}$, the transition function is given as in Theorem~\ref{thm:stack} by changing the number of elements in tuples $Q$ and by allowing for nondeterminism. The acceptance condition is chosen to be by the following final states:
\begin{align*}
F=\bigl\{\brks{x,s_1 \blk^{k_1},\ldots,s_m\blk^{k_m}}\in Q\mid o^\M(x)(s)=1, s_i\in\Gamma^{n-k_i}\bigr\}.
\end{align*}
It then follows along the same lines as in the proof of Theorem~\ref{thm:stack} that for every $w\in A^*$, $\tsem{x_0}_{\M}(w)(\gamma_0,\ldots,\gamma_0)=1$ iff $w$ is accepted by $M$ with $\brks{x_0,\gamma_0,\ldots,\gamma_0}$ as the initial configuration.

In order to show the second part of the claim, assume that $M$ is a NPDQRT with $m$ stacks, a transition function~\eqref{eq:NPDQRT}, an initial state $q_0\in Q$, a set of accepting states $F\subseteq Q$ and an initial stack symbol $\blk$. According to~\cite{BookGreibach70}, we assume that $M$ is \emph{real-time}, i.e.\ there
are no internal transitions. 

We define a nondeterministic $\BBT$-automaton over $m$ stacks with $X=Q$ and with stack symbols $\Delta$ as follows: for any $q\in X$, $s_i\in\Delta^*$, $a\in A$, $o^\M(q)(s_1,\ldots,s_m)=1$ iff $q\in F$ and
\begin{flalign*}
&&t^\M(q,a)(s_1,\ldots,s_m) =&\; \emptyset& \text{(if $s_i=\eps$ for some $i$)}\\
&&t^\M(q,a)(\gamma_1 s_1,\ldots,\gamma_m s_m)=&\;\{\brks{q',s'_1 s_1,\ldots,s'_ms_m}\mid\\
&&&\quad\! \brks{q',s_1',\ldots, s_m'}\in\delta(q,a,\gamma_1,\ldots\gamma_m)\}& \text{(otherwise)}
\end{flalign*}
A similar argument as in Theorem~\ref{thm:stack} then shows that for
every $w\in A^*$, $q\in Q$ and $s\in\Delta^*$ an accepting
configuration is reachable from $\brks{q,s_1,\ldots,s_m}$ by $w$ iff
$\sem{q}_{\M}(w)(s_1,\ldots,s_m)=1$.
\end{proof}

\noindent
Theorem~\ref{thm:bound} shows, on the one hand, that the coalgebraic formalization of nondeterministic pushdown automata as nondeterministic $\BBT$\dash automata over one stack is adequate in the sense that it recognizes the same class of languages. On the other hand, it indicates the boundaries of the present model: it seems unlikely to capture languages beyond $\mathsf{NTIME}(n)$ (e.g.\ all recursive ones) by a computationally feasible class of $\BBT$\dash automata. This is in
agreement with the early work on (quasi-)real-time recognizable languages~\cite{BookGreibach70}, which underlies the proof of Theorem~\ref{thm:bound}. We return to this issue in Section~\ref{sec:cps} where we provide an extension of the present semantics that allows us to capture language classes up to recursively enumerable ones.   

We conclude this section with a corollary of Theorem~\ref{thm:bound} and Proposition~\ref{prop:test}. It is well known that equivalence of context-free languages is undecidable; in fact, it is $\Pi_1^0$-complete (the non-halting problem for arbitrary Turing machine can be encoded as an equality of certain context-free languages~\cite{Hartmanis67}).
We will use this to prove a similar completeness result for the
equivalence of reactive expressions. We say that a $\Sigma$-algebra
$B$ over a set of generators $B_0$ is \emph{effectively presented
  (over $\Sigma$ and $B_0$)} if $\Sigma$ and $B_0$ are recursive sets
and the set
\begin{displaymath}
  \{ (t,s) \mid \text{$t,s$ are closed terms over $\Sigma$, $B_0$ with $t^B = s^B$}\} 
\end{displaymath} 
is decidable (recall Notation~\ref{not:terms}(2)).
The language equivalence problem for reactive expressions is then the
following decision problem: given recursive sets $\Sigma$ and $B_0$,
an effectively presented $\BBT$-algebra $B$, and two reactive expressions $e_1$
and $e_2$ in $\Exp{\Sigma}{B_0}$,
decide if $e_1 \sim e_2$ (cf.~\eqref{eq:partial}).

\begin{corollary}\label{cor:nre}
The language equivalence of reactive expressions is $\Pi_1^0$-complete.
\end{corollary}
\begin{proof}
  The fact that language equivalence of reactive expressions is
  in $\Pi_1^0$, i.e.\ co-r.e.,\ follows from
  Proposition~\ref{prop:test}: if two reactive expressions $e$ and $u$
  are not language equivalent, we can eventually detect this by
  finding a suitable word $w\in A^*$ for which
  $o(\partial_w(e))\neq o(\partial_w(u))$. Here we rely on our effectiveness assumption, 
  for, by the definitions~\eqref{eq:partial}, both $o(\partial_w(e))$ and $o(\partial_w(u))$ are terms over $\Sigma$ and $B_0$ 
  evaluated over $B$. 
  To prove $\Pi_1^0$-hardness, let us show how to reduce the
  equivalence problem of context-free languages to the current
  equivalence problem of reactive expressions. Given two context-free
  languages $L_1$ and $L_2$ recognized by two pushdown automata over a
  stack alphabet $\Gamma = \{\gamma_1, \ldots, \gamma_n\}$, we provide
  an instance of our problem with $B \subseteq 2^{\Gamma^*}$ being the
  $\BBT$-algebra from Definition~\ref{defn:mstack} for the
  nondeterministic stack theory $\BBT$ over one stack. We need to
  prove that $B$ is effectively presented. First note that both
  $\Sigma$ and $B_0$ are finite, specifically, $B_0$ is the
  two-element set $\{\top,\bot\}$. Indeed, $\Sigma$ consists of the
  operation symbols $pop$, $push_i$, $i = 1, \ldots, n$, $+$ and
  $\emptyset$, and we recall from
  Definition~\ref{defn:mstack} that $B$ consists of precisely those
  predicates $p$ over ${\Gamma^*}$ for each of which there is a $k$
  such that for every $w \in \Gamma^k$ and $u \in \Gamma^*$,
  $p(w u)=p(w)$. Hence, each predicate $p$ in $B$ can be finitely
  represented, e.g.~by the list of words $w$ in $\Gamma^k$ with
  $p(w) = \top$. In order to check that $t^B = s^B$ for a given pair
  of terms $t, s$ over $\Sigma, B_0$ we first compute the two
  predicates $t^B, s^B$ and then verify that they are equal. Indeed,
  for the latter we only need to verify $(t^B)(w)=(s^B)(w)$ for
  finitely many $w\in\Gamma^*$, which is a decidable problem, and
  for the former we need to verify that the algebra operations on $B$
  are computable. Using the definition of $\alpha^\M$ in
  Definition~\ref{defn:mstack} and the interpretation of the
  nondeterministic stack theory over $\BBT$, which is a
  submonad of the store monad transform
  $\PFin(X \times \Gamma^*)^{\Gamma^*}$ (cf.~the proof of
  Proposition~\ref{prop:stens}), we verify that for every $p, q, p_i$,
  $i = 1, \ldots n$, in $B$ we obtain that the semantics $pop^B$, $push_i$,
  $+^B$ and ${\emptyset}^B$ are computable.

  By Theorem~\ref{thm:bound}(2), we have two nondeterministic stack
  $\BBT$-automata $\M_1$ and $\M_2$ and states $x_1$, $x_2$,
  respectively, in them such that $L_i = \left\{ w\in A^* \mid \sem{x_i}_{\M_i} (w) (\gamma_1) = \top
    \right\}$ with $i = 1,2$.

  Further, by Theorem~\ref{thm:kleene}, we obtain reactive expressions
  $e_i$ in $\Exp{\Sigma}{B_0}$ such that $\sem{e_i} =
  \sem{x_i}_{\M_i}$ with $i = 1,2$. Thus, we have $L_1=L_2$ iff
  $\lambda w.\,\sem{e_1}(w)(\gamma_1)=\lambda
  w.\,\sem{e_2}(w)(\gamma_1)$. Of course, the latter is not equivalent
  to $\lambda w.\,\sem{e_1}(w)(s)=\lambda w.\,\sem{e_2}(w)(s)$ for \emph{all}
  $s\in\Gamma^*$. However, it is easy to modify $e_1$ and $e_2$ to obtain
  this property: let for $i=1,2$,
  \begin{align*}
    e_i' = pop(pop(\emptyset,\ldots,\emptyset,push_1(e_i)),\emptyset,\ldots,\emptyset).
  \end{align*}
  Then, clearly, $\sem{e_1'}=\sem{e_2'}$ iff
  $\lambda w.\,\sem{e_1}(w)(\gamma_1)=\lambda
  w.\,\sem{e_2}(w)(\gamma_1)$.
\end{proof}
\takeout{%
Note that the formulation of Corollary~\ref{cor:nre} technically rules out examples
like probabilistic automata, because $B=[0,1]$ is not recursive. Of course, the
corresponding examples are not lost, because in practice we are still only interested
in automata that have an effective presentation, which strictly saying amounts to
taking as $B$ a suitable subset of constructive reals.}%

One can also consider the language equivalence problem of reactive
expressions for a fixed $\Sigma$ and $B$; a very similar argument than
the one in the previous proof then shows that the language equivalence
of reactive expressions for the algebra $B \subseteq 2^{\Gamma^*}$ of
Definition~\ref{defn:mstack} is $\Pi_1^0$-complete.

However, for other $\Sigma$ and $B$ (coming from a type of
$\BBT$-automaton), the language equivalence of reactive expressions is
decidable, e.g.~for finite $\Sigma$ and $B$ this follows from
Theorem~\ref{thm:kleene} and Proposition~\ref{prop:fin_b}, for the
identity monad $\BBT$ and any recursive set $B$ (in this case
$\BBT$-automata are simply Moore automata with output in $B$), for
the finite powerset monad $\BBT = \PFin$ and $B= 2$ (for which
$\BBT$-automata are classical nondeterministic automata), or for the
monad $\BBT$ assigning to a set the set of formal linear combinations
with coefficients from a field $B= k$ (for which
$\BBT$-automata are weighted automata over $k$). Identifying further
monads $\BBT$ and algebras $B$ for which the language equivalence for
reactive $\BBT$-expressions is decidable is an interesting question
for future work.

\iffull
\section{Context-free Languages and Valence Automata}
Throughout this section we assume that $R$ is a semiring finitely
generated by the set $R_0$; $B$ is an $R$-semimodule finitely
generated by the set $B_0$; and $\BBT_R$ is the semimodule monad for
$R$.

By Proposition~\ref{prop:freyd}, a nondeterministic $\BBT$-automaton over one stack is a specific case of a weighted $\BBT$-automaton~(Definition~\ref{defn:waut}). In this form it is rather similar to \emph{valence automata}, another example of a machine previously studied in the literature~(e.g.~\cite{RenderKambites09,Kambites09}). We present the corresponding algebraic theories side by side and explain how valence automata can be formalised as $\BBT$-automata. 
\begin{example}[Nondeterministic stack theory]\label{expl:sex} 
The \textit{nondeterministic stack theory} is obtained by tensoring $\CE_{\PFin}$, i.e.~the theory of commutative, idempotent monoids, with the stack theory. The result is a semimodule theory over an idempotent semiring $R$ presented by the generators $o_{i}$, $u_{i}$, $i = 1, \ldots, |\Gamma|$ and $e$ and the following relations:
\begin{flalign*}
\quad u_io_i = 1 && u_io_j = 0 && u_ie = 0 && 
o_1u_1 + \ldots + o_nu_n + e = 1&&
eo_i = 0 && ee = e&&  (i\neq j)
\end{flalign*}
The corresponding unary operations of the semimodule theory are
denoted by $pop_i=\overline{o_i}: {1 \to 1}$,
$push_i=\overline{u_i}: 1\to 1$ and $empty = \overline{e}: 1 \to 1$
(cf.~the notation of Definition~\ref{defn:srmon}). It is
straightforward to relate the nondeterministic stack theory and the
presented semimodule theory by giving two translations defining the
operations of one theory in terms of operations of the other. First
the unary operations $pop_i$ and $empty$ of the semimodule theory
determine $pop$:
\begin{align*}
pop(x_1,\ldots,x_n,y) = pop_1(x_1) + \ldots + pop_n(x_n) + empty(y).
\end{align*}
Conversely, $pop_i$ and $empty$ can be defined from $pop$:
\begin{align*}
  pop_i(x)  = pop(\emptyset\comma\ldots\comma x\comma\ldots\comma\emptyset\comma\emptyset) &&
  empty(x)  = pop(\emptyset\comma\ldots\comma\emptyset\comma x)
\end{align*}
($x$ is on the $i$-the position in the sequence on the left).
It is then straightforward to prove that the axioms of the
nondeterministic stack theory and semimodule theory for $R$,
respectively, are satisfied for the operations as defined by the
translations.\smnote{Details are in handwritten notes in the svn.}
\end{example}
\begin{example}[Nondeterministic monoid action theory]\label{expl:nd_act}
  The \emph{nondeterministic monoid action theory} is obtained by tensoring the theory $\CE_{\PFin}$ with the theory of $M$-actions of the monoid $(M, \cdot, 1)$ (see Example~\ref{ex:mon}). As shown in~\cite{HylandLevyEtAl07}, the corresponding monad $\BBT$ maps a set $X$ to $\PFin(M \times X)$ and has the unit $\eta_X: x \mapsto \{(1,x)\}$ and the Kleisli lifting given by extending a map $f: X \to \PFin(M\times Y)$ to $f^\klstar: \PFin(M \times X) \to \PFin(M \times Y)$ with 
\[
f^\klstar(S) = \{\, (m\cdot n, y) \mid \exists x.\,(m,x) \in S\land (n,y) \in f(x)\,\}.
\]
Note that the theory corresponding to $\BBT$ is the semimodule theory for the semiring $R = \PFin(M)$ with addition given by $\cup$ with unit $\emptyset$ and multiplication given by $S \cdot S' = \{ m \cdot n \mid m \in S, n \in S'\}$ for any finite subsets $S$ and $S'$ of $M$ with unit~$\{1\}$.   

Now let us fix a monoid $(M, \cdot, 1)$. The idea of valence automata over $M$ is to use the monoid structure to model various kinds of stores (stack(s), counter(s), etc.). Recall e.g.~from~\cite{RenderKambites09,Kambites09} that a valence automaton over $M$ is a tuple $\CA = (X, M, A, \delta, q_0, F)$ where $X$ is a finite set of states, $\delta$ is finite subset of $X \times A^* \times M \times X$ of transitions, $q_0 \in X$ is an initial state and $F  \subseteq X$ a set of final states. This induces a transition relation $\Rightarrow$ on $X \times A^* \times M$ as usual by defining $(p, w, m) \Rightarrow (q, wu, mn)$ if there exists $(p, u, n, q)\in\delta$. The language accepted by a given valence automaton $\CA$ is 
\[
L(\CA) = \{\, w \in A^* \mid \exists q \in F.\, (p, \eps, 1) \Rightarrow (q, w, 1) \,\}
\]
We call $\CA$ \emph{$\epsilon$-free} if $\delta$ does not contain tuples of the form $(p,\eps,n,q)$. Note that for any $\epsilon$-free valence automaton there is an equivalent one that only contains single letters in its transitions, for every transition $(p, a_1a_2 \cdots a_n, m,q)$ can be replaced by transitions 
\[
(p, a_1, 1, p_1), (p_1, a_2, 1, p_2), \ldots, (p_{n-1}, a_n, m, q).
\]
An $\epsilon$-free valence automaton $\CA$ in which every transition contains only single letters can be regarded as a $\BBT$-automaton~\eqref{eq:aut} for the nondeterministic monoid action theory over $M$. Indeed, let $B = \PFin(M \times 1) \cong \PFin(M)$ be the free $\BBT$-algebra on $1$ and let us define $o^\M: X \to \PFin(M)$ by $o^\M(q) = \{1\}$ if $q \in F$ and $o^\M(q) = \emptyset$ else; the transitions function $\delta$ produces $t^\M: X \to \PFin(M \times X)^A$. Using Lemma~\ref{lem:ind} it is easy to prove that $\{w \in A^* \mid 1 \in \sem{q_0}_\M\}$ is the language accepted by~$\CA$.%
\end{example}
\begin{example}[Nondeterministic polycyclic theory]\label{expl:poly}%
A relevant special case of the previous example is when $M$ is a \emph{polycyclic monoid}~\cite{Lawson99}. This means that $M$ is the monoid over a set of generators $\lightning,g_1,\ldots,g_k,\ldots, g_1^{\mone},\ldots,g_k^{\mone}$ and satisfying the relations
\begin{align*}
\lightning g_i = g_i \lightning = \lightning, && g_i g_i^{\mone}=1,&& g_i g_j^{\mone}=\lightning\qquad (i\neq j).
\end{align*}
The number $k$ is called the \emph{rank} of $M$. We call the theory of $\PFin(M \times (-))$ the \emph{nondeterministic polycyclic theory}.
\end{example}
\noindent 
The technical distinction between the nondeterministic stack theory and the nondeterministic polycyclic theory is minor. On the one hand, the nondeterministic stack theory uses the zero $0$ of the semiring to model failure in computing the right inverse, while the nondeterministic polycyclic theory has its own zero $\lightning$, which coexists with $0$. On the other hand, emptiness detection is explicitly available for stacks (using $e$) but not for polycyclic monoids. 

It is well-known that valence automata over polycyclic monoids of rank at least~$2$ recognize context-free languages and so do nondeterministic stack $\BBT$-automata. We would like to give a uniform proof of this fact applying both to Example~\ref{expl:sex} and to Example~\ref{expl:poly}.

First, observe that if the semiring $R$ is idempotent then any $R$-semimodule (equivalently, $\BBT_R$-algebra) $B$ can be partially ordered by putting $b\leq c$ iff $b+c=c$. 

\begin{definition}
  Given a $\BBT_R$-automaton $\M: X \to B \times (T_R  X)^A$, and an initial state $x_0\in X$ we define the language \emph{recognized by $b\in B$} by
\begin{displaymath}
L_b(\M) = \{w\in A^*\mid\sem{x_0}_\M(w)\geq b\}.
\end{displaymath}
\end{definition}
Note that both  semirings arising from Examples~\ref{expl:sex} and~\ref{expl:poly} are idempotent (since addition is given by union of sets). For nondeterministic stack $\BBT$-automata we typically choose as $b\in B \subseteq 2^{\Gamma^*}$ the predicate distinguishing the initial stack configuration, e.g.\ $b(w)=\top$ iff $w$ is the initial stack symbol. For valence automata over $M$ we take $B=\PFin(M)$ and $b=\{1\}$. Then the above definition of accepted languages instantiates as expected. 

Recall that the language of balanced parentheses, or \emph{Dyck language} is a language $\CD_n\subseteq\CA_n=\{(_1,)_1,\ldots,(_n,)_n\}^*$ consisting of string of parentheses balanced in the standard sense.
The following result is a reformulation of the classical Chomsky-Sch\"utzen\-berger theorem.
\begin{theorem}\label{thm:chsh}
Let $\alpha$ be a monoid morphism from $\mathcal{A}_2$ to the multiplicative structure of some idempotent semiring $R$ such that 
\begin{enumerate}
 \item for some $b_0,b_1\in B$, $\alpha(w)\cdot b_0\geq b_1$ iff $w$ is balanced;\label{a:cch1}
 \item for any $c_1,c_2$ if $c_1+c_2\geq b_1$ then either $c_1\geq b_1$ or $c_2\geq b_1$.\label{a:cch2}
\end{enumerate} 
Then for any context-free language there is a $\BBT_R$-automaton recognizing it by~$b_1$. 
\end{theorem}
\begin{proof}
Let us denote $\Omega_n=\{(_1,)_1,\ldots,(_n,)_n\}$ so that $\mathcal{A}_n =
\Omega_n^*$. First note that from ${\alpha: \CA_2 \to R}$ that we postulated
we can obtain a monoid morphism $\alpha': \CA_n \to R$ for every $n$ with the same
property~\eqref{a:cch1}. Indeed, following~\cite{Book75}, we define a monoid morphism
$\gamma:\CA_n\to\CA_2$ sending every $(_n$ to $(_1^n(_2$ and every
$)_n$ to $)_2)_1^n$ and having the property that $\CD_n=\gamma^{\mone}(\CD_2)$, which means that if
$\gamma(w)\in\CA_2$ is balanced then $w$ is also balanced. Since the converse
is obvious, the composition $\alpha'=\alpha\cdot \gamma:\CA_n\to R$ inherits from 
$\alpha$ the property that $\alpha'(w)\cdot b_0\geq b_1$ iff $w$ is
balanced. 

Let $\CL$ be any context-free language. By Theorem~\ref{thm:kleene} and Proposition~\ref{prop:aexp}, it suffices to prove that there is an additive reactive expression $e$ such that 
\[
\CL = \{w \in A^* \mid \sem{e}(w) \geq b_1\}.
\]
By the Chomsky-Sch\"utzenberger theorem, we have
$\CL = \beta(\CR\cap\CD_n)$ for some regular language~$\CR$ over
$\Omega_n$ and some monoid morphism $\beta:\CA_n\to A^*$, and,
according to the above argument, in what follows let us regard
$\alpha$ as a morphism from $\CA_n$ to $R$. We use the version of the
Chomsky-Sch\"utzenberger theorem from~\cite{Okhotin12} where it is
shown that if $\CL$ does not contain one-letter words then $\beta$ can
be chosen \emph{non-erasing}, i.e.\ ${\eps\not\in\beta(g)}$ for all
${g\in\Omega_n}$. The assumption that $\CL$ does not contain
one-letter words does not restrict generality, for if we could show
for $\CL'=\CL\setminus A$ and an expression~$e$ that
$\CL'=\{w\in A^*\mid \sem{e}(w)\geq b_1\}$ then of course we would
have
\begin{align*}
\CL=\left\{w\in A^*\mid \bigl\llbracket e+\sum\nolimits_{a\in \CL\cap A} a.b_1\bigr\rrbracket(w)\geq b_1\right\}.
\end{align*}
Henceforth we assume that $\CL\cap A=\emptyset$ and $\beta$ is non-erasing.

As we know from Propositions~\ref{prop:fin_b} and~\ref{prop:aexp}, $\CR$ can be given by an additive reactive expression over the boolean semiring $R=B=\{0,1\}$. We replace in this expression every occurrence of the form $g.-$ where $g\in \Omega_n$ by $a_1.\,\ldots \,a_k.\,\alpha(g)\cdot (-)$ where $a_1\cdots a_k=\beta(g)$ and every occurrence of $1\in B_0=\{1\}$ by $b_0$. The resulting expression $e$ is a reactive additive expression for the semimodule monad $\BBT_R$. Note that the assumption that $\beta$ is nonerasing ensures that $e$ remains guarded. It is then easy to check that
\begin{align}\label{eq:chsh}
\sem{e}(w) \geq r\cdot b_0 \text{~~~~if~~~~} \exists u\in\CA_n.\, r=\alpha(u)\land w=\beta(u);
\end{align}  
indeed, given $u=g_1 \ldots g_k\in\CA_n$ with $g_i\in\Omega_n$ such 
that $r=\alpha(u)$ and $w=\beta(u)$, by definition, $\sem{e}(w) \geq r\cdot b_0$ iff 
$o(\partial_{\beta(g_1)\cdots\beta(g_k)}(e))\geq \alpha(g_1)\cdots\alpha(g_k)\cdot b_0$,
which follows from  the definition of $e$, specifically, from the way
$g.-$ is replaced.
Suppose, $w\in\CL=\beta(\CR\cap\CD_n)$. Then there is $u\in\CD_n$ such that $w=\beta(u)$. By assumption~\eqref{a:cch1}, $\alpha(u)\cdot b_0\geq b_1$ and by~\eqref{eq:chsh}, $\sem{e}(w) \geq \alpha(u)\cdot b_0$. Therefore, $\sem{e}(w)\geq b_1$.

For the converse, suppose $\sem{e}(w)\geq b_1$ and show that $w\in\CL$. Note that $\sem{e}(w)$ is representable as a finite sum $\alpha(u_1)\cdot b_0+\cdots+\alpha(u_k)\cdot b_0$ in such a way that $w=\beta(u_i)$ and $u_i\in\CR$ for all $i$. By assumption~\eqref{a:cch2}, $\alpha(u_j)\cdot b_0\geq b_1$ for some $j$ and therefore by assumption~\eqref{a:cch1}, $u_j$ is balanced. Since $w=\beta(u_j)$, $u_j\in\CR$ and $u_j\in\CD_n$, we obtain $w\in\CL$. %
\end{proof}
\begin{example}
  Let us check that conditions of Theorem~\ref{thm:chsh} apply to Examples~\ref{expl:sex} with $|\Gamma|>1$ and~\ref{expl:poly} with $k>1$.
    \begin{cenumerate}
    \item For the nondeterministic stack theory, let us consider
      $B\subseteq 2^{\Gamma^*}\cong\PSet(\Gamma^*)$ as in
      Definition~\ref{defn:mstack} (for $m=1$). It is not difficult to
      work out that the action of $R$ on $B$ satisfies for every given
      $f: \Gamma^* \to 2$ the following laws
      \begin{align}\label{eq:Bmod}
        e \cdot f (u) = \begin{cases}
          f(\eps) & \text{if $u = \eps$}  \\ 
          0 & \text{else},
        \end{cases} &&
        o_i \cdot f(u) =  \begin{cases}
          f(v) & \text{if $u = \gamma_iv$} \\
          0 & \text{else},
        \end{cases} &&
        u_i \cdot f(u) = f(\gamma_iu); 
      \end{align}
      in fact, this holds because $e\cdot (-)$, $o_i \cdot (-)$ and
      $u_i\cdot (-)$ are the unary operations $empty^B$, $pop_i^B$ and
      $push_i^B$, respectively, by using the definition of $empty$, $pop_i$ and
      $push_i$ from $push$ and $pop$ (see Example~\ref{expl:sex}), and
      by using how $pop^B: B^{n+1} \to B$ and $push^B: B \to B$ act
      ensuing the definition of the algebra structure $\algstruc$ on $B$
      (see Definitions~\ref{defn:mstack}, and~\ref{defn:stack}).

      \medskip
      \indent
      We take $\alpha: \mathcal{A}_2 \to R$ sending $(_i$ to $u_i$,
      $)_i$ to $o_i$ for $i = 1, 2$ and $b_0 = b_1 =
      \{\eps\}$. Condition~\eqref{a:cch2} holds because $\{\eps\}$ is
      an atom of the Boolean algebra $\PSet(\Gamma^*)$ (noting that
      $+$ on $B$ is union of languages over~$\Gamma$). 
      Condition~\eqref{a:cch1} means that $w$ is balanced
      iff $\alpha(w)\cdot \{\eps\} \supseteq \{\eps\}$. This is easy
      to verify: on the one hand, if $w$ is balanced, then $\alpha(w)$
      can be reduced to $1$ by successively replacing every
      $\alpha((_i)_i) = u_io_i$ by $1$, and therefore for such $w$,
      $\alpha(w)\cdot \{\eps\} = \{\eps\}$; on the other hand, if $w$
      is not balanced, by replacing $\alpha((_i)_i) = u_io_i$ with $1$
      we eventually obtain that $\alpha(w)$ either (i) contains a
      factor $u_io_j$ with $i\neq j$, or (ii) contains a factor
      $o_iu_j$, or (iii) is a nonempty product of $u_i$'s, or (iv) is
      a nonempty product of $o_i$'s. In the cases (i)--(iii), we see
      that $\alpha(w) \cdot \{\eps\}$ is $\emptyset$ using the
      relations from Example~\ref{expl:sex} and the equations
      in~\eqref{eq:Bmod}. In the remaining case, $\alpha(w)$ is a
      (nonempty) product of the $o_i$, and hence
      $\alpha(w)\cdot \{\eps\} \supseteq \{\eps\}$ would imply a
      contradiction:
      $(e\alpha(w))\cdot \{\eps\} = 0 \cdot \{\eps\} = \emptyset
      \supseteq e\cdot \{\eps\} = \{\eps\}$.

    \item For the polycyclic theory we take $\alpha: \CA_2 \to \PFin(M)$ sending $(_i$ to $\{g_i\}$ and~$)_i$ to $\{g_i^{\mone}\}$ for $i = 1,2$ and $b_0 = b_1 = \{1\}$. The verification of conditions~\eqref{a:cch1} and~\eqref{a:cch2} here is analogous, in particular, for~\eqref{a:cch1} one readily checks that $\alpha(w)=1$ iff $w$ is balanced.
    \end{cenumerate}
\end{example}
Contrasting~\cite{Kambites09} we cannot replace the polycyclic monoid in Example~\ref{expl:poly} by a free group and conclude by Theorem~\ref{thm:chsh} that automata over free groups recognize context-free languages, for the relevant construction would essentially depend on internal transitions, which we do not~allow.

As we have seen by Examples~\ref{expl:sex} and~\ref{expl:nd_act}, for any 
$\Sigma$-theory we can automatically generate its nondeterministic variant by tensoring
with $\CE_{\PFin}$ and this has a sensible interpretation in terms of $\BBT$-automata. 
One may wonder if it is possible to convert a given $\BBT$-automaton to a $\BBT\tensor\PFin$-automaton 
(which is necessarily a weighted $\BBT\tensor\PFin$-automaton, by Proposition~\ref{prop:freyd}) 
preserving the semantics. The answer turns out to be affirmative under a natural assumption on the
$\BBT$-algebra component $B$.

Let us first establish the following general result. It follows
from~\cite[Proposition~5.1]{BonsangueEA15};\sgnote{Reference not
  found. Does it really follow from anywhere?}\smnote{The paper is in
  the svn. Prop.~5.1 there {\bf is} our lemma!} we include a proof for the convenience of the reader. 
\begin{lemma}\label{lem:2sem}
Let $\kappa:\BBT\to\BBS$\/ be a monad morphism, and let $\M:X\to
B\times TX^A$ and $\M_*:X\to B\times SX^A$ be a $\BBT$- and an
$\BBS$-automaton over $X$, respectively, such that
\begin{align*}
o^{\M}=o^{\M_*}, && \algstruc = \algstrucs\cdot \kappa_B, && \kappa_X \cdot t^{\M} = t^{\M_*}
\end{align*}
(in particular this implies that $B$ is simultaneously a $\BBT$- and an $\BBS$-algebra). Then the language semantics of $\M$ and $\M_*$ agree, i.e. 
  $
  \sem{x}_{\M} = \sem{x}_{\M_{*}}
  $
for any $x\in X$.
\end{lemma}
\begin{proof}
The proof amounts to showing commutativity of the following diagram:
\begin{equation*}
\begin{tikzcd}[column sep=huge, row sep=normal]
X
  \rar["\eta^{\BBT}_X"]
  \ar[rr, bend left=20, "\eta^{\BBS}_X"] 
  \dar["\M"'] & 
T X
  \rar[r, "\kappa_X"]
  \ar[rr,"\widehat \M^\sharp", bend left=20]
  \ar[ld, "\M^\sharp"'] 
& 
SX
  \ar[ld, "\M_{*}^\sharp"']
  \rar[r, "\widehat\M_{*}^\sharp"] & 
B^{A^*}
  \dar[d, "\iota"]\\
B\times (TX)^A
  \rar[r, "\id\times\kappa_X^A"] & 
B\times (SX)^A
  \ar[rr, "\id\times(\widehat \M_{*}^\sharp)^A"]  
&& 
B\times (B^{A^*})^A
\end{tikzcd}
\end{equation*}
The left-hand triangle commutes by the definition of $\M^\sharp$, the
right-hand part by the finality of $B^{A^*}$ (recall
from~\eqref{diag:hatm} that $\widehat \M_{*}^\sharp$ denotes the
unique coalgebra morphism) and the upper left-hand triangle since
$\kappa$ is a monad morphism. The upper right-hand triangle commutes
by uniqueness of the final map from $TX$ to $B^{A^*}$ as soon as we
establish commutativity of the middle parallelogram.

To see the latter we will use the freeness of the $\BBT$-algebra
$(TX, \mu_X)$ (see Section~\ref{sec:mon}) in the upper left-hand
corner, i.e.~we shall show that all morphisms in this part are
$\BBT$-algebra morphisms and that this part commutes when precomposed
with $\eta_X^\BBT$. Indeed, the latter follows from the fact that the
upper left-hand triangle commutes and since clearly
\begin{displaymath}
\M_* = \bigl(
 X \xrightarrow{~~\M~~} B \times (TX)^A \xrightarrow{~\id \times (\kappa_X)^A~} B \times (SX)^A
\bigr).
\end{displaymath}
Now to see that all morphisms in the middle part are $\BBT$-algebra
morphism, recall first that the monad morphism
$\kappa: \BBT \to \BBS$ induces a functor $\bar\kappa$ from the
category of $\BBS$-algebras to the category of $\BBT$-algebras given
on objects by $(Y, t) \mapsto (Y, t \cdot \kappa_Y)$ and being
identity on morphism. Clearly, this functor maps
$(B, \algstrucs)$ to $(B, \algstruc)$. Now let $\beta$ and $\beta^*$,
denote the algebraic structures on $B \times (TX)^A$ and
$B \times (SX)^A$, respectively, which are componentwise given by the
structures of the free algebras $(TX, \mu_X)$ and by $\algstruc$ and
$\algstrucs$ on $B$, respectively. Now consider the four morphisms of
the middle parallelogram of our diagram: (1)~$\kappa_X: TX \to SX$ is easily
seen to be a $\BBT$-algebra morphism from the free $\BBT$-algebra
$(TX, \mu^\BBT_X)$ to the $\BBT$-algebra $(SX, \mu^{\BBS}_X \cdot \kappa_{SX})$
(since $\kappa$ is a monad morphism) and therefore
(2)~$\id \times (\kappa_X)^A$ is a $\BBT$-algebra morphism from
$(B \times (TX)^A, \beta)$ to
$(B \times (T_B X)^A, \beta^* \cdot \kappa_{B \times (S X)^A})$;
(3)~$\M^\sharp$ is by definition a $\BBT$-algebra morphism
and~(4)~$\M_*^\sharp$ is an $\BBS$-algebra morphism and hence by
applying the functor $\bar\kappa$ we see it is also a $\BBT$-algebra
morphism as desired. %
\end{proof}
We immediately obtain the following corollary.
\begin{corollary}\label{cor:aut_tens}
Let $B$ be a $\BBT\tensor\PFin$-algebra with structure $\algstrucs: (T\tensor\PFin) B\to B$.
Then $B$ is also a $\BBT$-algebra under $\algstrucs\cdot \kappa_B: TB\to B$ where $\kappa : \BBT \to \BBT \tensor \PFin$ is the left tensor injection. Let $\M$ be any 
$\BBT$-automaton~\eqref{eq:aut} with $\algstruc=\algstrucs\cdot \kappa_B$ and form the $\BBT \tensor \PFin$-automaton $\M_*$ with 
  $o^{\M_*}=o^\M$, $t^{\M_*}=\kappa_X \cdot t^{\M}$ %
and the given $\algstrucs$. Then the semantics of $\M$ and $\M_*$ agree; in symbols: $\sem{x}_{\M}=\sem{x}_{\M_*}$ for every state $x\in X$.
\end{corollary}
Effectively, Corollary~\ref{cor:aut_tens} states that for every $\BBT$
we can understand a $\BBT$-automaton as a special nondeterministic
automaton, i.e.~a $\BBT\tensor \PFin$-automaton, provided that its
output $\BBT$-algebra $B$ additionally carries the structure of a
commutative idempotent monoid which commutes with the operations of
$\BBT$ (in the sense of satisfying the tensor laws), e.g.\ this applies
to submonads $\BBT$ of the state monad over a store
$S$ and output algebras $B$ which are subalgebras of $2^S$ (see e.g.\
Example~\ref{defn:stack}).

\section{\texorpdfstring{CPS-transforms of $\BBT$-automata and r.e.-languages}{CPS-transforms of $T$-automata and r.e.-languages}}\label{sec:cps}
\sloppypar
Theorem~\ref{thm:bound} suggests that the present language semantics
is unlikely to produce languages beyond $\mathsf{NTIME}(n)$ under a
computationally convincing choice of the components of a
$\BBT$-automaton~\eqref{eq:aut}. The approach suggested by the
classical formal language theory is to replace $A$ with the set
$A_{\tau}=A\cup \{\tau\}$, where $\tau$ is a new \emph{unobservable
  action}\footnote{We prefer to use $\tau$ instead of the more
  standard $\eps$ to avoid confusion with the empty word.}, but in
lieu of the formal power series $B^{A^*_\tau}$ we use $B^{A^*}$ as the semantic domain. This new \emph{observational semantics} is supposed to be obtainable from the standard one by eliminating the unobservable actions.   

\iffull
We argue briefly, why the general assumptions about the structure of a $\BBT$-automaton are not sufficient to define the observational semantics. Consider an automaton $\M:X\to B\times X^{A_\tau}$ with $A=\{a\}$ and $B = \{b_0,b_1\}$. The underlying monad is the identity monad and $\algstruc$ is the identity morphism. Let $X=\{x_0,x_1\}$, $o_\M=\{\brks{x_0,b_0},\brks{x_1,b_1}\}$, $t_\M=\{\brks{x_0,a,x_0},\brks{x_0,\tau,x_1},\brks{x_1,a,x_1},\brks{x_1,\tau,x_1}\}$. Removal of $\tau$-transitions leads to a nondeterministic automaton having two $a$-transitions from~$x_0$ to states marked with $b_0$ and with $b_1$ by $o_\M$, which cannot be determinized unless we assume the structure of a commutative idempotent monoid (i.e.~a $\PFin$-algebra structure) on $B$. A similar argument applied to  looped internal transitions shows that $B$ must support countable iterations of the monoid operation.  

Using these assumptions on $B$, our idea is to use
Lemma~\ref{lem:2sem} to transform a given $\BBT$-automaton~$\M$ to
some $\BBS$-automaton $\M_*$ for which $\tau$-transitions can be
eliminated in a generic way. After eliminating the $\tau$-transitions
we then obtain an $\BBS$-automaton $\M_v$, and finally we define the
observational semantics of $\M$ as the standard language semantics of
$\M_v$. Note that Corollary~\ref{cor:aut_tens} does not offer a
sufficiently good candidate for $\M_*$, because we would have to
assume that $B$ is a $\BBT\tensor\PFin$-algebra, which would rule out
too many interesting instances, e.g.~Examples~\ref{ex:alter} and~\ref{ex:segal}. We therefore obtain~$\M_*$ by a technique borrowed from higher-order programming language semantics and known as \emph{continuation passing style (CPS)} transformation~\cite{Plotkin75}. \SG{Before I forget: if $B$ is a schizophrenic object, the hom-functor $\Hom(-,B)$ generates a duality. Then the CPS transformation may have something to do with recognizability by monoids.}

Let $\alpha: TB \to B$ be a $\BBT$-algebra. We denote by $\BBT_B$ the
continuation monad (see Example~\ref{ex:mon}) with $T_B X = B^{B^X}$. We define
$\kappa:\BBT\to\BBT_{B}$ by sending $p\in TX$ to $\kappa_X(p)=\lambda f.\,
(\alpha \cdot Tf(p))\in T_B X$, which yields a monad morphism; in fact, it is
well known that for every monad $\BBT$ on a category with powers the above
assignment of $\kappa$ to $\alpha$ is part of a bijective correspondence between
Eilenberg-Moore algebras on $B$ and monad morphisms from $\BBT$ to $\BBT_B$ (see
e.g.~\cite[Theorem 3.2]{Kock70_2}). %

\begin{construction}
  Given a $\BBT$-automaton~\eqref{eq:aut}, let $\kappa: \BBT \to \BBT_B$ be the monad morphism given by $\kappa_X(p)=\lambda f.\, (\algstruc \cdot Tf(p))$ and let   
  \[
    o^{\M_{*}}=o^{\M},\qquad  t^{\M_{*}}=\kappa_X\cdot t^{\M}, 
    \qquad \algstrucs =\lambda t.\, t(\id),
  \] 
which yields a $\BBT_B$-automaton\footnote{We abuse terminology here since $\BBT_B$ is not finitary (see~Remark~\ref{rem:fincoref}).} $\M_{*}:X\to B\times (T_B X)^A$.  It is easily seen that $\algstrucs: T_B B \to B$ is a $\BBT_B$-algebra and $\algstruc=\algstrucs\cdot\kappa_B$. We call $\M_{*}$ the \emph{CPS-transform} of~\eqref{eq:aut}. 
\end{construction}

The following is another a corollary of Lemma~\ref{lem:2sem}. 
\begin{corollary}\label{cor:aut_cont}
The language semantics of a $\BBT$-automaton and of its CPS-trans\-form agree; more precisely, for every $\BBT$-automaton~\eqref{eq:aut} and a state $x \in X$, ${\sem{x}_{\M} = \sem{x}_{\M_{*}}}$.
\end{corollary}
Corollary~\ref{cor:aut_cont} implies Proposition~\ref{prop:fin_b} announced previously in Section~\ref{sec:react}. 
\begin{proof}[Proof of Proposition~\ref{prop:fin_b}]
If $B$ in~\eqref{eq:aut} is finite then, by definition, $T_B X$ is also finite. Thus, the generalized powerset construction performed on the CPS-transform $\M_*$ yields a Moore automaton. Thus, for every $x \in X$, $\sem{x}_\M = \sem{x}_{\M^*}$ is a regular formal power series.
\end{proof}
We now proceed with the definition of the observational semantics. In order to do this we shall make use of algebras for the \emph{countable multiset monad} $\BBM$. Its monad structure will not be needed; however, we recall its definition for the convenience of the reader.
\begin{remark}
  For the countable multiset monad $\BBM$, $MX$ consists of countable multisets on $X$,~i.e.
\[
MX = \{f: X \to \nat_\infty \mid \text{$f$ has countable support}\},
\]
where $\nat_\infty = \nat + \{\infty\}$ with the operations of
addition and multiplication extended to $\infty$ in the expected
way. The unit of $\BBM$ is given by
$\eta_X(x) = \delta_x: X \to \nat_\infty$ with $\delta_x(x) = 1$ and
$\delta_x(y) = 0$ otherwise. For any map $h: X \to MY$ the Kleisli lifting
$h^{\klstar}: MX \to MY$ acts as follows:
\[
  h^{\klstar}(f)(y) = \sum\nolimits_{x \in X} f(x) \cdot h(x)(y).
\]
\end{remark}
An $\BBM$-algebra is, equivalently, a commutative monoid with infinite summation satisfying the expected laws. We call such a monoid \emph{$\omega$-additive}. 
\iffull
For an $\omega$-additive monoid we denote by $\emptyset$ the neutral element, by $a+b$ the binary sum and by $\sum_{i=1}^\infty a_i$ the countable sum. 
\fi
\begin{definition}[$\omega$-additive $\BBT$-automata]
A $\BBT$\dash automaton~\eqref{eq:aut} is \emph{$\omega$-additive} if $B$ (besides being $\BBT$-algebra) is an $\omega$-additive monoid.
\end{definition}
It is easy to see that the $\omega$-additive monoid structure extends from $B$ to $T_BX$ pointwise:  
\begin{lemma}\label{lem:add}
If $B$ is an $\omega$-additive monoid and a $\BBT$-algebra then for
every set $X$, $T_B X$ is an $\omega$-additive monoid.
\end{lemma}
\begin{proof}
This follows from the fact that $B$ carries an Eilenberg-Moore
algebra structure for the countable multiset monad
$\BBM$. Equivalently, we have a monad morphism $m: \BBM \to \BBT_B$
(see~\cite[Theorem 3.2]{Kock70_2}). Thus, by forming $(T_B X,
\mu^{\BBT_B}_X \cdot m_{T_B X})$ we obtain an Eilenberg-Moore algebra
structure for $\BBM$ on $T_B X$, i.e., $T_B X$ is an $\omega$-additive
monoid. 
\end{proof}
The $\omega$-additive monoid structure on $T_B X$ allows us to define for any given $\BBT$-automaton over the alphabet $A_\tau$ a $\BBT_B$-automaton over $A$. To this end, we first form the CPS-transform of the given $\BBT$-automaton and then use infinite summation to get rid of unobservable actions $\tau$: 

\begin{construction}
Given a $\BBT$-automaton $\M: X \to B \times (TX)^{A_\tau}$, we construct $\M_{v}:X\to B\times (T_B X)^A$ with $\algstrucv=\algstrucs=\lambda t.\,t(\id)$ and with $t^{\M_v}$, $o^{\M_v}$ defined as
\begin{align*}
t^{\M_v}(x_0,a)=&\;\sum\nolimits_{i=1}^{\infty}
\letTerm{x_1\leteq t^{\M_*}(x_0,\tau);\ldots;x_{i-1}\leteq t^{\M_*}(x_{i-2},\tau)}{t^{\M_*}(x_{i-1},a)},
\\[-.3ex]
o^{\M_v}(x_0)=&\; o^{\M_*}(x_0)+\sum\nolimits_{i=1}^{\infty}
\bigl(\letTerm{x_1\leteq t^{\M_*}(x_0,\tau);\ldots}{t^{\M_*}(x_{i-1},\tau)}\bigr)(o^{\M_*}).
\end{align*}
(Note that $\letTerm{x_1\leteq t^{\M_*}(x_0,\tau);\ldots}{t^{\M_*}(x_{i-1},\tau)}$ is an element of $T_B X = B^{B^X}$, i.e.~a function $B^X \to B$ which can be applied to $o^{\M_*} \in B^X$.)
\end{construction} 
Intuitively, for $t^{\M_v}(x_0,a)$ we accumulate the effects underlying the $\tau$-transitions preceding the first $a$-transition; for $o^{\M_v}(x_0)$ we accumulate the effects along any sequence of $\tau$-transition leading to an accepting state detected by $o^{\M_*}$. 

We define the observational semantics for $\M$ to be the language semantics for $\M_v$.
\begin{definition}[Observational semantics]
  Given a $\BBT$-automaton~\eqref{eq:aut} over input
  alphabet $A_{\tau}$, its \emph{observational semantics} is defined as 
  $
  \sem{-}_\M^\tau = \sem{-}_{\M_v}.
  $
\end{definition}
In order to instantiate $\sem{-}^\tau_\M$ to concrete cases, we need to characterize
in a way similar to Lemma~\ref{lem:ind}. Before we state and prove it we make some
auxiliary observations.
\begin{remark}
  Since $\kappa: \BBT \to \BBT_B$ is a monad morphism we have
  that for every $f: X \to TY$: 
  \begin{equation}\label{eq:kappa}
    \kappa_Y \left(\letTerm{x \leteq p}{f(x)}\right) = \letTerm{x \leteq
      \kappa_X(p)}{\kappa_Y \cdot f(x)}.
  \end{equation}
\end{remark}
\begin{remark}
  We shall need two properties of the
  $\omega$-additive monoid structure on~$T_B X$. 
  \begin{cenumerate}%
    \renewcommand{\labelenumi}{(\arabic{enumi})}
  \item Kleisli substitution distributes over sums: 
    \begin{equation}\label{eq:sumkl}
      \letTerm{y \leteq \sum\nolimits_{i=1}^\infty p_i}{f(y)} = 
      \sum\nolimits_{i=1}^\infty\letTerm{y \leteq p_i}{f(y)}.
    \end{equation}
    Indeed, this equation expresses that the outside of the following
    diagram commutes for every $f: X \to T_B Y$ (here we abbreviate
    $T_B$ as $T$, and recall that $M$ denotes the countably supported
    multiset monad):
\begin{equation*}
\begin{tikzcd}[column sep=normal, row sep=normal] 
MTX
  \rar["m_X"]
  \dar[d, "Mf^\klstar"'] &
TTX
  \rar[r, "\mu_X"]
  \dar[d, "Tf^\klstar"] & 
TX 
  \dar[d, "f^\klstar"]\\
MTY
  \rar["m_Y"] &
TTY
  \rar[r, "\mu_Y"] &
TY
\end{tikzcd}
\end{equation*}
And this diagram clearly commutes by the naturality of the monad
morphism $m: \BBM \to \BBT_B$, and since~$f^\klstar$ is a $\BBT$-algebra morphism.
  \item Similarly, sums commute with the $\BBT_B$-algebra structure
    $\algstrucs$, i.e.~the following equation holds for every
    countable family of elements $p_i \in T_B B$:
    \begin{equation}\label{eq:aMstar}
      \algstrucs\Bigl(\sum\nolimits_{i=1}^\infty p_i\Bigr) = \sum\nolimits_{i=1}^\infty \algstrucs(p_i);
    \end{equation}
    in other words, $\algstrucs:T_B B\to B$ is a morphism of $\omega$-additive monoids. Indeed, this follows from the commutativity of the following diagram (again we abbreviate $T_B$ by~$T$):
\begin{equation*}
\begin{tikzcd}[column sep=normal, row sep=normal]
MTB
  \rar["m_{TB}"]
  \dar["M\algstrucs"'] &
TTB
  \dar["T\algstrucs"]
  \rar["\mu_B"] &
TB
  \dar["\algstrucs"] \\
MB
  \rar["m_B"] &
TB
  \rar["\algstrucs"] &
B
\end{tikzcd}
\end{equation*}
  \end{cenumerate}
\end{remark}
The following Lemma is obtained by combining~Lemma~\ref{lem:ind} with the above
properties~\eqref{eq:kappa}--\eqref{eq:aMstar}.

\begin{lemma}\label{lem:indtm}
Given a $\BBT$-automaton~\eqref{eq:aut}, $x_0\in X$ and $u\in A^*$ then 
\begin{align*}
\sem{x_0}^\tau_\M(\eps)=&~o^\M(x_0)~+ \sum\nolimits_{i=1}^{\infty} \algstruc\left(
  \letTerm{x_1 \leteq t^\M(x_0, \tau);\ldots;x_i \leteq t^\M(x_{i-1}, \tau)}{\eta^\BBT_B \cdot o^\M(x_i)}
\right)
\\[.5ex]
\sem{x_0}^\tau_\M(au) =&~\sum\nolimits_{i=1}^{\infty} \algstruc\left(
  \letTerm{x_1 \leteq t^\M(x_0, \tau);\ldots;x_i \leteq t^\M(x_{i-1}, a)}{\eta^\BBT_B \cdot \sem{x_i}^\tau_\M(u)}\right)
\end{align*} 
\end{lemma}

\begin{example}%
  We consider two concrete instances of our observational semantics. %
  \begin{cenumerate}
    \renewcommand{\labelenumi}{(\arabic{enumi})~~}
  \item Nondeterministic stack $\BBT$-automata $\M$ over $A_\tau$, i.e.~where $\BBT$ is the tensor product of the stack monad and $\PFin$, are in bijective correspondence with ordinary pushdown-automata (i.e.~nondeterministic ones with $\epsilon$-transitions). In fact, a similar construction to the one performed in the proof of Theorem~\ref{thm:bound} allows one to obtain for any given $\M$, $x_0 \in X$ and $\gamma_0 \in \Gamma$ a push-down automaton that accepts the language 
    \[
      \{ w \in A^* \mid \sem{x_0}_\M^\tau (w)(\gamma_0) = \top\}.
    \]
    Conversely, every pushdown automaton $M$ yields a nondeterministic stack $\BBT$-automaton such that the above language is the language accepted by $M$. It follows that the class of these languages is precisely the class of context-free languages over $A$. 
  \item Coming back to Example~\ref{expl:nd_act} let us consider valence automata again, but now \emph{with} $\epsilon$-transitions. Given any valence automaton $\CA = (X, M, A, \delta, q_0, F)$ we can again assume w.l.o.g.~that its transitions are labelled with a single letter or $\epsilon$. Then we can regard $\CA$ as a $\BBT$-automaton $\M$ over $A_\tau$ for the nondeterministic monoid action theory over $M$. Using Lemma~\ref{lem:indtm} it is not difficult to prove that $\{w \in A^* \mid 1 \in \sem{q_0}_\M^\tau\}$ is the language accepted by $\CA$.
  \end{cenumerate}
\end{example}

We now proceed to define a class of $\BBT$-automata that correspond to classical Turing machines in the sense that the observational semantics yields precisely all recursively enumerable languages.
\begin{definition}[Tape $\BBT$-automaton]\label{dfn:ta}
  A \emph{tape $\BBT$-automaton} is a $\BBT$-automaton~\eqref{eq:aut} for
  which
  \iffull
  \begin{itemize}
  \item $\BBT$ is the tape monad over $\Gamma$ (see Definition~\ref{defn:tape});
  \item $B$ is the set of predicates over $\int \times
    \Gamma^\int$ consisting of all those $p \in 2^{\int \times
    \Gamma^\int}$ for each of which there is a $k$ such
  that $p(i, \sigma) = p(i, \sigma')$ and $p(i,\sigma_{+j})=p(i+j,\sigma)$ whenever $\sigma \equiv \sigma'\pmod{[i-k,i+k]}$;
  \item $\algstruc: TB \to B$ is given by evaluation; it restricts the morphism
      $T(2^S) = (2^S \times S)^S \xrightarrow{\mathsf{ev}^S} 2^S$,
    where $S = \int \times \Gamma^\int$.
  \end{itemize}
  \else
  $\BBT$ is the tape monad over $\Gamma$; $B$ is the set of predicates over $\int \times \Gamma^\int$ consisting of all those $p \in 2^{\int \times \Gamma^\int}$ for each of which there is a $k$ such that $p(i, \sigma) = p(i, \sigma')$ and $p(i,\sigma_{+j})=p(i+j,\sigma)$ if $\sigma =_{i \pm k} \sigma'$; and $\algstruc: TB \to B$ is given by evaluation as in~Definition~\ref{defn:stack}.
  \fi
\end{definition}
The argument showing that $B$ is indeed a $\BBT$-algebra is completely analogous to the one for stack $\BBT$-automata (Definition~\ref{defn:stack}).

Tape $\BBT$-automata over $A_\tau$ are essentially deterministic $2$-tape Turing machines with input alphabet $A$, 
where the first tape is read-only and traversed in on direction only as the machine 
runs. Thus, we obtain that tape automata recognize all the recursively enumerable languages.
\begin{theorem}\label{thm:re}
  For every tape $\BBT$-automaton~$\M$ over $A_\tau$, $\Gamma$ with
  $|\Gamma| \geq 2$ containing a special blank symbol $\blk$, and
  every state $x \in X$ the following language is recursively
  enumerable:
  \[
    \{w \in A^* \mid \sem{x}^\tau_\M(w)(0,\sigma_\blk) = \top\},
  \]
  where $\sigma_\blk$ is the constant function
  returning~$\blk$. Conversely, every recursively enumerable language
  can be represented in this way.
\end{theorem}
In order to prove this theorem, we will 
relate tape automata and a special form of Turing machines called 
\emph{online Turing machines}. The idea of an online Turing machine is a rather
old one~\cite{Hennie66} and essentially amounts to equipping a standard (offline) Turing
machine with an additional input tape which can only be read in one direction and
not modified. From the coalgebraic point of view online Turing machines naturally
extend finite state machines and push-down automata.
\begin{definition}[Online Deterministic Turing Machine (ODTM)]
  An \emph{online deterministic Turing machine} is a six-tuple $M = (Q, A,
  \Gamma, \delta, q_0, F)$, where $Q$ is a finite set of states, $A$ is
  the action (or input) alphabet, $\Gamma$ is the tape alphabet (assumed
  to contain the special \emph{blank} symbol $\blk$), $q_0$
  is the initial state, $F \subseteq Q$ is a set of final (or
  accepting) states and 
  \[
  \delta: Q  \times (A \cup \{\,\tau\,\}) \times \Gamma \to Q \times \Gamma \times \{\,\L,\N,\R\,\}
  \]
  is the transition function.
\end{definition}
The difference to an ordinary TM is that transitions do not only
depend on the tape contents but also on an input in the form of an
action $a \in A$ given by the user from the outside during runtime of
the machine, and there are also \emph{internal} transitions,
i.e.~where a silent action $\tau$ triggers the transition. Hence, a
\emph{configuration} of an ODTM $M$ is an element of
$Q \times A^* \times (\int \times \Gamma^\int)$
consisting of the current state $q \in Q$ the remaining input actions
$w \in A^*$ and a pair $(i,\sigma)$ consisting of the current position $i$ of the read/write head and tape
content $\sigma:\int \to \Gamma$. \emph{Computations} (or \emph{runs}) are then
defined in the usual way as sequences of configurations starting from
the initial configuration $(q_0, w, (0, \sigma_\blk))$ where $w \in
A^*$ is the input word and $\sigma_\blk$ denotes the constant function on $\blk$. Note that internal transitions leave the
remaining input actions untouched while otherwise the head symbol is
removed from $w \in A^*$ in a configuration.
\begin{remark}
The above definition is essentially the one from~\cite{Aanderaa74}. A nondeterministic
variant of this definition has been recently employed by Baeten et al.~\citeyear{BaetenLuttikEtAl11} under the name \emph{reactive Turing machine}
with the aim to equip TM's with a notion of interaction and so bridge the gap 
between classical computation and concurrency theory. In particular, the standard 
equivalence relation for reactive Turing machines is \emph{bisimilarity} rather than 
\emph{language equivalence} we study here. 
\end{remark}

\begin{definition}[Language of a ODTM]
  Let $M$ be an ODTM. The formal language accepted by $M$ is the set
  of words $w \in A^*$ such that there exists a computation from the
  initial configuration to a configuration $(q, \eps, (n, \sigma))$ with
  $q \in F$. 
\end{definition}
More informally, a word is accepted by $M$ if there is a computation
that consumes all the letters in the input work $w$ and leads to an accepting
state. Note that due to the internal actions there may be several 
accepting computations of a word. So an ODTM is only deterministic in
the sense that in every configuration there can be no two different
moves consuming the same input letter. But an internal transition can
happen nondeterministically in any configuration. 

That ODTM's are an appropriate model of computations is stated by the
following lemma.

\begin{lemma}\label{lem:ODTM-lem}
  The class of languages accepted by ODTM's is the class of
  semi-decidable languages.
\end{lemma}

\begin{proof}[Proof of Theorem~\ref{thm:re}]
  We give for a tape automaton $\M$ as in the statement of the theorem an
  equivalent ODTM and vice versa. 

  (a)~Given $\M$, we define an ODTM $M$. For every $x \in X$ and $a \in
  A$ let $k_{x,a}$ be the minimal natural number as in Definition~\ref{defn:tape} for 
  $t^\M (x,a) = \langle r, z, t\rangle \in TX$.
  Analogously, let $l_x$ be the minimal natural number according to the second clause of Definition~\ref{dfn:ta} for 
  $o^\M(x): \int \times \Gamma^\int \to 2$.
  The state set of $M$ consists of the states $X$ of $\M$ times a
  finite memory that can store a finite portion of $M$'s tape and is
  of the form
  \[
    \{-n, \ldots, 0, \ldots, n\} \times \Gamma^{2n+1},
    \qquad \text{where $n = \max\{l_x, \max \{k_{x,a} \mid a \in A\}\}$}.
  \]
  We say that a memory content $(0,\bar\sigma)$ \emph{restricts}
  $(i,\sigma) \in \int \times \Gamma^\int$ if
  $\bar\sigma(j) = \sigma(i+j)$ for all
  $j = -n, \ldots, 0, \ldots, n$. The final states of $M$ are those
  states $x \in X$ together with memory contents $(0, \bar\sigma)$
  that restrict $(i,\sigma)$ with $o^\M(x)(i,\sigma) = \top$; that this
  is well-defined follows from Definition~\ref{defn:tape}. We now
  describe informally how $M$ simulates $\M$. Since $\M$ can access
  several symbols from the tape at once we need to simulate
  transitions of $\M$ by several steps of $M$. These steps will make
  sure that the contents of $M$'s finite memory always restricts its
  tape contents. Hence, a transition of $\M$ given by
  $t^\M(x,a)(i,\sigma) = (x', i', \sigma')$ is simulated by the
  following steps of $M$ (where $M$ starts in state $x$ with the
  memory contents $(0,\bar\sigma)$ restricting $M$'s tape content
  $(i,\sigma)$):

  \begin{enumerate}[wide]
    \renewcommand{\labelenumi}{(\arabic{enumi})}
  \item $M$ performs a transition that consumes the input letter $a$ and
    changes the state to $x'$ and the memory content to the appropriate value $(j, \bar\sigma')$ that reflects the values of $i'$ and $\sigma'$, i.e.~$j = i'-i$ and $\bar\sigma'(\ell) = \sigma'(i+\ell)$ for every $\ell = -n, \ldots, 0, \ldots, n$ (this is possible by Definition~\ref{defn:tape});
  \item now $M$ replaces the $2n+1$ tape cells around the current position of the read/write head according to $\bar\sigma'$ from the memory content and then the read/write head's position is changed according to $j$ (this uses a finite number of additional auxiliary states);
  \item finally, the memory is overwritten with the $2n+1$ tape symbols around the new position of the read/write head so that the computation of the $\M$-transition ends in state $x'$ with a memory content $(0, \bar\sigma)$ restricting the new tape content $(i',\sigma')$.
  \end{enumerate}
  Note that all the above points except~(1) are realized by internal
  transitions of $M$. 

  Now we need to prove that $M$ accepts a word $w \in A$ from the
  initial state $x_0$ (with memory content $(0, \sigma_\blk)$ iff
  $\sem{x_0}^\tau_\M (w) (0, \sigma_\blk) = \top$. We will prove more
  generally that for every state $x_0$, we have
  $\sem{x_0}^\tau_\M(w)(z_0,\sigma_0) = \top$ iff there exists an
  accepting $M$-computation from state $x_0$ starting with tape
  content $(z_0, \sigma_0)$. 

  Before we proceed with the proof recall that the $\BBT$-algebra
  structure $\algstruc: TB \to B$ is given by evaluation. It follows that
  for every map $f: X \to TB$ the uncurrying of $\algstruc \cdot f: X \to B
  \subseteq 2^S$ is
  \begin{align*}
    X \times S \xrightarrow{~~~f'~~}  
    B \times S \subseteq 2^S \times S \xrightarrow{~~\mathsf{ev}~~} 
    2,
  \end{align*}
  where $S = \int \times \Gamma^\int$ and $\mathsf{ev}$ is the evaluation map.

  Now we prove the desired statement by induction on $w$. For the base
  case observe that, by Lemma~\ref{lem:indtm},
  $\sem{x_0}^\tau_\M(\eps)(z_0,\sigma_0) = \top$ iff $o^\M(x_0)(i_0, \sigma_0) = \top$ or
  there exists an $i \geq 1$ such that 
  \[
  \begin{array}{c}
    \algstruc\bigl(\letTerm{x_1 \leteq t^\M(x_0, \tau);\ldots;x_i \leteq
        t^\M(x_{i-1}, \tau)}{\eta^\BBT_B \cdot o^\M(x_i)}\bigr)(z_0, \sigma_0) = \top.
  \end{array}
  \]
  In the first case $x_0$ is a final state of $M$ and so the empty
  ($0$-step) computation of $M$ is an accepting $M$-computation of
  $\eps$. In the second case, let $i$ be such that the above equation
  holds. Equivalently, the following morphism
  \begin{align*}
    X \times S \xrightarrow{~~~} 
    ~\cdots~\xrightarrow{~~~} X \times S \xrightarrow{~~o^\M\times S~~}
    B \times S \subseteq 2^S \times S
    \xrightarrow{\mathsf{~~ev~~}} 2,
  \end{align*}
  where the unlabelled arrows form the $i$-fold composition of the
  uncurrying of $t^\M(-,\tau): X \to (X \times S)^S$, maps $(z_0,
  \sigma_0)$ to $1$. So equivalently, we have $x_1, \ldots, x_i$ and
  tape configurations $(z_k, \sigma_k)$, $1 \leq k \leq i$, such that
  $t^\M(x_k, \tau)(z_k, \sigma_k) = (x_{k+1}, z_{k+1}, \sigma_{k+1})$
  for all $0 \leq k < i$, and $o^\M(x_i)(z_i,\sigma_i) = \top$.
  Equivalently, we have an $M$-computation that performs steps
  (1)--(3) above $i$ times (simulating $\tau$-steps of $\M$) and ends
  in the accepting state $x_i$ with tape content $(z_i, \sigma_i)$.

  In the induction step of our proof let $w = au$. By
  Lemma~\ref{lem:indtm}, we have $\sem{x_0}^\tau_\M(au)(z_0, \sigma_0)
  = \top$ iff there exists an $i \geq 1$ such that 
  \begin{displaymath}
      \algstruc\bigl(
        \letTerm{x_1 \leteq t^\M(x_0, \tau);\ldots;x_i \leteq t^\M(x_{i-1}, a)}{\eta^\BBT_B \cdot \sem{x_i}^\tau_\M(u)}
      \bigr)(z_0,\sigma_0) = \top.
  \end{displaymath}
  By a similar argument as in the base case, this is equivalent to the
  existence of states $x_1, \ldots, x_i$ and tape content $(z_k,
  \sigma_k)$, $1 \leq k \leq i$, such that $t^\M(x_k, \tau)(z_k,
  \sigma_k) = (x_{k+1}, z_{k+1}, \sigma_{k+1})$ for all $0 \leq k <
  i-1$, $t^\M(x_{i-1},a) (z_{i-1},\sigma_{i-1}) = (x_i, z_i,
  \sigma_i)$ and $\sem{x_i}^\tau_\M(u)(z_i,\sigma_i) = \top$. The last
  conditions corresponds, by induction hypothesis, bijectively to an
  accepting $M$-computation from state $x_i$ with initial tape content
  $(z_i, \sigma_i)$. And the rest corresponds bijectively to an
  $M$-computation that consists of $i$-iterations of steps (1)--(3)
  simulating $i-1$ many $\tau$-steps and one $a$-step of the given
  tape automaton $\M$ starting in state $x_0$ with tape
  content $(z_0, \sigma_0)$ and ending in state $x_i$ with tape content
  $(z_i,\sigma_i)$. Putting these two parts together, we obtain the
  desired bijective correspondence to an accepting $M$-computation from
  state $x_0$ with initial tape content $(z_0, \sigma_0)$.

  (b)~Conversely, given an ODTM $M = (Q, A, \Gamma, \delta, q_0, F)$ we construct an equivalent
  tape automaton~$\M$. We take $Q$ as the set of states and
  we let
  \[
  o^\M(q)(z,\sigma) = \top \iff q \in F\text{~~~and~~~}  t^\M(q,a)(z,\sigma) = (q', z',\sigma'),
  \]
  where $q'$ and $(z',\sigma')$ are the state and the tape content, respectively, of $M$ after
  performing an $a$-transition in state $q$ with tape content
  $(z,\sigma)$. For internal transitions, $t^\M(q, \tau)$ is
  defined analogously. 

  We need to prove that $M$ accepts a word $w \in A$ iff
  $\sem{q_0}^\tau_\M (w) (0, \sigma_\blk) = \top$. More generally, one
  proves that for every state $q_0$ and tape content
  $(z_0,\sigma_0)$ of $M$ one has $\sem{q_0}^\tau_\M(w)(z_0,\sigma_0) =
  1$ iff there exists an accepting $M$-computation from state $q_0$
  with initial tape content $(z_0, \sigma_0)$. This is proved by
  induction on $w$ once again. The details are similar (but slightly
  easier) than in part~(a) of our proof, and so we leave them as an
  easy exercise for the reader.
\end{proof}

\section{Conclusions and Future Work} In the present paper, we have
presented the first steps towards a uniform theory of effectful state
machines combining Moore automata with computational monads. We have
given a coalgebraic account of several types of state machines with
effects (such as manipulation of a store, their accepted languages and
syntactic expressions to specify them). We have presented several
results of our theory including a generic Kleene-style theorem
(Theorem~\ref{thm:kleene})%
\iffull\ and one-direction of a Chomsky-Sch\"utzenberger-style theorem
(Theorem~\ref{thm:chsh})\fi. We have also given the first treatment of
Turing machines in a coalgebraic setting: the observational language
semantics of tape automata yields precisely the recursively enumerable
languages.

There are several possible directions for future work. A converse to
Theorem~\ref{thm:chsh} is of interest. In addition, we plan to derive
a sound calculus of reactive expressions
extending~\cite{BonsangueMiliusEtAl13} and explore the boundaries for
completeness. Such a calculus will depend on the monad $\BBT$ and its
algebra $B$; in fact, while currently we only need the signature
$\Sigma$ and the algebra $B$ for our results, the axioms of the theory
presenting $\BBT$ will become laws of the calculus. Note that
completeness is only possible for specific choices of $\BBT$ and $B$,
for it follows from Corollary~\ref{cor:nre} that for the
nondeterministic stack theory and $B$ from
Definition~\ref{defn:mstack} a finite complete axiomatization is not
possible.

Another avenue is capturing further language and
complexity classes, such as the context-sensitive languages, using
$\BBT$-automata. This will prospectively result in standard tools such as
bisimulation proof methods becoming available for those classes of
machines and their language semantics. Hence, further investigations
into such proof principles are of interest.

\begin{acks}
We thank the anonymous reviewers for their very careful reading of our manuscript and for their suggestions to improve the presentation. 

Sergey Goncharov and Stefan Milius acknowledge support by the \grantsponsor{DFG}{German Research Foundation (DFG)}{https://www.dfg.de/} 
under Grants No.:~\grantnum{DFG}{GO~2161/1-2} and~\grantnum{DFG}{MI~717/5-2} respectively.
Alexandra Silva's work is partially supported by the ERC Starting Grant ProFoundNet 
No.:~\grantnum{ERC}{679127} and EPSRC Standard Grant CLeVer No.:~\grantnum{EPSRC}{EP/S028641/1}. 
\end{acks}
  
\bibliographystyle{ACM-Reference-Format}
\bibliography{monads,chomsky}

\clearpage
\appendix
\allowdisplaybreaks

\section{Omitted Proofs}

\subsection{Proof of Proposition~\ref{prop:test}}
Let $f$ and $g$ be the transition structures of $X$ and $Y$,
respectively.  Since both $\widehat f$ and $\widehat g$ are
$\Lfun$-coalgebra morphisms, we have $o(z) = o(\widehat f(z))$ and
$\widehat f(\partial_a(z)) = \partial_a(\widehat f(z))$ for every
$a\in A$ and similarly for $\widehat g$. By an easy induction, the
latter equation yields $\widehat f(\partial_w(z)) =
\partial_w(\widehat f(z))$ for every $z \in X$ and $w \in A^*$.
Therefore,
\begin{align*}
  o(\partial_w(x))&\,=o(\widehat f(\partial_w(x))) = o(\partial_w(\widehat f(x))) = \widehat f(x)(w),\\
  o(\partial_w(y))&\,=o(\widehat g(\partial_w(y))) = o(\partial_w(\widehat g(y))) = \widehat g(y)(w),
\end{align*}
where the last equations easily follow from the definitions of $o$ and
$\partial_w$ on $\nu L = B^{A^*}$.

Now note that $x\sim y$ iff $\widehat f(x)=\widehat g(y)$ and the latter
holds iff $\widehat f(x),\widehat g(y):A^*\to B$ are equal on every
$w\in A^*$. Thus we conclude that
$o(\partial_w(x))=o(\partial_w(y))$ iff $x\sim y$ as desired.
\qed

\subsection{Proof of Lemma~\ref{lem:subalg}}
For every set $X$ take the factorization of $\alpha_X \cdot i_X$ into a surjective map $\alpha_X': T'X \to P'X$ followed by an injective map (inclusion) $j_X: P'X \to PX$. Using the diagonal fill-in property of image factorizations, it is easy to verify that $\alpha'$ and $j$ form natural transformations. Define $\eta_X': X \to P'X$ as the composition of $\eta_X: X \to T'X$ and $\alpha_X': T'X \to P'X$ and $\mu_X': P'P'X \to P'X$ as the unique diagonal fill-in below (here $*$ denotes the usual horizontal composition of natural transformations):
\begin{equation*}
\begin{tikzcd}[column sep=large, row sep=normal]
T'T'X \rar["(\alpha' * \alpha')_X"] \dar["\alpha_X'\cdot\mu_X"']
  & 
P'P'X \dar["\mu_X^P\cdot(j * j)_X"] \ar[dl, dashed, "\mu_X'"']\\
P'X \rar["j_X"] & PX
\end{tikzcd}
\end{equation*}
Indeed, $(\alpha' * \alpha')_X = T'\alpha'_X \cdot \alpha_{P'X}'$ is surjective since $T'$ preserves surjections, and the outside square clearly commutes (using that $\alpha\cdot i$ is a monad morphism): 
\[
\mu_X^P \cdot (j * j)_X \cdot (\alpha'*\alpha')_X = \mu_X^P \cdot ((\alpha\cdot i) * (\alpha\dot i))_X = (\alpha\cdot i)_X \cdot \mu_X = j_X \cdot \alpha_X' \cdot \mu_X.
\]
Using the unique diagonal fill-in property, it is now an easy exercise to verify that $\eta'$ and $\mu'$ are natural, that $(P', \eta', \mu')$ satisfies the monad laws and that $\alpha'$ and $j$ are monad morphisms. 
\takeout{ %
Let for any $X$, $P'X=\alpha(T'(X))$. We then define $\eta:X\to P'X$ as the composition of $\eta:X\to T'X$ and $\alpha'_X:T'X\to P'X$. Then for any $f:X\to P'Y$ and any $p\in P'X$ let $f^{\klstar}(p)=\alpha'(g^\klstar(q))$ where $q\in T'X$ is such that $\alpha'(q)=p$ and $g:X\to T'Y$ is such that $f=\alpha g$. Such $q$ and $g$ exist because $\alpha'$ is componentwise epic by definition and $f$ is uniquely defined because for any $h:X\to T'Y$ and any $r:T'X$ if $q=\alpha'(r)$ and $f=\alpha' h$ then $\alpha'(g^\klstar(q))=(\alpha g)^\klstar(\alpha q)=(\alpha h)^\klstar(\alpha r)=\alpha'(h^\klstar(r))$. By definition $\alpha'$ is a monad morphism. }%
\qed

\subsection{Proof of Lemma~\ref{lem:stack_sem}}
Using the semantics of $pop$, for any $1\leq i\leq n$ and any $w\in\Gamma^*$, 
\begin{align*}
\tsem{p_i}(w)
=\tsem{pop(p_1,\ldots,p_n,p)}(\gamma_i w)
=\tsem{pop(q_1,\ldots,q_n,q)}(\gamma_i w)
=\tsem{q_i}(w).
\end{align*}
Analogously, one proves $\tsem{p}(\eps)=\tsem{q}(\eps)$. 

In order to prove $\tsem{p}(w) = \tsem{q}(w)$ for all words
$w \in \Gamma^*$, we use that neither $p$ nor $q$ contain $pop$,
i.e.~both of them are nested applications of $push$ (with various
indices) to some variables. Using the above semantics it is easy to
calculate that for any $w\in\Gamma^*$,
\[
\tsem{p}(w)=\brks{x_1,u_1 w}\qquad\text{and}\qquad\tsem{q}(w)=\brks{x_2,u_2 w}
\] 
for some $x_1, x_2 \in X$ and $u_1, u_2 \in \Gamma^*$ that do not depend on $w$. By substituting $w$ with $\eps$ and using $\tsem{p}(\eps) = \tsem{q}(\eps)$ we obtain $x_1=x_2$ and $u_1=u_2$. It follows that $\tsem{p}(w)=\tsem{q}(w)$ holds for all $w \in \Gamma^*$ as desired.
\qed

\subsection{Proof of Lemma~\ref{lem:pm}}
We need to prove that $\theta(i) = \theta'(i)$ for all $i \in J$. If $J =
\emptyset$, we are done. So let $i\in J$. If $i \in I$, we are done
since $\theta \equiv \theta' \pmod{I}$. Otherwise we have $i \not\in I$ and obtain
\[
  \theta(i) = \rho(i) = \rho'(i) = \theta'(i)
\]
by using the third, first, and last of the given equivalences.
\qed

\subsection{Proof of Lemma~\ref{lem:tape_rule}}
Indeed we have
\begin{flalign*}
  &&s =&\; \tmread_k(\tmwrite_{1,k}(s),\ldots,\tmwrite_{n,k}(s))&&\by{{\bf (mv-l)}, {\bf(mv-r)}, {\bf(rd-wr)}}\\
  &&=&\; \tmread_k(\tmwrite_{1,k}(t),\ldots,\tmwrite_{n,k}(t))&&\by{premises}\\
  &&=&\; t&&\by{{\bf (mv-l)}, {\bf(mv-r)}, {\bf(rd-wr)}}
\end{flalign*}
\qed

\subsection{Proof of Lemma~\ref{lem:tape_more}}
Equation~\eqref{eq:tape_more1} is shown as follows:
\begin{flalign*}
&&\tmwrite_{i,k}(&\tmwrite_{j,k}(x))\\
&&=&\;\tmmove_k(\tmwrite_{i}(\tmmove_{\mkah}(\tmmove_k(\tmwrite_{j}(\tmmove_{\mkah}(x))))))&&\by{definition}\\
&&=&\;\tmmove_k(\tmwrite_{i}(\tmwrite_{j}(\tmmove_{\mkah}(x))))&&\by{\textbf{(mv-l)},~\textbf{(mv-r)}}\\
&&=&\;\tmmove_k(\tmwrite_{j}(\tmmove_{\mkah}(x)))&&\by{\textbf{(wr-wr)}}\\
&&=&\;\tmwrite_{j,k}(x).&&\by{definition}
\end{flalign*}
Analogously one obtains~\eqref{eq:tape_more3} using~\textbf{(wr-rd)}. Let us show~\eqref{eq:tape_more2}:
\begin{flalign*}
&&\tmwrite_{i,k}(&\tmwrite_{j,{k'}}(x))\\
&&=&\;\tmmove_k(\tmwrite_{i}(\tmmove_{\mkah}(\tmmove_{k'}(\tmwrite_{j}(\tmmove_{\mkah'}(x))))))&&\by{definition}\\
&& =&\;\tmmove_k(\tmwrite_{i}(\tmmove_{k'-k}(\tmwrite_{j}(\tmmove_{\mkah'}(x)))))&&\by{\textbf{(mv-l)},~\textbf{(mv-r)}}\\
&& =&\;\tmmove_{k'}(\tmmove_{k-k'}(\tmwrite_{i}(\tmmove_{k'-k}(\tmwrite_{j}(\tmmove_{\mkah'}(x))))))&&\by{\textbf{(mv-l)},~\textbf{(mv-r)}}\\
&& =&\;\tmmove_{k'}(\tmwrite_{j}(\tmmove_{k-k'}(\tmwrite_{i}(\tmmove_{k'-k}(\tmmove_{\mkah'}(x))))))&&\by{\textbf{(wr-mv)}}\\
&& =&\;\tmmove_{k'}(\tmwrite_{j}(\tmmove_{k-k'}(\tmwrite_{i}(\tmmove_{\mkah}(x)))))&&\by{\textbf{(mv-l)},~\textbf{(mv-r)}}\\
&& =&\;\tmmove_{k'}(\tmwrite_{j}(\tmmove_{\mkah'}(\tmmove_{k}(\tmwrite_{i}(\tmmove_{\mkah}(x))))))&&\by{\textbf{(mv-l)},~\textbf{(mv-r)}}\\
&& =&\;\tmwrite_{j,k'}(\tmwrite_{i,k}(x)).&&\by{definition}
\end{flalign*}
Finally, let us show~\eqref{eq:tape_more4}. To this end, apply
$\tmwrite_{j,k'}$ to both sides of the identity and simplify the
result. For the left-hand side of the equation we obtain
\begin{flalign*}
&&\tmwrite_{j,k'}(&\tmwrite_{i,k}(\tmread_{k'}(r_1,\ldots,r_n)))\\
&&=&\;\tmwrite_{i,k}(\tmwrite_{j,k'}(\tmread_{k'}(r_1,\ldots,r_n)))&&\by{\eqref{eq:tape_more2}}\\
&&=&\;\tmwrite_{i,k}(\tmwrite_{j,k'}(r_j)),&&\by{\eqref{eq:tape_more3}}
\end{flalign*}
and for the right-hand side,
\begin{flalign*}
&&\tmwrite_{j,k'}(&\tmread_{k'}(\tmwrite_{i,k}(r_1),\ldots,\tmwrite_{i,k}(r_n)))\\
&&=&\;\tmwrite_{j,k'}(\tmwrite_{i,k}(r_j))&&\by{\eqref{eq:tape_more3}}\\
&&=&\;\tmwrite_{i,k}(\tmwrite_{j,k'}(r_j)).&&\by{\eqref{eq:tape_more2}}
\end{flalign*}
We are now done by Lemma~\ref{lem:tape_rule}, since the desired
equation holds when $\tmwrite_{j,k'}$ is applied to both sides for
every $k' \in \int$.
\qed

\subsection{Full Proof of Proposition~\ref{prop:aexp}}
(1)~Let $\Sigma$ be the signature of the $\Sigma$-theory of\/ $\BBT$.  
First, we observe that $\AExp{\Sigma}{B_0}$ clearly carries a $\Sigma$-algebra structure. Moreover, it also carries an $L$-transition structure.
In order to define it we first define an auxiliary normalization
function $\norm$ on (not necessarily closed) additive expressions as follows:
{
\begin{flalign*}
\quad\norm(f(e_1, \ldots, e_n)) = f(\norm(e_1), \ldots, \norm(e_n))\qquad (f\neq +)&&
\norm(p+q) = p\qquad (\norm(q)=\emptyset)\quad\\[.5ex]
\quad\norm(p+q) = \norm(p)+\norm(q)\qquad (\norm(p)\neq\emptyset,~\norm(q)\neq\emptyset)&&
\norm(p+q) = q\qquad (\norm(p)=\emptyset)\quad\\[-4.5ex]
\end{flalign*}
\begin{gather*}
\norm(\mu x.e) = \mu x.\, \norm(e)\qquad
\norm(a.e) = a. \norm(e)\qquad
\norm(p) = p\qquad (\text{$p$ a variable or $p\in B_0$}) 
\end{gather*}

}
Then we inductively define the $L$-transition structure on $\AExp{\Sigma}{B_0}$:
\begin{flalign*}
o(b) =&~b^B& 			o(\mu x.\, e) =&~o(e[\mu x.\, e/x])& o(a_i.e) =&~{\emptyset}^B\\
\partial_{a_i}(b) =&~\emptyset& \partial_{a_i}(\mu x.\, e) =&~\partial_{a_i}(e[\mu x.\, e/x])&
\partial_{a_i}(a_i.e) =&~\norm(e),~\partial_{a_i}(a_j.e)=\emptyset && (i\neq j)
\end{flalign*}\vspace{-4.5ex}
\begin{flalign*}
o(f(e_1, \ldots, e_n)) = f^B(o(e_1),\ldots, o(e_n))&&
\partial_{a_i}(f(e_1, \ldots, e_n)) = \norm(f(\partial_{a_i}(e_1),\ldots, \partial_{a_i}(e_n)))
\end{flalign*}
Our usage of $\norm$ here is merely a technical trick to keep the proof elementary. Note that the clauses for $\mu x.\, e$ are well-founded due to guardedness. 

We record the following simple properties of $\norm$:
\begin{align}
\norm(\norm(p)) =&\;\norm(p)\label{eq:alpha_id1}\\
\norm(p+q) =&\;\norm(\norm(p)+\norm(q))\label{eq:alpha_id2}\\
\norm(e[\mu x.\,t/y]) =&\; \norm(e)[\norm(\mu x.\,t)/y]\label{eq:subst}\\
\norm(\partial_a(p))=&\;\partial_a(\norm(p))\label{eq:alpha_d}
\end{align}
where $p,q\in\AExp{\Sigma}{B_0}$. Identity~\eqref{eq:alpha_id1}
follows by structural induction over $p$. The only nontrivial case is
$p=p_1+p_2$ with $\norm(p_1)\neq\emptyset$ and
$\norm(p_2)\neq\emptyset$ (note that in the third step below we use
that, by induction, $\norm(\norm(p_i))=\norm(p_i)\neq\emptyset$):
\begin{flalign*}
&&\norm(\norm(p))=&\;\norm(\norm(p_1+p_2))\\
&&=&\;\norm(\norm(p_1)+\norm(p_2))&\by{def.~of~$\norm$}\\
&&=&\;\norm(\norm(p_1))+\norm(\norm(p_2))&\by{def.~of~$\norm$,~\eqref{eq:alpha_id1}}\\
&&=&\;\norm(p_1)+\norm(p_2)&\by{\eqref{eq:alpha_id1}}\\
&&=&\;\norm(p_1+p_2)&\by{def.~of~$\norm$}\\
&&=&\;\norm(p).
\end{flalign*}
Identity~\eqref{eq:alpha_id2} then follows
from~\eqref{eq:alpha_id1} by case distinction: it is obvious if
$\norm(p)=\emptyset$ or $\norm(q)=\emptyset$, otherwise
$\norm(p+q)=\norm(\norm(p+q))=\norm(\norm(p)+\norm(q))$. Identity~\eqref{eq:subst}
is a restricted form of substitution lemma, which can as usual be
established by induction over the context~$e$ and the proof relies
both on~\eqref{eq:alpha_id1} and~\eqref{eq:alpha_id2}. Note, however
that in our setting it does not hold more generally, e.g.\ with
$e=b+y$,
$\norm(e[\emptyset/y])=b\neq
b+\emptyset=\norm(e)[\norm(\emptyset)/y]$.
Finally, identity~\eqref{eq:alpha_d} follows from the previous identities by induction over $p$, 
in particular, the most difficult case $p = \mu x.\,e$ requires~\eqref{eq:subst}:
\begin{flalign*}
&&\norm(\partial_a(p)) &= \norm(\partial_a(\mu x.\,e))&\\
&&&=\norm(\partial_a(e[p/x])) &\by{def.~of $\partial_a$}\\
&&&=\partial_a(\norm(e[p/x])) & \by{ind.~hypothesis} \\
&&&=\partial_a(\norm(e)[\norm(p)/x]) & \by{\eqref{eq:subst}}\\
&&&=\partial_a(\mu x.\,\norm(e)/x]) & \by{def.~of~$\norm$} \\
&&&=\partial_a(\norm(p)).&\by{def.~of $\partial_a$}
\end{flalign*}
Another case of interest in proving~\eqref{eq:alpha_d} is $p = p_1 + p_2$ under
$\norm(p_1) \neq \emptyset \neq \norm(p_2)$:
\begin{flalign*}
&&  \partial_a(\norm(p_1 + p_2)) & = \partial_a(\norm(p_1) + \norm(p_2)) & \by{def.~of $\norm$}\\
&&  & = \norm(\partial_a(\norm(p_1)) + \partial_a(\norm(p_2))) & \by{def.~of $\partial_a$}\\
&&  & = \norm(\norm(\partial_a(p_1)) + \norm(\partial_a(p_2))) & \by{ind.~hypothesis} \\
&&  & = \norm(\partial_a(p_1) + \partial_a(p_2)) & \by{\eqref{eq:alpha_id2}} \\
&&  & = \norm(\norm(\partial_a(p_1) + \partial_a(p_2))) & \by{\eqref{eq:alpha_id1}} \\
&&  & = \norm(\partial_a(p_1 + p_2)). & \by{def.~of $\partial_a$}
\end{flalign*}

(2)~By Definition~\ref{defn:tsem}, the above $L$-coalgebra structure on
$\AExp{\Sigma}{B_0}$ induces a language semantics; again we
write $\sem{e}$ for the formal power series denoted by
$e \in \AExp{\Sigma}{B_0}$. Let us show that this semantics agrees
with the semantics of $\Exp{\Sigma}{B_0}$, that is
$\sem{e}=\sem{\otr(e)}$ with $e\in\Exp{\Sigma}{B_0}$ and
$\otr: \Exp{\Sigma}{B_0} \to \AExp{\Sigma}{B_0}$ defined inductively
as follows:
\begin{align*}
\otr(f(e_1, \ldots, e_n)) =&~\norm(f(\otr(e_1), \ldots, \otr(e_n))),& \otr(x) =&~ x,\\
\otr(\mu x.\,a_1.e_1\pitchfork\ldots\pitchfork a_{n}.e_n\pitchfork s)
=&~\mu x.\,\norm(a_1.\otr(e_1)+\ldots+ a_{n}.\otr(e_n)+ \otr(s)),& \otr(b) =&~ b.
\end{align*}
Note that $s$ in the bottom left equation is an arbitrary term in the theory of\/
$\BBT$ according to the $\beta$-clause of the grammar in
Definition~\ref{def:react}. In fact, the above assigments define
$\otr$ on expressions containing free variables and according to the
$\gamma$ and $\beta$-clauses of Definition~\ref{def:react}. By case
distinction it is straightforward to prove that for every (not
necessarily closed) $e$ we have
\begin{align}\label{eq:alpha_tr}
\norm(\otr(e)) = \otr(e).
\end{align}
Moreover, we have the following property
\begin{align}\label{eq:tr_subst}
\otr(e[t/x]) = \norm(\otr(e)[\otr(t)/x]).
\end{align}
The proof of the latter is essentially straightforward but quite tedious.
\smnote{But see the handwritten notes in the svn!} 
In order to show the desired equation $\sem{e}=\sem{\otr(e)}$,
by Proposition~\ref{prop:test}, it suffices to check that $\otr$ is an
$L$-coalgebra homomorphism, i.e.
\begin{flalign*}
&& \partial_{a_i}(\otr(e)) =\otr(\partial_{a_i}(e))  \quad\text{and}\quad o(\otr(e)) = o(e) &&\text{($a_i \in A$, $e\in\Exp{\Sigma}{B_0}$)}
\end{flalign*}
This again follows by induction over the number of clauses recursively applied to define $o(e)$ and $\partial_{a_i}(e)$
and the proof relies on~\eqref{eq:alpha_id1}--\eqref{eq:alpha_d}. E.g.\ for $e = f(e_1, \ldots, e_n)$ we calculate
\begin{flalign*}
&&\partial_{a_i}(\otr(f(e_1,\ldots,e_n))) =&~ \partial_{a_i}(\norm(f(\otr(e_1), \ldots, \otr(e_n)))) & \by{def.~of $\otr$} \\
&&=&~ \norm(\partial_{a_i}(f(\otr(e_1), \ldots, \otr(e_n)))) & \by{\eqref{eq:alpha_d}} \\
&&=&~ \norm(\norm(f(\partial_{a_i}(\otr(e_1)), \ldots, \partial_{a_i}(\otr(e_n))))) & \by{def.~of $\partial_{a_i}$} \\
&&=&~ \norm(f(\partial_{a_i}(\otr(e_1)), \ldots, \partial_{a_i}(\otr(e_n)))) & \by{\eqref{eq:alpha_id1}} \\
&&=&~ \norm(f(\otr(\partial_{a_i}(e_1)), \ldots, \otr(\partial_{a_i}(e_1)))) & \by{ind.~hypothesis} \\
&&=&~ \otr(f(\partial_{a_i}(e_1), \ldots, \partial_{a_i}(e_n))) & \by{def.~of $\otr$} \\
&&=&~ \otr(\partial_{a_i}(f(e_1,\ldots,e_n))), & \by{def.~\eqref{eq:partial} of $\partial_{a_i}$ on $\Exp{\Sigma}{B_0}$}
\end{flalign*}
\begin{flalign*}
  &&o(\otr(f(e_1,\ldots,e_n)))
  =&~ o(\norm(f(\otr(e_1),\ldots,\otr(e_n)))) & \by{def.~of $\otr$}\\
  &&=&~ o(f(\norm(\otr(e_1)), \ldots, \norm(\otr(e_n)) &\by{def.~of $\norm$}\\
  &&=&~ o(f(\otr(e_1),\ldots,\otr(e_n))) & \by{\eqref{eq:alpha_tr}}\\
  &&=&~ f^B(o(\otr(e_1)),\ldots,o(\otr(e_n))) & \by{def.~of $o$}\\
  &&=&~ f^B(o(e_1),\ldots,o(e_n)) & \by{induction hypothesis}\\
  &&=&~ o(f(e_1,\ldots,e_n)).  & \by{def.~of $o$}\\
\end{flalign*}

The remaining clauses do not cause any trouble and are handled in a
similar fashion. For example, for
$e= \mu x.\,a_1.e_1\pitchfork\ldots\pitchfork a_{n}.e_n\pitchfork s$
we have by the definition of $o$
\[
o(\mu x.\,a_1.e_1\pitchfork\ldots\pitchfork a_{n}.e_n\pitchfork s) = s^B.
\]
Starting at the right-hand side we have
\begin{flalign*}
&&  o(\otr(&\mu x.\,a_1.e_1+\ldots+ a_{n}.e_n+ s) \\
&& & = o(\mu x.\,\underbrace{\norm(a_1.\otr(e_1)+\ldots+a_n.\otr(e_n) + \otr(s))}_{\text{\small $t$}}) & \by{def.~of $\otr$} \\
&& & = o(\norm(a_1.\otr(e_1)+\ldots+a_n.\otr(e_n) + \otr(s))[\mu x.\,t/x]).  & \by{def.~of $o$}
\end{flalign*}

If $\norm(\otr(s)) = \emptyset$ then the latter evaluates to
\[
o(a_1.\norm(\otr(e_1))[\mu x.\, t/x] + \ldots + a_n.\norm(\otr(e_n))[\mu x.\, t/x]) = {\emptyset}^B = s^B,
\]
using the definition of $\norm$ for the first equation, and the
fact that $\norm(\otr(s)) = \emptyset$ implies $s=\emptyset+\cdots+\emptyset$
for the second equation.

If $\norm(\otr(s)) \neq \emptyset$ then, analogously,
\begin{flalign*}
&& o(\norm(&a_1.\otr(e_1)+\ldots+a_n.\otr(e_n) + \otr(s))[\mu x.\,t/x])\\
&& & = o(a_1.\norm(\otr(e_1))[\mu x.\, t/x] + \ldots + a_n.\norm(\otr(e_n))[\mu x.\, t/x] + \norm(\otr(s))) & \by{def.~of $\norm$} \\
&& & = (\otr(s))^B = s^B,
\end{flalign*}
where the last step is established by an easy induction (over terms
$s$ according to the $\beta$-clause in Definition~\ref{def:react}). %

Finally, we calculate:
\begin{flalign*}
&&\otr(\partial_{a_i}(&\mu x.\,a_1.e_1\,\pitchfork\ldots\pitchfork a_{n}.e_n\pitchfork b))\\ 
&&=&~\otr(e_i[\mu x.\,a_1.e_1\pitchfork\ldots\pitchfork a_{n}.e_n\pitchfork b/x])& \by{def.~of $\partial_{a_i}$}\\
&&=&~\norm(\otr(e_i)[\mu x.\,\norm(a_1.\otr(e_1)+\ldots+ a_{n}.\otr(e_n)+ b)/x])& \by{\eqref{eq:tr_subst}}\\
&&=&~\norm(\partial_{a_i}(\mu x.\,\norm(a_1.\otr(e_1)+\ldots+ a_{n}.\otr(e_n)+ b)))& \by{def.~of $\partial_{a_i}$}\\ %
&&=&~\partial_{a_i}(\norm(\mu x.\,\norm(a_1.\otr(e_1)+\ldots+ a_{n}.\otr(e_n)+ b)))& \by{\eqref{eq:alpha_d}}\\
&&=&~\partial_{a_i}(\otr(\mu x.\,a_1.e_1\pitchfork\ldots\pitchfork a_{n}.e_n\pitchfork b)).& \by{def.~of $\otr$}
\end{flalign*}

(3)~In order to prove the desired converse in the statement of the
proposition, we define a translation map
$\tr\colon \AExp{\Sigma}{B_0} \to \Exp{\Sigma}{B_0}$. To that end,
we first define an auxiliary map $\bar o$ on every expression
according to~\eqref{eq:addexp} that is guarded in each of its
variables; $\bar o$ works similarly as $o$ but without interpreting
$\emptyset$, $f$ and $b$ in $B$, whence delivering a term in the
theory of $\BBT$ according to the $\beta$-clause of
Definition~\ref{def:react}:
\[
\begin{array}{r@{~}c@{~}l@{\qquad}r@{~}c@{~}l}
  \bar o(b) &=& b & \bar o(\mu x.\,e) &=& \bar o(e[\mu x\,e./x]) \\
  \bar o(a.e) &=& \emptyset & \bar o(f(e_1, \ldots, e_n)) &=& f(\bar o(e_1), \ldots, \bar o(e_n))
\end{array}
\]
Then $\bar o(e)$ is well-defined by guardedness of $e$. Similarly, we
define auxiliary maps $a^{\mone}$ completely similarly as
$\partial_{a}$; however, $a^{\mone}$ can be applied to expressions $e$
containing free variables but which are still guarded in each of their
variables. That means we do not (need to) define $a^{\mone}$ on variables~$x$. Now we define $\tr$ (on not necessarily closed expressions) as follows:
\begin{align*}
\tr(x) =&~x,\\
\tr(b) =&~\mu x.\,a_1.\emptyset \pitchfork\ldots\pitchfork a_{n}.\emptyset \pitchfork b, \\
\tr(a_i.e) =&~\mu x.\,a_1.\emptyset \pitchfork\ldots\pitchfork a_i .\tr(e) \pitchfork\ldots\pitchfork a_{n}.\emptyset \pitchfork \emptyset, \\
\tr(f(e_1, \ldots, e_n)) =&~f(\tr(e_1), \ldots, \tr(e_n)),\\
\tr(\mu x.\,e) =&~\mu x.\,a_1.\tr(a_1^{\mone}(e))\pitchfork\ldots\pitchfork a_{n}.\tr(a_n^{\mone}(e))\pitchfork \bar o(\mu x.\,e).
\end{align*}
Before we proceed we first need a substitution lemma similar to~\eqref{eq:tr_subst}: %
\begin{equation}\label{eq:tr_subst2}
\tr(e[t/x]) = \tr(e)[\tr(t)/x].
\end{equation}
We deduce $\sem{e}=\sem{\tr(e)}$ for any $e\in\AExp{\Sigma}{B_0}$ from
\[
o(\tr(e)) = o(e) \quad\text{and}\quad
\partial_a(\tr(e)) = \tr(\partial_a(e))
\ \text{~for every $a \in A$}.
\]
We have, e.g.\ for $e=\mu x.\,t$,
\begin{flalign*}
&&\partial_{a_i}(\tr (e))
=\;&\partial_{a_i}(\mu x.\,a_1.\tr(a_1^{\mone}(t))\pitchfork\ldots\pitchfork a_{n}.\tr(a_n^{\mone}(t))\pitchfork o(e))\\
&&=\;& \tr(a_i^{\mone}(t))[\tr(e)/x]\\
&&=\;& \tr(a_i^{\mone}(t)[e/x]) & \by{\eqref{eq:tr_subst2}}\\
&&=\;& \tr(\partial_{a_i}(t[e/x])) & \by{guardedness}\\
&&=\;& \tr(\partial_{a_i}(e)).
\end{flalign*}
The remaining cases are verified routinely.\smnote{See the notes in the svn for these cases!}
\qed

\subsection{Proof of Lemma~\ref{lem:ind}}
Recall that the transition structure $\iota$ in~\eqref{diag:hatm} arises from
  $o: B^{A^*} \to B$ and $\partial_a: B^{A^*} \to B^{A^*}$ with
  $o(\sigma) = \sigma(\eps)$ and $\partial_a(\sigma) = \lambda w.\,
  \sigma(aw)$. Thus, we obtain the semantics map
\begin{displaymath}
\sem{-}_\M = \bigl(X \xrightarrow{~~\eta_X~~} TX \xrightarrow{~~\widehat
    \M^\sharp~~} B^{A^*}\bigr).
\end{displaymath}
The commutativity of~\eqref{diag:hatm} can now equivalently be restated as the two equations 
\[
  o(\sem{x}_\M) = o^\M(x), \qquad
  \partial_a (\sem{x}_\M) =  \widehat\M^\sharp(t^\M(x,a))\qquad
  \text{for every $x  \in X$ and $a \in A$}.
\]
The left equation implies the left of~\eqref{eq:sem_ind} since
$o(\sem{x}_\M) = \sem{x}_\M(\eps)$. For the second statement notice
first that by the freeness of $TX$ we have that $\widehat\M^\sharp$ is
the unique $\BBT$-algebra morphism extending $\sem{-}_\M$. Thus, we
have
  \begin{align*}
  \widehat \M^\sharp = \alpha \cdot T\sem{-}_\M:TX\to B^{A^*},
  \end{align*}
  where $\alpha$ is the $\BBT$-algebra structure on $B^{A^*}$. Observe
  that $\alpha: T(B^{A^*}) \to B^{A^*}$ is given pointwise, i.\,e.~$\alpha$ is the unique morphism satisfying 
  \begin{align*}
  \ev_u \cdot \alpha = 
  \bigl(
    T(B^{A^*})\xrightarrow{~~T\ev_u~~} TB \xrightarrow{~~\algstruc~~} B \bigr),
  \end{align*}
  for every $u \in A^*$, where $\ev_u: B^{A^*} \to B$ is the obvious evaluation at $u
  \in A^*$: $\ev_u(f) = f(u)$. It follows that for every word
  $u \in A^*$ we have
  \begin{align*}
  \widehat \M^\sharp (-)(u) = \bigl(
    TX 
    \xrightarrow{~~T(\ev_u \cdot \sem{-}_\M)~~} TB \xrightarrow{~~\algstruc~~} B \bigr);
   \end{align*}
    indeed we have:
    \begin{flalign*}
      \widehat\M^\sharp(-)(u) =  \ev_u \cdot \widehat\M^\sharp 
       =  \ev_u \cdot \alpha \cdot T\sem{-}_\M 
       =  \algstruc \cdot T\ev_u \cdot T\sem{-}_\M 
       =  \algstruc \cdot T(\ev_u \cdot \sem{-}_\M)
    \end{flalign*}
and therefore
    \begin{flalign*}
      &&\sem{x}_\M(au) = &\, \partial_a(\sem{x}_\M)(u) & \by{definition of $\partial_a$}\\
      && = &\, \widehat\M^\sharp(t^\M(x,a))(u) & \by{\eqref{diag:hatm}} \\
      && = &\, (\algstruc \cdot T(\ev_u \cdot \sem{-}_\M)) (t^\M(x,a)).
    \end{flalign*}
    The last line is the desired right-hand side of the right equation in~\eqref{eq:sem_ind}.
\qed

\subsection{Proof Details of Proposition~\ref{prop:stens}}
\begin{enumerate}
\item Let us show the equivalence of \eqref{eq:stack-trans-ext}
  and~\eqref{eq:stack-trans}.

The implication~\eqref{eq:stack-trans-ext} $\impl$~\eqref{eq:stack-trans} 
is obvious. For the converse one, let $k$ be as in~\eqref{eq:stack-trans}, let  
$s,u\in \Gamma^*$ and let $|s|\geq k$. Then $s = s'w$ for suitable $s'\in\Gamma^k$,
$w\in\Gamma^*$, and
\begin{align*}
 p(su) =\;&p(s'wu)\\
=\;&\letTerm{\brks{x,s''}\leteq p(s')}{\eta_{X\times\Gamma^*}\brks{x,s''wu}}\\
=\;&\letTerm{\brks{x,s''}\leteq(\letTerm{\brks{x,s''}\leteq p(s')}{\eta_{X\times\Gamma^*}\brks{x,s''w}})}{\eta_{X\times\Gamma^*}\brks{x,s''u}}\\
=\;&\letTerm{\brks{x,s''}\leteq(\letTerm{\brks{x,s''}\leteq p(s'w)}{\eta_{X\times\Gamma^*}\brks{x,s''}})}{\eta_{X\times\Gamma^*}\brks{x,s''u}}\\
=\;&\letTerm{\brks{x,s''}\leteq p(s)}{\eta_{X\times\Gamma^*}\brks{x,s''u}}.
\end{align*}
We next check that~\eqref{eq:stack-trans} does indeed identify a submonad
of $(T(-\times\Gamma^\star))^{\Gamma^\star}$. First,
for any $x\in X$, $p=\eta_X(x)$ satisfies~\eqref{eq:stack-trans} with $k=0$. 
Then, for every $f:X\to (T(Y\times\Gamma^*))^{\Gamma^*}$, such that for every 
$x\in X$, $f(x)$ satisfies~\eqref{eq:stack-trans} with some~$k_x$, and for every 
$p:\Gamma^*\to T(X\times\Gamma^*)$, satisfying~\eqref{eq:stack-trans} with some $k$, we 
must show that $f^\klstar(p)$ also satisfies~\eqref{eq:stack-trans}. Note that
for $s\in\Gamma^k$, 
\begin{align*}
f^\klstar(p)(su)
=&\;\letTerm{\brks{x,s'}\leteq p(su)}{f(x)(s')}\\
=&\;\letTerm{\brks{x,s'}\leteq (\letTerm{\brks{x,s'}\leteq p(s)}{\eta_{X\times\Gamma^*}\brks{x,s'u}})}{f(x)(s')}\\
=&\;\letTerm{\brks{x,s'}\leteq p(s)}{f(x)(s'u)}.
\end{align*}
Since by assumption, $\BBT$ is finitary, for some finite $X'\subseteq X$ and 
$m\in\nat$, $p(s)\in T(X'\times\Gamma^*)$. By~\eqref{eq:stack-trans-ext}, 
for $\hat k=\max\{ k_x\mid x\in X'\}$, and $u\in \Gamma^{\hat k}$, we continue as follows:
\begin{align*}
f^\klstar(p)(suw)
=&\;\letTerm{\brks{x,s'}\leteq p(s)}{f(x)(s'uw)}\\
=&\;\letTerm{\brks{x,s'}\leteq p(s);\brks{y,s''}\leteq f(x)(s'u)}{\eta_{Y\times\Gamma^*}\brks{y,s''w}}\\
=&\;\letTerm{\brks{y,s''}\leteq (\letTerm{\brks{x,s'}\leteq p(s)}{f(x)(s'u)})}{\eta_{Y\times\Gamma^*}\brks{y,s''w}}\\
=&\;\letTerm{\brks{y,s'}\leteq f^\star(p)(su)}{\eta_{Y\times\Gamma^*}\brks{y,s'w}}.
\end{align*}
That is, we have proven~\eqref{eq:stack-trans} for $f^\klstar(p)$ with $k+\hat k$.

\item The calculation showing that~\eqref{eq:pop-pop-sigma} is sound wrt to $\CE\tensor\CT$ is as follows: 
\begin{align*}
  pop(&x_1,\ldots,x_n,f(y_1,\ldots,pop(z_1,\ldots,z_n,z),\ldots, y_m))\\
  =&\; pop(x_1,\ldots,x_n,f(pop(push_1(y_1),\ldots,push_n(y_1),y_1),\ldots,\\
& \hspace{2.8cm} pop(z_1,\ldots,z_n,z),\ldots,\\
& \hspace{2.8cm} pop(push_1(y_m),\ldots,push_n(y_m),y_m)))\\
  =&\; pop(x_1,\ldots,x_n,pop(f(push_1(y_1),\ldots,z_1,\ldots,push_1(y_m)), \ldots,\\
& \hspace{3.1cm} f(push_1(y_1),\ldots,z_n,\ldots,push_1(y_m)), f(y_1,\ldots,z,\ldots,y_m)))\\
  =&\; pop(x_1,\ldots,x_n,f(y_1,\ldots,z,\ldots, y_m)).
\end{align*}
\item 
In the completeness part we used the fact that for normal $s=f(s_1,\ldots,s_m)$ and 
$t=g(t_1,\ldots,t_l)$, such that each of the $s_1,\ldots,s_m,t_1,\ldots,t_l$ is 
either a variable or has an operation of the stack theory at the~top, if 
$s_{j}=pop(\ldots,s')$ for some $j \in \{1, \ldots, m\}$ then the equations
\begin{equation*}
  \begin{array}{r@{\,}c@{\,}l}
  s & = & pop(push_1(s),\ldots, push_n(s), f(s_1,\ldots,s',\ldots,s_m)),
  \\
  t &= & pop(push_1(t),\ldots, push_n(t), t).
\end{array}
\end{equation*}
belong to $\CE\tensor\CT$. This is shown as follows.
Using the laws of $\BBR$ we have that
\begin{flalign*}
&&s &=\; f(s_1, \ldots, s_j, \ldots, s_m)\\
&&& =\; f(s_1, \ldots, pop(\ldots, s'), \ldots, s_m) \\
&&& =\; f(pop(push_1(s_1),\ldots, push_n(s_1),s_1), \ldots,  pop(\ldots, s'), \ldots,\\
&&& \phantom{\ = f(} pop(push_1(s_m),\ldots, push_n(s_m),s_m)) & \by{\bf (pop-push)}\\
&&& =\; pop(f(push_1(s_1), \ldots, push_1(s_m)), \ldots, \\*
&&& \phantom{\ = pop(}f(push_n(s_1), \ldots, push_n(s_m)), f(s_1, \ldots, s', \ldots, s_m)). & \by{tensor law}
\end{flalign*}
Now substitute the last term for the right-hand $s$ in 
\[
s = pop(push_1(s), \ldots, push_n(s), s)
\]
and use {\bf (pop-pop)} and {\bf (pop-push)} to conclude 
\begin{equation*}
  \begin{array}{r@{\,}c@{\,}l}
  s & = & pop(push_1(s),\ldots, push_n(s), f(s_1,\ldots,s',\ldots,s_m)),
  \\
  t &= & pop(push_1(t),\ldots, push_n(t), t).
\end{array}
\end{equation*}
\end{enumerate}

\subsection{Proof of Lemma~\ref{lem:indtm}}
  We proceed by induction over the argument $w \in A^*$ of
  $\sem{x_0}^\tau_\M$. For $w = \eps$:
\begin{flalign*}
\qquad    \sem{x_0}^\tau_\M (\eps) & = \sem{x_0}_{\M_v} (\eps) && \by{definition of $\sem{-}^\tau_\M$} \\[1ex]
    & = o^{\M_v}(x_0) && \by{Lemma~\ref{lem:ind}} \\
    & = \makebox[0pt][l]{$o^{\M_*}(x_0) + \sum\nolimits_{i=1}^{\infty}\left(\letTerm{x_1 \leteq t^{\M_*}(x_0, \tau);\ldots}{t^{\M_*}(x_{i-1}, \tau)}\right)(o^{\M_*})$} \\
    &&&  \by{definition of $o^{\M_v}$} \\
    & = \makebox[0pt][l]{$o^{\M}(x_0) + \sum\nolimits_{i=1}^{\infty}\kappa_X\bigl(\letTerm{x_1 \leteq t^{\M}(x_0, \tau);\ldots}{t^{\M}(x_{i-1}, \tau)}\bigr)(o^{\M})$} \\
    & & & \by{repeated application of~\eqref{eq:kappa}}\\
    & & & \by{with $o^{\M_*} = o^\M$, $t^{\M_*} = \kappa_X \cdot t^\M$} \\
    & = \makebox[0pt][l]{$o^{\M}(x_0) + \sum\nolimits_{i=1}^{\infty}(\algstruc \cdot To^\M)\bigl(\letTerm{x_1 \leteq t^{\M}(x_0, \tau);\ldots}{t^{\M}(x_{i-1}, \tau)}\bigr)$}\\
    & & & \by{definition of $\kappa_X$} \\
    & = \makebox[0pt][l]{$o^{\M}(x_0) + \sum\nolimits_{i=1}^{\infty} \algstruc \bigl(\letTerm{x_1 \leteq t^{\M}(x_0, \tau);\ldots}{}
         x_i \leteq t^{\M}(x_{i-1}, \tau); \eta^\BBT_B\cdot o^\M(x_i)\bigr)$}\\
    & && \by{property of $\mathsf{do}$-notation}
\end{flalign*}
  For the induction step we consider $w = au$ and compute:
{\allowdisplaybreaks
  \begin{flalign*}
\qquad \sem{x_0}^\tau_\M(au)
    &= \sem{x_0}_{\M_v} (au) && \by{definition of $\sem{-}^\tau_\M$} \\[1ex]
    & =  \makebox[0pt][l]{$\algstrucv \bigl(\letTerm{y \leteq t^{\M_v}(x_0, a)}{\eta^\BBT_B \cdot \sem{y}_{\M_v}(u)} \bigr)$} \\
    & & & \by{Lemma~\ref{lem:ind}} \\
    & =  \makebox[0pt][l]{$\algstrucs \bigl(\letTerm{x_i \leteq \bigl(\sum\nolimits_{i=1}^\infty \letTerm{x_1\leteq t^{\M_*}(x_0,\tau);\ldots}{t^{\M_*}(x_{i-1},a)}\bigr)}{\eta^{\BBT_B}_B \cdot \sem{x_i}_{\M_v}(u)}\bigr)$} \\
    & & & \by{definition of $t^{\M_v}$, since $\algstrucv = \algstrucs$,} \\
    & & & \by{and renaming $y$ to $x_i$} \\
    & =  \makebox[0pt][l]{$\algstrucs \bigl(\sum\nolimits_{i=1}^\infty \letTerm{x_1\leteq t^{\M_*}(x_0,\tau);\ldots; x_i \leteq t^{\M_*}(x_{i-1},a)}{\eta^{\BBT_B}_B \cdot \sem{x_i}_{\M_v}(u)}\bigr)$} \\
    & & & \by{\eqref{eq:sumkl}} \\
    & =  \makebox[0pt][l]{$\algstrucs \bigl(\sum\nolimits_{i=1}^\infty \kappa_B\bigl(\letTerm{x_1\leteq t^{\M_*}(x_0,\tau);\ldots; x_i \leteq t^{\M_*}(x_{i-1},a)}{\eta^{\BBT}_B \cdot \sem{x_i}_{\M_v}(u)}\bigr)\bigr)$} \\
    & & & \by{\eqref{eq:kappa} and since $\kappa \cdot \eta^{\BBT} = \eta^{\BBT_B}$} \\
    & =  \makebox[0pt][l]{$\sum\nolimits_{i=1}^\infty \algstrucs \cdot \kappa_B \bigl(\letTerm{x_1\leteq t^{\M_*}(x_0,\tau);\ldots;x_i \leteq t^{\M_*}(x_{i-1},a)}{\eta^{\BBT}_B \cdot \sem{x_i}_{\M_v}(u)}\bigr)$} \\
    & & & \by{\eqref{eq:aMstar}} \\
    & =  \makebox[0pt][l]{$\sum\nolimits_{i=1}^\infty \algstruc \bigl(\letTerm{x_1\leteq t^{\M_*}(x_0,\tau);\ldots; x_i \leteq t^{\M_*}(x_{i-1},a)}{\eta^{\BBT}_B \cdot \sem{x_i}_{\M_v}(u)}\bigr)$} \\
    & & & \by{since $\algstrucs \cdot \kappa_B = \algstruc$} \\
    & =  \makebox[0pt][l]{$\sum\nolimits_{i=1}^{\infty} \algstruc\bigl(\letTerm{x_1 \leteq t^\M(x_0, \tau);\ldots;x_i \leteq t^\M(x_{i-1}, a)}{\eta^\BBT_B \cdot \sem{x_i}^\tau_\M(u)}\bigr)$}\\
    & & & \by{definition of $\sem{-}^\tau_\M$.} \hspace*{2.3cm} \text{\qed}
  \end{flalign*}
}

\subsection{Proof of Lemma~\ref{lem:ODTM-lem}}
  We show that an ordinary TM can be simulated by an ODTM and vice
  versa. 

  (a)~Given an ODTM $M$ it can be simulated by a nondeterministic TM
  $\bar M$ with two tapes as follows: the first (input) tape of $\bar
  M$ stores the input word $w \in A^*$ which is processed read-only
  from left to right, and the second tape of $\bar M$ corresponds to
  the tape of $M$. The NTM $\bar M$ simulates $M$ as follows: in each
  step $\bar M$ nondeterministically either performs an internal
  action of $M$ or reads one symbol from the first tape (then moving
  the head to the right by one position on this tape). In addition,
  $\bar M$ has a special accepting halting state $q_f$, and it can
  nondeterministically decide to move to that state from every accepting
  state of $M$ whenever a blank symbol is read on the first tape; this
  allows $\bar M$ to halt and accept if $M$ is in any accepting
  configuration after consuming its input. It is then clear that $\bar
  M$ and $M$ accept the same language. We conclude that the language accepted by any
  ODTM is semi-decidable.

  (b)~Conversely, suppose we have a deterministic TM with input
  alphabet $A$. Then $M$ can be simulated by an ODTM $\bar M$. The
  computation of $\bar M$ has two phases: in the first phase $\bar M$
  consumes its entire input and writes it on its tape. During this
  phase no internal transitions happen. The first phase ends as soon
  as $\bar M$ performs its first internal action, which starts
  the second phase. In this phase $\bar M$ only performs internal
  actions in the sense that all transitions consuming an input symbol
  $a \in A$ lead to a non-accepting state that is never left
  again. At the beginning of the second phase $\bar M$ then moves the
  head to the first input symbol (if any) on its tape. It then starts
  a simulation of the DTM $M$ using internal transitions only. Whenever $M$
  halts in a (non-)accepting state, then $\bar M$ moves to a
  (non-)accepting state that it never leaves again. Again, $\bar M$
  clearly accepts the same language as $M$. Thus, it follows that
  every semi-decidable language is accepted by an ODTM. 
\qed
\end{document}

